\documentclass[acmsmall]{acmart}

\AtBeginDocument{%
  }

\setcopyright{rightsretained}
\copyrightyear{2025}
\acmYear{2025}
\acmDOI{XXXXXXX.XXXXXXX}

\usepackage{xspace}
\usepackage{url}
\usepackage{adjustbox}
\usepackage{tcolorbox}

\theoremstyle{definition}
\newtheorem{definition}{Definition}

\newcommand{\ie}{\emph{i.e.,}\xspace}
\newcommand{\eg}{\emph{e.g.,}\xspace}

\begin{document}


\title{A Survey of Real-World Recommender Systems: Challenges, Constraints, and Industrial Perspectives}

\author{Kuan Zou}
\orcid{0009-0004-0353-1904}
\email{zouk0002@e.ntu.edu.sg}
\author{Aixin Sun}
\orcid{0000-0003-0764-4258}
\email{axsun@ntu.edu.sg}
\affiliation{%
  \institution{Nanyang Technological University}
  \city{Singapore}
  \country{Singapore}
}

\begin{abstract}
Recommender systems have generated tremendous value for both users and businesses, drawing significant attention from academia and industry alike. However, due to practical constraints, academic research remains largely confined to offline dataset optimizations, lacking access to real user data and large-scale recommendation platforms. This limitation reduces practical relevance, slows technological progress, and hampers a full understanding of the key challenges in recommender systems. 
In this survey, we provide a systematic review of industrial recommender systems and contrast them with their academic counterparts. We highlight key differences in data scale, real-time requirements, and evaluation methodologies, and we summarize major real-world recommendation scenarios along with their associated challenges. We then examine how industry practitioners address these challenges in \textit{Transaction-Oriented Recommender Systems} and \textit{Content-Oriented Recommender Systems}, a new classification grounded in item characteristics and recommendation objectives. Finally, we outline promising research directions, including the often-overlooked role of user decision-making, the integration of economic and psychological theories, and concrete suggestions for advancing academic research. 
Our goal is to enhance academia's understanding of practical recommender systems, bridge the growing development gap, and foster stronger collaboration between industry and academia.
\end{abstract}

\ccsdesc[500]{Information systems~Recommender systems}

\keywords{Industrial Recommender Systems, Practical Challenges, Academic–Industry Gap}

\maketitle

\section{Introduction}
\label{sec:intro}

In recent years, the widespread adoption of mobile App/Internet has brought about a surge in fragmented user sessions and more frequent interactions with online services. The explosion of content has made it increasingly important for information to proactively find its target audience. As a result, recommendation technologies have emerged as a crucial solution.

We have witnessed this trend across various domains. E-commerce platforms such as Amazon~\cite{xian2021ex3,chen2024shopping} and Alibaba~\cite{10.1145/3459637.3482292,jiang2022triangle}, content platforms like YouTube~\cite{yi2023online,shao2024optimizing}, HBO Max~\cite{edizel2024towards} and Netflix~\cite{tong2023navigating}, and music services such as Spotify~\cite{baran2023accelerating,nazari2022choice} have all experienced significant growth driven by various recommendation strategies. More notably, short video platforms such as Douyin and Tiktok~\cite{farias2023correcting}, Kuaishou~\cite{pan2023learning,deng2024mmbee}, and Instagram Reels~\cite{yao2024user} have reshaped mobile Internet consumption habits, with recommender systems at their core.
Furthermore, it is worth noting that performance-driven online advertising, one of the most important revenue streams in the Internet economy, is fundamentally built upon recommendation technologies. In this sense, recommender systems have become not only indispensable to today’s digital ecosystem, but also a technological cornerstone in the history of Internet development.

Recommender systems (RecSys) have been studied for decades. Even today, we continue to see a steady stream of novel approaches being published at top-tier conferences and journals. However, whether these  methods from academic research can be directly applied or adopted with minimal effort in real-world industrial settings remains a contentious issue. More fundamentally, the objectives pursued by academia and industry often differ in nature.
Academic research typically emphasizes \textit{methodological innovation}. A model that demonstrates superior algorithmic performance on public datasets, measured by  precision, recall, or NDCG, is often deemed a valuable contribution. In contrast, industrial applications are driven by \textit{business-oriented metrics}: whether a method leads to tangible or implicit gains in revenue, user engagement, or system efficiency. This creates a very different evaluation landscape.
Industrial deployment requires careful consideration of engineering constraints, system cost, and performance trade-offs. Moreover, beyond offline validation, models must undergo rigorous A/B testing in online environments to verify real-world effectiveness. This gives rise to a fundamental tension: academic research often emphasizes advancing algorithmic performance under substantially relaxed constraints, whereas industrial systems prioritize holistic business impact within a complex landscape of operational, financial, and organizational constraints.
These differences are further amplified by disparities in data scale, real-time responsiveness, deployment constraints, and even baseline definitions. As a result, academic and industrial communities frequently differ in how they define and assess the \textit{effectiveness} of recommender systems.

To help academic researchers better understand the practical challenges of recommender systems  in real-world deployments, and to guide developers entering industry to grasp the most advanced techniques, we conduct \textit{a comprehensive review of industrial recommendation papers} published over the past five years (2020 - 2024).
Our survey summarizes the common challenges faced by industrial recommender systems, including large-scale data processing, real-time constraints, response latency, and the cost of training and inference. In addition, we decompose the ecosystem into concrete application domains under two main categories: Transaction-Oriented RecSys (\eg e-commerce) and Content-Oriented RecSys (\eg video, news, and music), and analyze their unique business objectives and algorithmic strategies.
Finally, we identify promising avenues for future research in RecSys academic research, notably user behavior understanding and theory-guided optimization in production environments—objectives that we believe represent some of the most meaningful pursuits in the field.

\subsection{Differences from Previous Surveys}

As a relatively mature research field, recommender systems have seen numerous insightful surveys. Some provide comprehensive discussions of specific technological paradigms~\cite{zhang2019deep} and methodological approaches~\cite{li2024recent}, while others thoroughly examine various application domains~\cite{lu2015recommender}. Such surveys greatly facilitate readers in quickly grasping particular advancements within the field. However, we observe a clear absence of systematic reviews focused specifically on \textit{real-world recommender systems}. Notably, the practical challenges, application scenarios, methodological considerations, and evaluation criteria in industry differ markedly from their academic counterparts. Yet, these industrial aspects and associated constraints have not been systematically analyzed in the existing literature. After all, most academic research on recommender systems is motivated or inspired by, and ultimately aims to inform, practical systems deployed in the real world.

We thus identify a critical gap:  a strong necessity to bridge the understanding between academia and industry. The research trajectories of academia and industry in recommender systems are increasingly divergent—a trend more pronounced here than in many other technological domains. Our work aims precisely at addressing this gap, offering a comprehensive analysis of industrial practices, challenges, and evaluation metrics  to foster alignment and inspire future innovations.

\subsection{Methods of Collecting Papers}

First, we collected papers on recommender systems from major conferences such as SIGIR, RecSys, WWW, WSDM, KDD, and CIKM, restricting the selection to those published in the past five years (2020–2024). We then used authorship as a key filter to identify work on industrial recommender systems, requiring that at least one of the authors be from industry. Academic authors alone typically lack the opportunity to build production-grade recommendation systems. However, we found that some papers authored entirely by industry researchers still relied solely on public datasets and offline experiments, without application in real-world production. Following the principle of practicality, we further filtered for papers whose methods were validated through \textit{online A/B testing}, a necessary component of applied recommender systems, ultimately narrowing the set to 272 papers.

Our survey is based on an in-depth exploration and categorization of them. In our classification, the sole criterion for assigning a paper to a specific application scenario is the business context in which its method was validated through online A/B testing. For example, even if a method demonstrates improvements on offline datasets from multiple domains (\eg e-commerce, movies, music), if its online A/B evaluation was conducted only in an e-commerce setting, we classify it under e-commerce recommender systems. 

Note that, although advertising technologies share many commonalities with recommender systems, their underlying logic is more complex and involves distinct considerations in industrial deployment. Therefore, we deliberately exclude advertising from the scope of this survey, resulting in a final set of 228 papers.

\subsection{Significance of this Survey}
\label{ssec:significance}

\begin{figure}
    \centering
    \includegraphics[trim = 6.5cm 5cm 6.5cm 5.1cm, clip,width=0.8\linewidth]{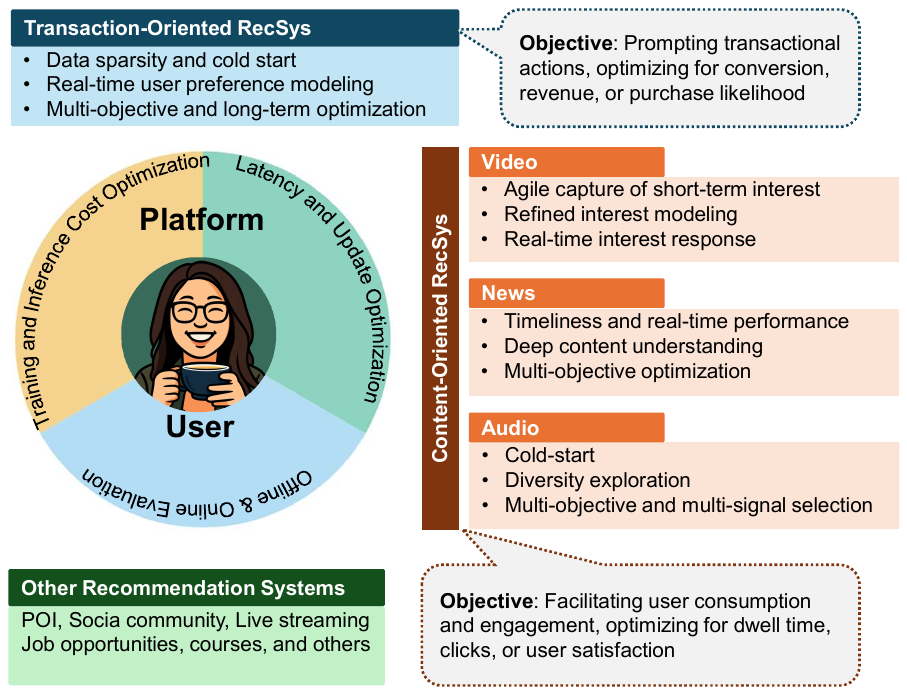}
    \caption{Overview of shared challenges/tasks across recommender platforms and the objectives of two main system types \ie transaction- and content-oriented RecSys, together with their primary challenges. Although some challenges look similar across different scenarios, solutions differ due to item characteristics. This figure also shows a user-centered view covering optimization at both the platform level and across application scenarios. }
    \label{fig:overview}
    \Description{An overview of RecSys types and challenges}
\end{figure}

This survey systematically summarizes the challenges of moving recommender systems from research labs to commercial applications and provides a comprehensive view of how industry practitioners achieve effective information matching at massive scales of data and users. Illustrated in Figure~\ref{fig:overview}, we  analyze how user behavior and data characteristics differ across various recommendation scenarios, revealing the underlying reasons for distinct optimization directions. To begin with, we classify the real-world recommendation systems into two main categories: 
\begin{tcolorbox}
[width=\linewidth, sharp corners=all, colback=white!95!black, boxrule=0.5pt]
\begin{definition}[\textbf{Transaction-Oriented Recommender System}]
A transaction-oriented recommender system is a system that generates item recommendations with the primary goal of \textit{prompting transactional actions} from users, optimizing for metrics such as conversion, revenue, or purchase likelihood.
\end{definition}
\end{tcolorbox}
Transaction-oriented RecSys is designed to encourage users to complete specific transactions, such as purchases, bookings, or subscriptions. Its primary goal is to maximize measurable business outcomes (\eg revenue, conversion rates, average order value) by predicting and suggesting items a user is likely to act on. These systems typically leverage historical transaction data, user behavior patterns, and predictive models to surface products or services with a high probability of conversion. E-commerce platforms are typical examples of transaction-oriented recommender systems.

\begin{tcolorbox}
[width=\linewidth, sharp corners=all, colback=white!95!black, boxrule=0.5pt]
\begin{definition}[\textbf{Content-Oriented Recommender System}]
A content-oriented recommender system is a system that generates item or content recommendations with the primary goal of \textit{facilitating user consumption and engagement}, optimizing for metrics such as dwell time, clicks, or user satisfaction rather than transactional outcomes.
\end{definition}
\end{tcolorbox}
Content-oriented RecSys is designed primarily to enhance user engagement and satisfaction by delivering relevant content for immediate consumption, such as news articles, videos, or music tracks. Unlike transaction-oriented systems, its core objective is not to drive purchases but to maximize metrics related to ongoing consumption and retention, including clicks, watch time, and reading duration. To achieve this, content-oriented RecSys typically combines rich content features with user preferences and behavioral signals, enabling personalized recommendations that adapt to evolving interests and promote sustained user interaction.

Under each category, we compile industrial methods focused on optimizing recommender systems, helping readers understand the practical gap between academic algorithms and real-world deployments. Consequently, our review will equip academic readers with deeper insights into the current industrial landscape of recommendation, thereby inspiring new theoretical research directions closely aligned with practical needs. Simultaneously, it offers industry practitioners a collection of practical techniques from a similar context that they can apply or refine to achieve greater commercial success.

Note that, recommendation systems are deeply integrated into various industries, way beyond online transactions and the content types discussed in this paper. However, the collected papers contain relatively few studies validated through A/B testing in scenarios such as POI~\cite{chen2021curriculum}, games~\cite{gou2024controllable}, live streaming~\cite{li2023stan}, social community~\cite{shi2023embedding,song2022friend,zhang2023constrained,kung2024improving,sagtani2023quantifying}, and job opportunities~\cite{hu2023boss,zheng2024mirror}; therefore, recommendations in these domains are not covered in our review.

In the remainder of this paper, Section~\ref{sec:overview} provides a general overview of recommender system applications in industry, including representative use cases, key business metrics, and explanations of why certain methodologies effectively address distinct data challenges.
Sections~\ref{sec:transaction} and~\ref{sec:content} detail the two major types of industry recommender systems: \textit{transaction-oriented}  and \textit{content-oriented} recommender systems. Both sections examine aspects such as data scale, real-time requirements, response latency, training and inference costs, and evaluation methodologies, highlighting the gaps that must be bridged for academic methods to achieve real-world applicability.
Section~\ref{sec:discussion} reflects on future research directions for recommender systems mainly from academic perspective. Finally, Section~\ref{sec:conclusion} concludes the survey, advocating for practical approaches to translate academic advances into tangible industry solutions, thereby advancing the field as a whole.

\section{Industrial–Academic Differences in Recommender Systems}
\label{sec:overview}

In academic research, recommender system problems are often abstracted into a simplified task of predicting user preferences based on past interactions. In this section, we first highlight the differences across application scenarios. We then use transaction-oriented recommender systems as an example to illustrate key industrial challenges beyond academic research.

\subsection{Differences across Application Scenarios} 

In the Internet industry, competition fundamentally centers on capturing user attention, attracting new users, increasing engagement time, and fostering loyalty, often with the ultimate goal of driving monetization and revenue. In transaction-oriented platforms such as online retail, e-commerce, and travel services, recommender systems typically aim to maximize \textit{Gross Merchandise Volume} (GMV)~\cite{yin2023heterogeneous,xie2021causcf}. Conversely, for content platforms like news and music, the primary goal of recommendation is to enhance user stickiness and loyalty, motivating users to spend more time on the platform~\cite{liu2022multi,li2024modem}. This increased engagement opens avenues for monetization through value-added services, such as memberships or increased advertisement impressions.

However, recommender systems must consider vastly different factors depending on their specific application scenarios. To elaborate, the relative characteristics of users and items differ significantly across domains. For example, e-commerce platforms handle large numbers of rapidly changing items, thus requiring recommendations that account for real-time inventory status to prevent unsuccessful transactions. For instance, \citet{10.1145/3459637.3481923} focus on building a recommendation system tailored for instant food delivery services. Their optimization strategy involves considering \textit{Order Fulfillment Cycle Time} to implement a personalized ranking approach. This method integrates order sequences and time-related features to more accurately estimate the duration from order placement to delivery. However, this measure is irrelevant to music recommendation systems. On the other hand, music  services place higher importance on interest exploration, diversity, user relationships, and group behavior~\cite{briand2021semi,baran2023accelerating}, aspects that are comparatively less critical in e-commerce.  Another illustrative example is item pricing, which is a crucial factor influencing user purchase decisions and a key feature that e-commerce recommender systems must carefully consider~\cite{10.1145/3459637.3482292}. In contrast, price is essentially irrelevant in music and news recommendation scenarios once the user has subscribed to the service. This illustrates that even within the broad category of recommender systems, distinct application domains encounter fundamentally different challenges. Such domain-specific differences help explain why recommendation technologies have evolved independently across verticals, developing largely in parallel with limited overlap.

Thus, only by systematically analyzing the business objectives and data characteristics of each scenario can we fully grasp the rationale behind specific recommendation methods and their underlying logic.

\subsection{Challenges beyond Academic Research}

It is evident that recommender systems in the industry encounter a greater number of practical operational challenges, which are absent in academic research. Here, we take the scenario of online transactions \eg e-commerce, as an example for ease of understanding. Challenges specific to content-oriented recommender systems \eg news and video, are discussed in Section~\ref{sec:content}.

\subsubsection{Large Data Scale} Industrial recommender systems are often built for user bases in the millions or billions, far exceeding the size of datasets used in academic experiments~\cite{chen2024macro,he2020contextual,10.1145/3583780.3615200,xu2023optimizing}. This difference has two major implications; the first one is balancing effectiveness and latency~\cite{liu2020autogroup}. In practice, a method that achieves significant improvements in recommendation quality but requires minutes to return results can severely harm the user experience, rendering it impractical in an industrial setting. This is why industrial systems often break the recommendation pipeline into multiple stages, such as recall, coarse ranking, fine ranking, and re-ranking, to ensure both efficiency and quality~\cite{zhao2024breaking,ma2022two,wang2023diversity,wang2024not}. Such multi-stage designs are rarely emphasized in academic studies. The other implication is the different evaluation standards. Given the massive scale of industrial data, even a 1\% improvement can translate into substantial commercial gains.

\subsubsection{Cost as a Critical Consideration.} In academia, the primary focus is on advancing methodologies, whereas in industry, the emphasis lies in overall profitability. Consequently, when developing recommender systems, industry practitioners pay close attention to system costs, and many approaches focus on cost optimization—a reflection of the commercial nature of industrial practice~\cite{10.1145/3627673.3679914,meisburger2023bolt}. At the same time, technologies can often be made cost-effective through various forms of optimization, whether methodological or engineering. A key advantage of academic research is its ability to propose novel solutions without being constrained by strict cost considerations.

\subsubsection{Real-Time Modeling as a Necessity.} In industry, real-time modeling is essential, whereas academia often lacks both the demand and the conditions to pursue such research. In e-commerce, user interests are highly dynamic, shifting rapidly over time, across contexts, or even in response to transient external stimuli. 
For example, \citet{xiao2024deep} noted that online shopping decision-making is brief and unstable, even a single scroll can instantly alter the intensity of a user’s current interest. Users may switch product categories within seconds and revise purchasing intentions when exposed to promotions or newly listed products.

Product conditions also fluctuate in real time: delivery speed~\cite{mondal2022aspire}, fulfillment cycles~\cite{10.1145/3459637.3481923}, prices~\cite{10.1145/3459637.3482292}, and inventory~\cite{pande2023personalized} can all update instantly, triggering further interest shifts. Consequently, the core challenge for transaction-oriented recommender systems is to detect and respond to instantaneous user interests within extremely short time windows~\cite{shen2022deep}. This imposes stringent requirements on the timeliness of interest modeling. Traditional offline models often struggle to keep pace. Practitioners must perform data collection, interest inference, recommendation computation, and result delivery in real or near-real time. Even latencies measured in milliseconds can render recommendations obsolete, as user interests may have already shifted~\cite{park2024slh}.

\subsubsection{Gaps Between Simulation-Based and Online Optimization.} Due to practical constraints, even within the same research directions, extensive academic studies often remain confined to simulated environments. Examples include the sparsity problem, the cold-start problem, and multi-objective optimization. Addressing these issues in real-world commercial settings differs substantially from experiments conducted on datasets.

In online transaction platforms, sparsity and cold-start issues often arise due to the frequent influx of newly listed products. These items may have little or no historical interaction data, and if a system relies heavily on such data, they can remain underexposed for extended periods, leaving much of their potential untapped. Traditional collaborative filtering-based methods struggle in this scenario, leading to the cold-start problem~\cite{10.1145/3511808.3557101,huai2023m2gnn}. Sparsity occurs because the user–product interaction matrix is typically large, yet each user interacts with only a small fraction of products. This scarcity of behavioral data makes it difficult for algorithms to accurately model user preferences~\cite{mei2022lightweight}. Long-tail items, with minimal interactions, further exacerbate the problem by making it difficult for the system to effectively detect and recommend them. As online transaction recommender systems scale, addressing cold-start and sparsity issues becomes critical, as both directly impact the system’s ability to discover new products and maintain scalability.

Online transaction platforms use recommendation technologies to deliver items (\eg products~\cite{qian2022intelligent}, services~\cite{xu2020recommender}, and insurance~\cite{bi2020heterogeneous}) that are more likely to lead to purchases, thereby enhancing overall commercial outcomes. Achieving these results, however, involves multiple components. Clicks are often the first step in user interaction, which is why \textit{Click-Through Rate} (CTR) is widely used to measure the accuracy of recommender systems~\cite{chen2022co,shen2022deep}. Yet, because purchasing incurs costs and users differ in purchasing power, even accurate recommendations may fail to convert if, for instance, prices are too high. To capture this dimension, metrics targeting \textit{Conversion Rate} (CVR) are often used~\cite{xu2020privileged,liu2020autogroup}. Moreover, recommending items that are appealing and low-priced may seem ideal, but if transaction values remain too small, the platform’s \textit{Gross Merchandise Volume} (GMV) suffers, undermining business objectives. As a result, GMV is also a critical consideration~\cite{chai2022user,zhang2023collaborative}. These examples highlight the inherent trade-offs in applying recommender systems to online transaction platforms, making multi-objective optimization essential~\cite{wang2020m2grl}. In recent years, balancing short-term transactions with long-term user interests has also emerged as a key area of discussion~\cite{xu2023optimizing}.

\subsubsection{Greater Emphasis on Long-Term Gains.} Industrial recommender systems aim not only for short-term revenue growth but also for shaping long-term commercial value potential~\cite{ha2021producer}. After all, the larger the user base and the stronger the user stickiness, the greater the potential for future monetization of a commercial platform. Consequently, in recent years, the industry has increasingly focused on enhancing long-term user satisfaction—such as improving user retention~\cite{celikik2022reusable,le2023cec}—which is a direction that aligns closely with the value proposition of recommender systems, yet lacks the experimental conditions in academia.

Grounded in industrial application scenarios and following the definitions in Section~\ref{ssec:significance}, Figure~\ref{fig:overview} lists the technical challenges of recommender systems at both the platform level and the application-specific level. All of these challenges ultimately stem from the goal of providing better user experience.

\section{Transaction-oriented Recommender Systems}
\label{sec:transaction}

In this section, we review solutions addressing the key challenges in transaction-oriented RecSys, or T-RecSys for short. Recall that a transaction-oriented recommender system is designed primarily to encourage users to complete specific transactions such as purchases and bookings. We organize our following discussion into three stages: when a user or item is new to the system (\textit{cold start and sparsity}), when the user is interacting with the system (\textit{real-time interest capture}), and when the system seeks to optimize outcomes over time (\textit{multi-objective and long-term optimization}).

\subsection{Data Sparsity and Cold Start in T-RecSys}

Most users interact with only a small subset of available items through clicks, purchases, or ratings. Consequently, the resulting user-item interaction matrix is highly sparse~\cite{chen2024macro,mei2022lightweight}, which is very similar to cold-start problems, because both new users and new products are lack of enough historical interaction data, the system cannot effectively recommend these new users or new products~\cite{huan2022industrial,spivsak2024interpretability,10.1145/3511808.3557101}.

The first solution leverages heterogeneous information networks to fuse diverse knowledge from user behaviors~\cite{yin2023heterogeneous,xu2020gemini,lu2022deep,baltescu2022itemsage}. For example, in the insurance domain, most users lack purchase history, limiting traditional methods like collaborative filtering or sequential models. \citet{bi2020heterogeneous} address this by transferring knowledge from a source domain (e-commerce) to insurance. They model user interests in the source with a GRU over purchase sequences, and build an insurance heterogeneous information network with users, products, attributes, and agents. A three-level attention mechanism (relational, node, and semantic) learns embeddings, which are then mapped from source to target by an MLP, enabling recommendations for cold-start users. Also, \citet{10.1145/3511808.3557101} tackle the cold-start problem by proposing a multimodal meta-learning approach. Their method stabilizes learning with multimodal side information (text, images, IDs) through multiple meta-learners whose predictions are fused via a learnable layer. They also introduce an embedding generator that uses multimodal data to create ID embeddings for new items, enhancing knowledge transfer and cold-start performance. 

The second type of method exploits abundant information from existing or external data sources to compensate for the shortage of user–item interactions in the target scenario.  Specific strategies include knowledge transfer~\cite{xu2024rethinking}, cross-domain or cross-scenario data utilization~\cite{10.1145/3627673.3680055,yang2024mlora,li2024scene}, efficient feature modeling~\cite{wilm2023scaling}, or mining of unobserved data~\cite{10.1145/3627673.3680067}.  For instance, implicit user feedback is widely considered an important feature to leverage.   \citet{qin2021bootstrapping} built a recommender system using installation and rating data, while  \citet{zhang2024unified} proposed a unified dual-intent transformation framework that utilizes query keywords from search scenarios as supervision signals to accurately infer users' implicit needs in recommendation scenarios. When no implicit information can be mined, \citet{huan2022industrial} proposed an offline-online framework. In the offline phase, representations of cold-start items are generated from historical user–item interaction data, and in the online phase, a lightweight cold-start model, combined with pre-generated item embeddings, is employed for real-time recommendations to predict the performance of newly launched products. 
At the same time, although the developers from industry aim to exploit information as fully as possible, practicality must also be considered. Thus, sample efficiency became important because it emphasizes better leveraging available labels, particularly when feedback is noisy or missing. For example, user signals (\eg clicks, installs, conversions) are abundant but often noisy and imbalanced. To address this, \citet{zhang2024self} proposed a self-distillation framework in which the main head trains on ground-truth labels while also generating pseudo-labels. An auxiliary head with two branches—one learning from soft teacher predictions and the other from ground-truth labels—employs a selector to merge signals, thereby balancing label quality, maintaining calibration, and avoiding additional serving cost. \citet{an2024ddcdr} combined knowledge distillation with representation disentanglement for cross-domain recommendation. A teacher model, jointly trained on source and target domains with an adversarial discriminator, provides domain-shared features. The student model then disentangles representations into shared features (distilled from the teacher) and domain-specific features (enhanced through contrastive learning), assigning higher weights to boundary samples to improve transferability.

With the rapid development of large language models (LLMs), numerous efforts have emerged that leverage LLMs to address sparsity and cold-start issues. For example, \citet{zhao2024breaking} proposed a complementary knowledge-enhanced recommender system that employs an entity extractor to identify conceptual terms from user and item information, generates entity pairs based on popularity and predefined strategies, and uses an LLM to construct complementary knowledge graphs, similar to~\cite{kekuda2024embedding}. In parallel, \citet{gong2023unified} introduced a unified foundation model that integrates search and recommendation tasks into commercial systems. Their approach leverages an LLM to extract domain-invariant textual features, improving cold-start recommendation performance. Solutions based on various architectures, such as macro-scale graph neural networks~\cite{chen2024macro}, graph collaborative filtering~\cite{10.1145/3627673.3680015}, and transformers~\cite{ye2023transformer}, have also been proposed and shown to address sparsity problems in real-world production environments.

\subsection{Real-time User Preference Modeling in  T-RecSys}

As aforementioned, users' interests change rapidly and can be influenced by factors such as time, transaction cycles, and geographical location. Improving the accuracy of real-time modeling is crucial for enhancing user experience and business success in T-RecSys~\cite{li2021path,lang2021architecture,park2024slh,sierag2022client,10.1145/3583780.3615218,fang2023alleviating,liu2024unified,pande2023personalized,10.1145/3459637.3481948}. However, the real-time modeling can be much task-dependent. We brief the solutions for a few different task settings. 

In mobile recommendation scenarios, context awareness is crucial for modeling the user's true intent, as in many other context-sensitive scenarios~\cite{chen2023contextual,ma2022two}.  For example, due to limited screen size, users typically only see the first few items, giving rise to two issues: (1) items ranked higher are more likely to be clicked \ie position bias, and (2) items ranked lower may not be seen at all but are mistakenly judged as “unpopular” \ie false exposure. To address this, \citet{he2020contextual} estimated the probability that each position is actually viewed by the user and incorporated these probabilities to better infer true user preferences. Experiments on the Taobao platform validated this approach in large-scale online tests.  

Besides the physically limits, the context of sequential applications needs to be carefully considered~\cite{10.1145/3627673.3679914,mei2022lightweight,lv2023deep,kouki2020lab,nie2024hybrid}. \citet{10.1145/3627673.3680037} proposed semantic fusion at the sequence level, by integrating text and ID sequences through frequency domain transformations. This approach models users’ multimodal sequential behaviors in real time, improving various technical metrics on commercial platforms. In addition to directly modeling temporal dependencies, some studies have explored leveraging users’ historical behavioral signals to uncover more stable preference features. For example, \citet{xian2021ex3} analyzed users’ past browsing and purchase histories to automatically identify the most critical item attributes in their decision-making process, and generated recommendation sets annotated with these differentiated attributes. Rather than modeling behavior sequences directly, this approach extracts higher-level preference cues from historical behaviors, complementing context-aware and sequence-based modeling. \citet{ren2023greenseq} proposed GreenSeq, a framework that uses neural architecture search to automatically design lightweight and efficient network architectures, reducing the computational resource consumption of sequential recommendation systems. This efficient network enables real-time, accurate modeling of user preferences while substantially lowering resource usage.

During user's interaction with the platform, users’ instant interests are often explicitly activated by their most recently clicked item, or a trigger. However, traditional recommendation models primarily rely on long-term behavior modeling, making it difficult to capture such momentary interests. To address this issue, \citet{shen2022deep} proposed a deep interest highlight network, which consists of three core components: (1) a user intent network that estimates the true intensity of user interest in the trigger item, (2) a fusion embedding module that adaptively combines the embeddings of trigger and target items based on the user intent network’s predictions, and (3) a hybrid interest extraction module that integrates soft sequence modeling and hard sequence modeling, to effectively highlight instant interests. Real-world deployment  validates its practical value for click-through rate prediction. Similarly in the trigger-induced recommendation scenario, \citet{xiao2024deep} further considered the dynamic evolution of user interests as the page scrolls; \citet{10.1145/3627673.3680065} incorporated latent intent and uncertainty modeling to mitigate the convergence problem caused by over-reliance on trigger items.

Graph structures~\cite{10.1145/3583780.3615200} can reveal complex relationships between users and items which is applied in real-time interest modeling. \citet{jiang2022triangle} treated triangular structures in item co-occurrence graphs as interest units to capture the latent motivations behind user clicks, while introducing more diverse and novel items. This approach not only alleviates issues of behavior sparsity and popularity bias but also expands the space for user interest exploration. Traditional graph convolution methods often rely on simple neighbor feature aggregation, which can overlook fine-grained relations such as user–user and item–item interactions. \citet{sun2020neighbor} explicitly modeled interaction relationships among neighbors while distinguishing the heterogeneity between user and item nodes, enabling a more accurate representation of users’ genuine intentions.  \citet{feng2020atbrg} proposed a knowledge graph-based relational graph to model users’ real-time historical behaviors (\eg click records) and capture relational information between these behaviors and target items. This approach constructs user- and item-specific subgraphs to dynamically mine user interests. 

From the business perspective, real-time processing is closely tied to the business characteristics of each platform, leading to diverse approaches tailored to specific transaction scenarios. For example, \citet{xu2020recommender} proposed a ride-hailing driver repositioning framework that leverages real-time data from drivers and passengers, including drivers’ current locations and statuses and the real-time distribution of passenger demand. In food delivery, \citet{zhang2023modeling} introduced a dual-periodic preference modeling method to capture users’ preferences for both restaurants and food across different time periods. By modeling these dual interaction preferences, the approach significantly improves the accuracy of real-time recommendations. Spatiotemporal features are also explored in~\cite{lin2023exploring} for online food recommendation service. In e-commerce, \citet{qian2022intelligent} designed an intelligent request strategy to dynamically capture real-time changes in user intent. Their AdaRequest framework optimizes request insertion by modeling real-time user behavior, estimating purchase uplift via causal inference, and dynamically planning requests. Deployed in Taobao’s system, AdaRequest enhanced commercial metrics such as GMV.  For online travel, \citet{10.1145/3511808.3557126} proposed a Stage-Aware Sequential Matching Network to capture real-time shifts in user preferences across different stages of the travel lifecycle, substantially improving the accuracy of travel recommendations.

Similarly, large language models can be leveraged to achieve semantic understanding and multimodal fusion of real-time contextual information. They can also extract multimodal signals from user behavior, enhancing the context-awareness of transaction-based recommender systems~\cite{wan2024larr,tian2024reland,10.1145/3627673.3680028}.

\subsection{Multi-objective and Long-term Optimization in  T-RecSys}

As mentioned earlier, recommendation systems are optimized while balancing multiple business objectives~\cite{zhang2023collaborative}. These objectives often conflict with one another, creating a situation where improving one goal may come at the expense of another. Engineers hence must optimize for multiple goals simultaneously to ensure the overall business interests are enhanced~\cite{xie2021causcf}. This includes algorithmic multi-objective optimization~\cite{wang2023diversity,wang2022recommending}, business special multi-objective optimization~\cite{mondal2022aspire,10.1145/3459637.3481923}, simultaneous optimization of engineering performance and algorithm quality~\cite{liu2020autogroup}, as well as the optimization of both short-term and long-term goals~\cite{10.1145/3627673.3680061,ha2021producer,wang2024not}. These challenges are among the most common issues faced in the industry. 

\citet{wang2020m2grl} proposed a multi-task multi-view graph representation learning framework for large-scale recommendation systems. This framework constructs multiple single-view graphs to learn node representations for different views, such as instances, categories, and stores, and models cross-view relationships through multi-view alignment. By adopting a multi-task learning paradigm, it jointly optimizes both the learning of intra-view representations and the modeling of cross-view relationships, addressing challenges in representation capacity and inductive bias that arise in multi-view data fusion methods. Trained on 57 billion samples on the Taobao platform, it addresses the conflicts between multiple objectives (such as click-through rate and purchase rate). Graph method has also been applied in long-term optimization~\cite{tu2023disentangled,10.1145/3627673.3680061}. For instance, \citet{wu2020airbnb} extended knowledge graphs and multi-objective optimization to balance user preferences with platform revenue. 

Another type of solution involves cross-domain modeling, co-training, or multiple task and stage strategies to simultaneously balance multiple objectives~\cite{tang2024touch,le2023cec}.  \citet{chen2022co} proposed a co-training framework to jointly train biased and unbiased models, explicitly leveraging popularity bias through feature decoupling and domain adaptation rather than simply eliminating it. They also designed a feature decoupling module that separates item attribute representations from popularity representations, addressing the popularity distribution shift problem. During online retrieval, they balanced retrieval results by adjusting the weights between biased and unbiased models to align with users' genuine interests, successfully achieving equilibrium across multiple business metrics such as clicks, views, and click-through rates.  \citet{zhang2023collaborative} proposed a collaborative cross-domain transfer learning framework to enable modeling cooperation across different business domains. This method evaluates the actual gain of source-domain samples on the target domain, dynamically selecting and weighting information flows to avoid the negative transfer problem caused by direct migration. Experiments on Meituan advertising scenarios show  improvements in both CTR and GMV. Cross-domain modeling not only alleviates the data sparsity problem but also balances optimization requirements across multiple domains in multi-objective scenarios. \citet{wang2023industrial} proposed a two-stage framework, quantify the novelty of products, and use a potential outcome model to determine when to provide personalized surprise recommendations. 

For optimizing short-term and long-term goals, reinforcement learning is the mainstream approach~\cite{xu2023optimizing,cai2023model,10.1145/3459637.3482292}. \citet{10.1145/3459637.3482292} proposed modeling cold-start recommendation as a long-term value optimization problem. In e-commerce scenarios, new items are often disadvantaged in immediate click-through rate–based ranking due to the lack of historical interactions, resulting in insufficient exposure and a “Matthew Effect,” where products with genuine growth potential may be buried in the early stages. To resolve this conflict, authors employed a reinforcement learning framework to balance short-term and long-term objectives: by modeling the lifecycle of items, predicting their potential cumulative returns, and incorporating this long-term value signal into the ranking process, high-potential items can receive more exposure during the cold-start phase. Additionally, transformer has also been widely used~\cite{10.1145/3459637.3481923,wilm2024pareto,celikik2022reusable}.

In the last, method effectiveness and engineering  efficiency is a unique research focus in industry~\cite{meisburger2023bolt,jian2023practical,liu2020autogroup}. \citet{joglekar2020neural}  revealed that in industrial recommender systems, the vast majority of parameters and resource consumption are concentrated in the embedding layer of the input, where embedding dimensions and vocabulary sizes are often set heuristically, leading to both resource inefficiency and suboptimal performance. Their method automatically searches for the optimal vocabulary size and embedding dimension for each feature under the same computational budget, thereby discovering superior configurations. They introduced a multi-size embedding mechanism: frequent or highly predictive feature values are allocated larger vector dimensions, while rare or weakly relevant features are represented with smaller vectors. In Google Play’s large-scale deployment, the final model reduced parameters by about 30\% while increasing app installations by 1.02\%, with long-term online experiments confirming the stability of its performance. These results demonstrate that industrial practice values not only model effectiveness but also engineering efficiency and scalability. 

\section{Content-Oriented Recommender Systems}
\label{sec:content}
A content-oriented recommender system generates  recommendations for item or content, with the primary goal of facilitating user immediate consumption and engagement. Because “content” encompasses many forms, there could be many different forms of recommendations like document recommendations~\cite{chen2020improving}, email and push notifications~\cite{lin2024bootstrapping}, social media posts~\cite{xie2021real}, and other recommendation scenarios~\cite{rangadurai2022nxtpost,yang2023graph,huang2021sliding}. Based on the distribution of collected papers and the principle of practical relevance, we focus our survey primarily on three types of content: video, news, and audio.

\subsection{Video Recommender Systems}

With the rapid development of the Internet and multimodal technologies, video has become a mainstream form of information dissemination and content consumption. Unlike transaction-oriented recommender systems, video recommendation systems face distinct and often more complex challenges.

\subsubsection{Challenges for Video Recommendation}

First, video recommendation systems face a fundamental conflict between interest diversity and recommendation precision. Users often maintain wide-ranging interests without consistent observable associations—for instance, one may simultaneously enjoy documentaries, dramas, news, and lifestyle vlogs. This diversity makes it difficult for systems to accurately identify a user’s current focus~\cite{klimashevskaia2023evaluating,grun2023transparently}. Overemphasis on homogeneous content risks user fatigue and boredom, while excessive pursuit of diversity may dilute relevance, reducing precision and degrading the overall experience. Balancing these two dimensions remains a central challenge~\cite{li2024modeling,li2024contextual}.

Second, feedback signals in video recommendation are complex and noisy. Clicks, watch time, likes, and comments differ in reliability, while negative samples may contain weak positives that only partially reflect interest. Such noisy, imbalanced signals complicate the accurate modeling of user preferences~\cite{zhang2024self}.

Third, video recommendation is highly context-dependent. User choices vary across scenarios: in public settings (e.g., family gatherings), they often select group-friendly content while concealing private interests~\cite{li2024modem}; in private contexts, choices align more closely with individual preferences. During fragmented scenarios such as commuting, users typically favor short-form videos over long content. Failing to account for such context shifts can drastically reduce recommendation accuracy and satisfaction~\cite{yang2024enhancing,chen2024missing,shao2024optimizing}.

In response to these challenges, video recommendation systems have evolved through several major technical stages, each shaped by unique difficulties and corresponding solutions.

\subsubsection{Agile Capture of Short-Term Interests}
User short-term interests shift rapidly, yet their viewing behaviors often follow sequential patterns. Sequence modeling and session-based methods capture these dynamics by leveraging historical viewing activities. Key metrics in this phase include Click-Through Rate (CTR), completion rate, and average viewing duration.

For short-sequence modeling, researchers have integrated multimodal and cross-domain signals~\cite{koneru2024enhancing,liu2023multitask,wu2023ruel,zhang2024co}. For example, \citet{chen2024multi} addressed the distribution gap between multimodal features and user interests in cold-start short-video recommendation, introducing trainable clustering IDs, modality encoders, and modality-intensity perception to enable content-driven recommendation independent of IDs. In e-commerce scenarios, \citet{lei2021semi} transferred product-domain behaviors into micro-video recommendation, improving intent modeling in short videos. \citet{wu2024learned} further linked short-term sequences to long-term satisfaction, showing that short-term interests can inform sustainable user engagement.

For long-sequence modeling, \citet{si2024twin} proposed TWIN-V2, a divide-and-conquer approach. Offline, it compresses long behavior sequences via hierarchical clustering to generate virtual clustered behaviors. Online, it applies cluster-aware target attention, adjusting weights by cluster size to balance short- and long-term interests. This method improved both CTR and viewing duration by efficiently detecting interest shifts.
\subsubsection{Refined Interest Modeling in Video Recommendation}
In video recommendation, user behaviors are diverse and mixed, making it difficult to model genuine interests~\cite{yao2024user,jiang2024prompt}. A common solution is multi-task learning ~\cite{liu2022multi,bai2024gradcraft,jeunen2024multi}, which captures multiple interest signals. For example, \citet{tang2020progressive} separated shared and task-specific knowledge through multi-level experts and progressive routing, reducing interference between tasks. \citet{zhang2022multi} combined reinforcement learning with multi-task fusion for refined interest modeling. In more complex settings, graph learning has been widely applied: \citet{zhang2023multi} integrated contrastive learning into expert representations for multi-region TV recommendation, while \citet{jiang2024mmgcl} employed meta-knowledge–enhanced multi-view graph contrastive learning to mitigate noise and sparsity.

Beyond modeling known preferences, practitioners also recommend novel content to surface emerging needs. \citet{li2020purs} introduced “unexpectedness” via multi-cluster interest modeling and attention mechanisms, balancing surprise with relevance to avoid filter bubbles. \citet{lin2022feature} proposed feature-aware representations to diversify recommendations and reduce fatigue, while \citet{li2024contextual} used context distillation to uncover latent interests. These works show that diversifying and introducing novelty not only enhances short-term engagement but also improves long-term user satisfaction.

Video recommendation further faces complexity from its multimodal and multi-scenario nature, spanning contexts like mobile vs. TV or single vs. shared accounts~\cite{deng2024mmbee,zhang2022scenario}. \citet{zhao2023m5} proposed a large-scale framework integrating multimodal data, multi-grained interests, and cross-scenario variation with a Split Mixture-of-Experts (SMoE) for dynamic preference fusion. \citet{lu2022deep} developed a unified representation learning method to align heterogeneous features into a shared space, preserving structure while reducing distributional shifts.

Another challenge is shared accounts, especially in TV/OTT (Over-The-Top)  settings. \citet{qin2023learning} combined session-aware co-attention with device-ID weak supervision to distinguish multiple viewers, while \citet{li2024modem} addressed shared-account behavior in large-screen devices. Additional advances include tree-based regression for fine-grained preference prediction~\cite{lin2023tree} and leveraging deep models with large language models for richer feature extraction and interest categorization~\cite{agrawal2023beyond}.
\subsubsection{Real-time Interest Response in Video Recommendation}
After modeling user interests, video recommender systems emphasize real-time feedback to capture rapid shifts in interest and contextual changes. Key challenges include quickly identifying current intent, incorporating signals such as likes, skips, and viewing duration, and balancing response time with recommendation quality~\cite{zhang2024enhanced,su2024rpaf}.

In short-video scenarios, where user interests are highly transient, real-time responsiveness is critical. \citet{gong2022real} deployed a lightweight on-device re-ranking model that dynamically adjusts sequences using real-time feedback (\eg scrolling, likes) and a context-aware adaptive beam search to handle video interdependencies. \citet{yi2023online} launched a real-time bandit system addressing the dual challenges of massive exploration spaces (millions of items) and real-time updates (millisecond-level latency). They introduced the Diag-LinUCB algorithm, leveraging sparse graphs and diagonal covariance matrices to enable distributed, real-time parameter updates at million-level queries per second (QPS). Other works also target scalable real-time responsiveness~\cite{gao2021learning,wang2024future,yuan2023hydrus,zhang2023shark}.

Evaluation metrics at this stage include response latency (lower is better), short-term interaction lift (e.g., likes, saves), and the timeliness of adapting to interest shifts. For example, \citet{shao2024optimizing} improved interactions and content creation, while \citet{ren2024non} enhanced long views, likes, and follows.

\subsubsection{Debiasing in Video Recommendation}
Video recommendation systems frequently require debiasing~\cite{zhao2023uncovering} to ensure accuracy, fairness, and a healthy content ecosystem. Common biases include exposure, position, and popularity biases, while user interests are further shaped by confounding factors such as video length or misleading titles. Feedback loops inherent in recommendation systems can amplify these issues.

\citet{zhan2022deconfounding} systematically studied duration bias, showing how video length skews watch-time prediction. They applied causal intervention (backdoor adjustment) to remove the confounding effect of duration on exposure while retaining its natural impact on watch time, thereby improving recommendation accuracy and engagement. Other studies addressed complementary biases: \citet{farias2023correcting} tackled creator-side interference, \citet{klimashevskaia2023evaluating} and \citet{grun2023transparently} focused on popularity bias calibration, and \citet{liu2024self} introduced an adaptive fairness-constraint framework balancing group-level and individual-level fairness.

Since watch time is a key metric yet highly biased, \citet{zhang2023leveraging} proposed distributional labels (watch-time percentile rank, effective/long views) to capture richer semantics. They further introduced a causal adjustment mechanism based on video-duration grouping to debias labels without modifying model structure, and employed multi-task learning to integrate predictions, improving interest modeling.

\subsection{News Recommender Systems}

Unlike transaction or video platforms, news platforms place much higher emphasis on \textit{timeliness}. News content is updated frequently, and users expect immediate access to the latest information.

\subsubsection{Challenges for News Recommendation}

\citet{kruse2023creating} noted that news cycles are fast and the relevance of articles decays quickly, while readers’ interests shift rapidly as topics change. Freshness is thus the top challenge: if users cannot access breaking news promptly, they may lose trust in the platform’s ability to deliver timely and relevant information~\cite{zhu2022spherical,cai2022reloop}. Moreover, trending stories typically have short lifecycles. When recommendations lag behind, presenting users with outdated coverage, engagement drops sharply~\cite{xiao2022training}.

Second, user interests on news platforms are highly diverse, spanning multiple categories (e.g., politics, sports, entertainment) and geographies (local, national, international). Delivering sufficiently broad content helps satisfy these varied interests~\cite{kruse2023creating,xi2024towards}.

Third, credibility and authority of sources carry greater weight than in video or e-commerce recommendations. \citet{xi2024towards} highlighted source identity as a key factor, alongside topic, region, style, and timeliness. Users strongly prefer content from authoritative publishers, which directly shapes perceptions of authenticity and reliability~\cite{horn2024more}. At the same time, platforms’ own editorial values influence what gets recommended~\cite{kruse2023creating}.

Fourth, content repeatability differs across domains. While transaction items can be repurchased and some videos rewatched, news articles lose value quickly once consumed or once their popularity declines. As \citet{kruse2023creating} emphasized, repeated exposure to the same news diminishes user satisfaction, since “read” content is no longer newsworthy.

Finally, news content is often topic-centric and clustered. Articles frequently develop around the same event or theme, and users tend to follow these evolving stories over time. This clustering requires recommendation systems to account for topic continuity while avoiding redundancy~\cite{shi2021wg4rec,horn2024more,kruse2023creating}.

\subsubsection{Timeliness and Real-time Performance in News Recommendation}

News has a strong timeliness factor: popular articles have short lifecycles, and readers’ interests shift quickly as the news agenda evolves. Thus, a news recommendation system must rapidly capture fresh content and adapt to changing user behaviors. 

Common strategies include high-frequency model updates, continuous learning, and online error correction. For instance, \citet{xiao2022training} accelerated fresh news integration by caching news vectors, removing redundant data, and simplifying representations, enabling pre-trained language models to train up to $100\times$ faster. \citet{qu2022single} improved update efficiency by evaluating feature embedding importance and pruning low-value dimensions, allowing hourly model refreshes. In parallel, \citet{kruse2023creating} employed a multi-model and A/B testing framework to dynamically prioritize trending news while maintaining editorial oversight in line with business needs.

\subsubsection{Deep Content Understanding in News Recommendation}

Because newly published news and new users lack interaction history, traditional collaborative filtering struggles to model them effectively. A common solution is to construct deep semantic representations of news text to better capture item characteristics. 

\citet{horn2024more} leverage LLMs to generate news embeddings, inferring a user’s immediate interests from article topics, people, and locations, thereby addressing the cold-start problem for non-logged-in users. \citet{xi2024towards} extracted user preferences and factual knowledge from LLMs and transformed them into embeddings via a mixture-of-experts network, integrating external world knowledge while avoiding inference latency. \citet{zhao2022improving} applied a conditional variational autoencoder to model the latent distribution of side information, generating warm-up embeddings for cold items that update continuously as interactions accumulate—without requiring additional data and compatible with diverse base models. Similarly, \citet{shi2021wg4rec} constructed a word graph encoding semantic similarity, co-occurrence, and co-click relations, linking words across articles to capture cross-platform and cross-domain topics—particularly effective for cold-user scenarios.

\subsubsection{Multi-objective Optimization in News Recommendation}
News recommendation systems must satisfy individual user needs while balancing broader objectives such as diversity, alignment with platform values, and content authority. Over-personalization risks creating information cocoons, undermining the platform’s role in informing users across a wide range of events. Similarly, over-reliance on algorithms without editorial oversight may weaken credibility and trust in news organizations~\cite{kruse2023creating}.

A common solution is multi-task learning. For example, \citet{bai2022contrastive} mitigate gradient conflicts across tasks by designing a contrastive shared network with alternating training, jointly optimizing CTR, reading time, and other metrics, yielding significant online gains. \citet{nie2022mic} unify three retrieval channels (user–user, user–item, and item–item) and enhance recall and diversity via cross-channel contrastive learning.

Diversity is often addressed through constrained optimization and re-ranking. \citet{kruse2023creating}, for instance, controlled category distribution and sources in ranking to preserve editorial values while maintaining CTR. \citet{horn2024more} introduced “immersion paths” and “novelty paths” to balance deep reading with cross-topic exploration.

Finally, multi-dimensional preference and knowledge enhancement extend multi-objective optimization to incorporate user preferences (topic, region, style, source) and content attributes. \citet{xi2024towards} leveraged factorized prompts to extract multi-dimensional preferences and factual knowledge from LLMs, encoding them into a unified format for ranking, thus optimizing core objectives while maintaining authority and diversity.

\subsection{Audio Recommender Systems}
We group audio scenarios, including podcasts and music, into a single category, as their user contexts, consumption habits, and item types are highly similar. Accordingly, this section focuses on exploring audio recommendation systems.

\subsubsection{Challenges in Audio Recommendation}

In audio recommendation systems, several cold-start scenarios commonly arise. User cold start occurs when newly registered users or those with minimal interaction history make it difficult to infer preferences for personalized recommendations. Similarly, newly released works and creators lack early interaction data, requiring the system to quickly connect new content with target audiences to mitigate the long-tail effect and improve content discovery and distribution efficiency.

Unlike other platforms that rely heavily on historical interactions, audio users especially in music, seek exploration, expecting fresh, engaging, and diverse recommendations. Repetitive or monotonous suggestions can lead to user fatigue, reducing satisfaction and platform stickiness; thus, maintaining diversity is a key task~\cite{lichtenberg2024ranking,lindstrom2024encouraging}.

The choice of implicit feedback signals significantly influences recommendation outcomes. For instance, in podcasts, subscription signals reflect long-term goals (\eg learning a language, getting news), whereas play signals indicate immediate consumption. Over-optimizing for short-term metrics such as plays may harm long-term user interests and disadvantage certain creators. Integrating multiple signals and applying calibration or similar methods is therefore crucial to balance short-term engagement with long-term user goals~\cite{li2023autoopt}.

\subsubsection{Cold-start in Audio Recommendation}

For the cold-start problem, external knowledge injection is a particularly effective approach, also applied in news recommendation. For instance, \citet{xi2024towards} proposed a framework, which uses large language models to infer underlying user preference factors (\eg genre, style) and factual knowledge about items. These insights are transformed into enriched vectors via a mixture-of-experts adapter and concatenated with existing model inputs. Deployment on a music platform significantly improved metrics such as song plays, device coverage, and total listening duration.

Deezer~\cite{briand2021semi} tackled user cold-start using a semi-personalized model that combines clustering with deep neural networks: users are first clustered based on external data, then a deep model predicts preferences, aligning recommendations more closely with individual tastes while avoiding overfitting to sparse data. Similarly, \citet{feng2021zero} addressed zero-interaction users via a two-tower architecture: one tower generates “virtual behaviors” using a cross-modal dual autoencoder, while the other performs ranking. Joint training enables personalized recommendations even without historical interactions, providing a practical solution for cold-start scenarios.

\subsubsection{Diversity Exploration in Audio Recommendation}
For the diversity exploration problem, which is highly valued by audio platforms, several approaches have been proposed. \citet{bendada2023scalable} developed a playlist continuation method that supplements user-created playlists with suitable songs from a large music library. Their approach first learns independent vector representations for each song, integrating intrinsic attributes and usage data, and then sequentially aggregates vectors of songs in a playlist to infer the most appropriate candidates. Various sequence modeling techniques—convolutional networks for local patterns and recurrent networks with attention for long dependencies—were evaluated. Designed for scalability, this framework operates efficiently on millions of playlists with tens of millions of songs. Experiments on  Deezer showed a ~70\% increase in users adding recommended songs, enhancing both experience and playlist diversity.

Amazon Music~\cite{lichtenberg2024ranking} proposed multinomial blending to handle multiple content types (\eg music and podcasts). This method tunes exposure budgets per content type while preserving personalized rankings within each type and accounting for long-term value and category diversity, boosting engagement for slow-burn content without harming CTR. Similarly, Douyin Music’s “interest clock”~\cite{zhu2024interest} promotes exploration by adjusting time-based recommendation weights, preventing users from staying in homogeneous content loops. Query recommendation pipelines and rankers oriented toward exploratory intent~\cite{lindstrom2024encouraging} provide additional practical solutions.

\subsubsection{Multi-objective Optimization and Multi-signal Selection in Audio Recommendation}
A common solution is multi-signal modeling and calibration. For instance, in podcast recommendations, subscriptions and plays reflect different user intents: long-term vs. immediate interests. To balance these signals, industry practices often jointly model them or apply calibration techniques~\cite{nazari2022choice}.

Additionally, to reduce the cost of large-scale online experiments, practitioners sometimes employ counterfactual evaluation methods~\cite{mcinerney2020counterfactual} to simulate long-term effects offline before selecting strategies for smaller-scale A/B tests. This approach accounts for sequential interactions and better aligns recommendations with long-term user experience.
\section{Reflections and Future Directions for RecSys Research}
\label{sec:discussion}

We begin by reviewing the shifting optimization priorities of industrial recommender systems, and then outline the broader technical roadmap. We then propose  promising directions where academic and industrial research can jointly make meaningful contributions.

\begin{figure}
    \centering
    \includegraphics[trim = 4.3cm 5.5cm 5cm 4.5cm, clip,width=0.65\linewidth]{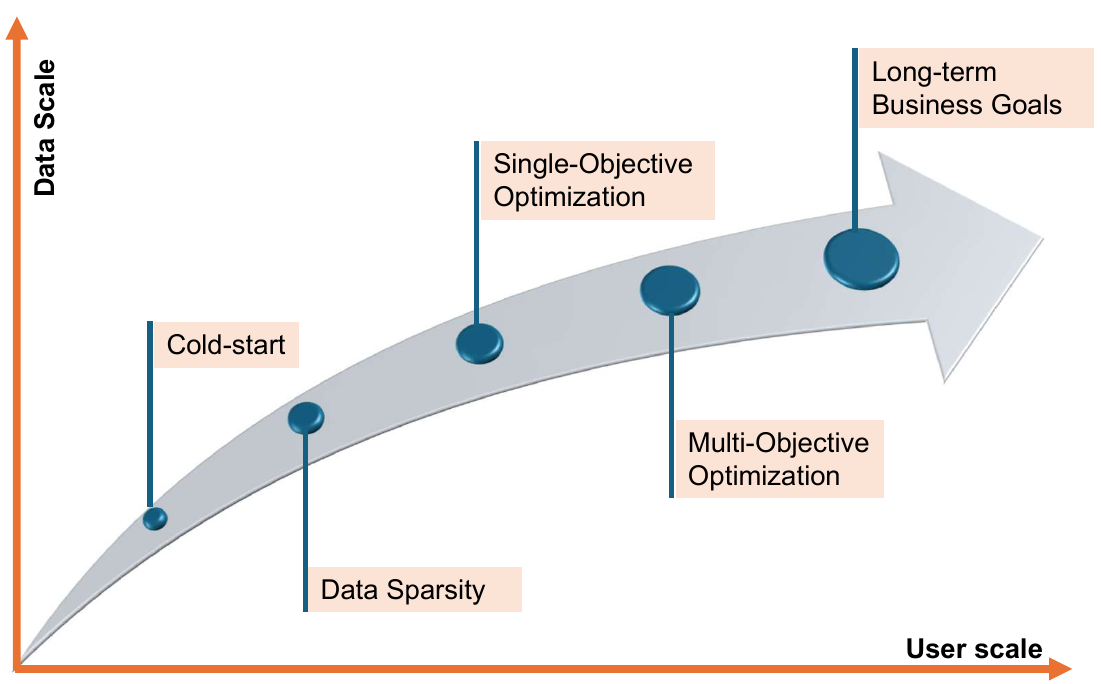}
    \caption{An illustration of the development of recommender systems in light of the increasing growth of both users and data (\eg items and user behavior records). }
    \label{fig:roadmap}
    \Description{The development of recommender systems}
\end{figure}

\subsection{Optimization Shifts in Industrial Recommender Systems}

Illustrated in Figure~\ref{fig:roadmap}, in the evolution of recommender systems, both the number of users and the volume of data (\eg items, features, and user interactions) have grown exponentially in a positively correlated manner. As business stages advance, the optimization focus has also shifted—industrial recommender technologies are propelled by the dual goals of serving users and creating business value. From a user-centered perspective: the cold-start stage emphasizes initial retention and conversion, laying the foundation of user trust. With the influx of new users, the challenge of sparsity emerges, requiring accurate modeling to deliver personalized results. Then, as businesses expand, behavior-based evaluation methods ensure that recommendation quality (\ie the single-objective at this stage) is measurable. Monetization, however, introduces complex trade-offs between technical and business metrics, making multi-objective optimization essential. Finally, long-term user value becomes the key to sustainable business growth, positioning long-term optimization as the next frontier. This progression depicts the overall development trajectory of industrial recommender systems. 

The evolution of recommendation technologies is closely tied to the scale and complexity of the recommendation tasks. As the volume of data increases and business scenarios become more complex, the methods employed become correspondingly more sophisticated. 

\subsection{The Technical Roadmap of Recommender Systems}

In the early development of recommender systems, \textit{heuristic methods} is widely applied.
These methods do not require complex models; instead, they rely on historical data, user behaviors, or simple rules to generate fast recommendations. Typical examples include collaborative filtering~\cite{xie2021causcf,10.1145/3627673.3680015} and rule-based recommendations~\cite{lichtenberg2024ranking}. Such methods are computationally efficient, easy to interpret, and particularly suitable for rapid deployment. They are sufficient to address many  industrial recommendation system requirements.

With the rapid growth of items and businesses, industry practitioners have widely adopted \textit{discriminative methods}, primarily relying on supervised learning optimized for goal-oriented and quantifiable metrics. These methods typically leverage explicit supervised signals (\eg click-through rate~\cite{sun2020neighbor,feng2021zero}, conversion rate~\cite{xu2020privileged}, dwell time~\cite{lichtenberg2024ranking}, GMV~\cite{chai2022user}) to directly optimize business objectives. They have proven particularly effective for large-scale prediction tasks such as CTR, CVR, and GMV optimization. Discriminative approaches encompass classical machine learning, deep learning, incremental and continual learning, as well as reinforcement learning, all of which aim to significantly enhance business performance in industrial recommender systems.

For further optimization, \textit{generative methods} have been explored to capture latent user interests and feature distributions by modeling intrinsic data patterns. These approaches employ various techniques including graph neural networks, generative adversarial networks, meta-learning, Transformer-based architectures, and large language models. Generative methods are particularly well-suited for addressing challenges such as cold-start scenarios~\cite{gong2023unified}, cross-domain recommendation~\cite{huai2023m2gnn}, new content discovery~\cite{baran2023accelerating}, and multimodal contexts~\cite{zhang2023multi}. While discriminative and generative approaches both stem from a problem-solving perspective, they give rise to distinct technical paradigms optimized for specific objectives. Notably, with the rise of large models, the paradigm itself has begun to shift: generative recommendation techniques, anchored in the scaling law, are emerging as the foundational backbone of next-generation recommender systems.

In the past two years, industrial exploration has begun to fundamentally reframe recommender systems as generative modeling tasks. The depth of this paradigm shift is comparable to the transformation in natural language processing from discriminative classifiers to foundation models. For example, \citet{zhai2024actions} introduced a generative recommendation framework that was the first to validate the principle of “bigger is better” in a large-scale industrial environment. Scaling the model to the trillion-parameter level, they observed no performance saturation, effectiveness continued to improve. To make such massive models practical, they proposed several key architectural innovations: (1) unifying heterogeneous user behaviors and features into a single long sequence, enabling direct learning akin to language models; (2) applying an intelligent compression mechanism to preserve critical information while discarding redundancy in ultra-long histories; and (3) adopting parallelization and shared computation for efficient inference across thousands of candidates. Collectively, these advances established the feasibility of foundation-scale models for recommendation and confirmed that recommender systems also follow the scaling law.

Building on this paradigm, subsequent works extended generative modeling across the recommendation pipeline. \citet{firooz2025360brew} reformulated ranking as next-token prediction over user–item sequences, unifying multiple recommendation tasks within a decoder-only foundation model. For retrieval, \citet{badrinath2025pinrec} replaced vector search with outcome-conditioned multi-token generation, directly producing candidate item IDs while aligning with business objectives such as clicks, saves, and conversions. On the ranking side, \citet{han2025mtgr} and \citet{huang2025towards} departed from score-based methods, treating recommendation as the generation of an ordered list—similar to how a language model writes a coherent sentence. Although list generation faces computational challenges, efficient generation strategies were introduced, enabling industrial deployment without latency overhead, and surpassing traditional ranking models.

Beyond retrieval and ranking, \citet{deng2025onerec} proposed OneRec, a single-stage end-to-end generative framework that eliminates the retrieve–rank cascade. By directly generating ordered lists from user histories and incorporating instruction tuning and preference optimization, OneRec achieves tighter alignment with user value. Meanwhile, \citet{lan2025next} offered an inversion of the problem: rather than generating the next item for a user, it generates the next user likely to interact with an item, which is particularly effective in cold-start scenarios.

Taken together, these efforts delineate a clear evolutionary trajectory: generative recommendation has moved from a problem-driven auxiliary solution to the foundation of large-scale industrial systems. The emerging trend emphasizes: (1) Unification of retrieval and ranking under a single generative framework, (2) Multi-objective alignment with diverse business goals, (3) Validation of scaling laws in recommender systems, and (4) Exploration of user-side generation for new scenarios.

Compared with discriminative models, which rely heavily on feature engineering and task-specific customization, generative approaches aspire to provide a unified, scalable, and generalizable foundation. Moreover, generative modeling can directly optimize long-term objectives such as watch time, conversion, and retention, rather than focusing solely on pointwise CTR prediction. This marks generative recommendation not as a marginal experiment, but as the emerging core paradigm of recommender systems in the current era.

\subsection{RecSys Research in Academia}

Recommender systems are inherently socio-technical in nature, sitting at the intersection of \textit{computing} (algorithms and large-scale systems), \textit{economics} (market mechanisms and traffic allocation), and \textit{psychology} (user cognition and behavior), among many other disciplines. This complexity means that simply refining model architectures or focusing on algorithmic and system-level improvements is insufficient to address the deeper challenges of the field. Short-term performance gains alone cannot ensure sustainable user value or ecosystem health. 

In the following, we highlight several long-overlooked issues and propose research directions aimed at bridging the persistent gap between academic research and industrial practice. Our goal is to encourage a more holistic perspective on recommender systems—one that not only advances technical frontiers but also fosters greater value creation for both users and platforms.

Due to practical constraints, academia often lacks the conditions necessary to conduct A/B testing, and thus most evaluations rely on offline experiments with simulated outcomes. The datasets used are typically small-scale and highly curated to facilitate benchmarking, with many already outdated and unable to accurately reflect recent user behavior patterns~\cite{BauerEvaluation}. Consequently, many proposed methods cannot be rigorously validated for their real value to users, leading to misaligned research incentives and a partial waste of academic resources. This gap has widened further with the recent trend of deploying large-scale foundation models in industry~\cite{khrylchenko2025scaling}, making it increasingly difficult for academic experiments—constrained by limited data, compute, and evaluation settings—to provide fair or meaningful comparisons with industrial systems.

Nevertheless, academia continues to play an indispensable role by providing the theoretical foundations and methodological innovations that underpin the sustainable growth of RecSys. While industry excels in large-scale deployment and optimization, academic research offers the rigor and creativity needed to explore new paradigms and guide long-term development. Building on our review of industrial recommender systems, we identify several key directions where academic efforts could generate high impact and help bridge the gap between theory and practice.

\subsubsection{A Deeper Interpretation of User Behavior}
Whether in industry or academia, most recommender systems still rely almost exclusively on users’ \textit{observed behaviors} (\eg clicks, views, purchases) to infer preferences. However, the decision-making processes that precede these observable actions are often neglected. For instance, surveys of thousands of users reveal that before a click occurs, users frequently experience a state of hesitation~\cite{zou2024hesitation}. A click, therefore, does not always signal genuine interest—it may simply reflect an attempt to gather more information. Similarly, when users click and then spend time consuming content they are not truly interested in, they often develop negative sentiments toward the system, a phenomenon referred to as \textit{tolerance}. Incorporating such pre-action and post-action cognitive phenomena into recommendation optimization holds promise for improving long-term outcomes such as user satisfaction and retention.

However, from the users’ perspective, psychological states are inherently complex and dynamic, with tolerance being only one of many possible reactions. We believe that investigating users’ decision-making states—by systematically incorporating concepts from psychology and cognitive science into recommender systems—can deepen practitioners’ understanding of users, thereby unlocking both greater user value and commercial value. For example, a user who watches a video to completion does not necessarily do so out of genuine interest; the behavior may instead be driven by innate curiosity about the unknown. In such cases, repeatedly recommending similar content may result in dissatisfaction rather than engagement. We contend that this line of research holds substantial practical significance and has the potential to open new directions for industrial recommender system design. 

\subsubsection{Theory-Guided Optimization of Industrial Recommendation}

Industrial recommender systems must balance multiple objectives, including profitability, recommendation accuracy, and system performance. Yet, mainstream practice often relies on empirical weight tuning of metrics such as CTR, CVR, GMV, and watch time. This trial-and-error approach lacks theoretical grounding, is unstable across settings, and makes development inefficient and uncertain.

We expect academia to provide systematic theoretical support for defining and optimizing multi-objective functions. In the long term, this means integrating factors such as user retention, cross-period utility, and satisfaction to move beyond short-term conversion toward long-term user value. In the short term, it involves formulating metrics like CTR, latency, and cost as constraints or secondary objectives, and applying methods such as multi-objective optimization, hierarchical optimization, and causal inference to achieve principled trade-offs.

Such research can shift industry practice from empirical tuning to theoretically grounded modeling, reducing tuning costs, improving efficiency, and achieving sustainable trade-offs across objectives. This not only leads to more robust algorithms but also fosters reusable optimization paradigms, laying the foundation for the long-term development of recommender systems.

\subsubsection{Problem Definitions or Methodologies More Aligned with Real-World Practice}

While academia drives theoretical advances, it often prioritizes “model elegance” without fully aligning with the practical constraints and problem definitions faced in industry. 

Academic research frequently assumes “unlimited computation” and single-objective optimization, overlooking latency, cost, and energy limitations. In contrast, industrial systems must return results within milliseconds and operate under strict GPU/CPU budgets. We suggest explicitly integrating latency, computational cost, and energy consumption into optimization problems and advancing research on accuracy–efficiency trade-offs.

Many benchmarks like MovieLens and Amazon feature low sparsity, mild cold-start issues, and stationary distributions, which differ markedly from real platforms~\cite{MovieLensIssue}. We recommend working on more recent public datasets that better capture realistic user behavior~\cite{MindDataset,Yambda5B} and explicitly modeling dataset shift and distribution drift to improve the transferability of academic methods to industry.

Academic studies often treat the platform as serving only users, whereas industrial systems must balance the interests of users, item providers, content creators, and advertisers, creating traffic allocation dilemmas. Incorporating mechanism design and game theory into research may help study stable multi-party recommendation mechanisms, while treating fairness and guaranteed exposure as hard constraints to reflect industrial realities.

Academic evaluations rely heavily on offline metrics in relaxed settings~\cite{JiDataLeakage,FreshlookSun23}, which often diverge from online A/B testing outcomes, and many researchers lack access to online experimentation. We propose developing theoretical guarantees for offline–online consistency and building universal simulation environments to replace traditional benchmarks, bringing academic experiments closer to real-world business decision-making while reducing research costs.

\subsection{RecSys Research in Industry}

Although thousands of papers on recommender systems are published annually, relatively few methods are directly deployable in production. As a result, practitioners often build systems from scratch and continuously optimize them. We therefore call for greater participation in pretraining and the release of production-tested, out-of-the-box recommender models, accompanied by effective tuning methodologies, practical implementation guidelines, and strategies for integrating new research into real applications. Such efforts could significantly reduce redundant development and save human resources in industry.

We also encourage the creation of datasets tailored to specific objectives, providing stronger evaluative power for targeted optimization. Transforming these datasets into user-friendly simulation environments would further enable researchers from both academia and industry to conduct meaningful studies, accelerating problem-solving and advancing the field.

Finally, optimizing for cost and performance at the system or platform level is a major component of industrial work, yet such research is rarely publicly available. Academia often lacks insight into user needs, business constraints, and real-world optimization strategies beyond purely technical aspects. Greater dissemination of research addressing these practical challenges would allow academic researchers to better understand industrial contexts and provide more meaningful, applicable guidance.
\section{Conclusion}
\label{sec:conclusion}

In this work, we review the applications of recommender systems in industry, covering both online transaction scenarios and content scenarios, including video, news, and audio. We highlight the business characteristics and objectives of these scenarios and, based on their respective technical challenges, categorize and analyze recent papers published at major conferences, presenting our perspectives. We also discuss future research directions, including potential theoretical advances from academia, industrial development trends, and opportunities for collaboration between the two communities.

Our survey argues that, constrained by practical conditions, academic research on recommender systems emphasizes algorithms, while industry must additionally balance system performance, cost, and effectiveness within a broader evaluation framework centered on commercial profitability. Looking ahead, we hope to see more practically meaningful research in recommender systems, particularly work that unites the strengths of academia and industry to jointly advance the flourishing development of the field.


\bibliographystyle{ACM-Reference-Format}
\bibliography{reference}


\begin{thebibliography}{193}


\ifx \showCODEN    \undefined \def \showCODEN     #1{\unskip}     \fi
\ifx \showISBNx    \undefined \def \showISBNx     #1{\unskip}     \fi
\ifx \showISBNxiii \undefined \def \showISBNxiii  #1{\unskip}     \fi
\ifx \showISSN     \undefined \def \showISSN      #1{\unskip}     \fi
\ifx \showLCCN     \undefined \def \showLCCN      #1{\unskip}     \fi
\ifx \shownote     \undefined \def \shownote      #1{#1}          \fi
\ifx \showarticletitle \undefined \def \showarticletitle #1{#1}   \fi
\ifx \showURL      \undefined \def \showURL       {\relax}        \fi
\providecommand\bibfield[2]{#2}
\providecommand\bibinfo[2]{#2}
\providecommand\natexlab[1]{#1}
\providecommand\showeprint[2][]{arXiv:#2}

\bibitem[Agrawal et~al\mbox{.}(2023)]%
        {agrawal2023beyond}
\bibfield{author}{\bibinfo{person}{Saurabh Agrawal}, \bibinfo{person}{John Trenkle}, {and} \bibinfo{person}{Jaya Kawale}.} \bibinfo{year}{2023}\natexlab{}.
\newblock \showarticletitle{Beyond labels: Leveraging deep learning and llms for content metadata}. In \bibinfo{booktitle}{\emph{Proceedings of the 17th ACM Conference on Recommender Systems}}. \bibinfo{pages}{1--1}.
\newblock


\bibitem[An et~al\mbox{.}(2024)]%
        {an2024ddcdr}
\bibfield{author}{\bibinfo{person}{Zhicheng An}, \bibinfo{person}{Zhexu Gu}, \bibinfo{person}{Li Yu}, \bibinfo{person}{Ke Tu}, \bibinfo{person}{Zhengwei Wu}, \bibinfo{person}{Binbin Hu}, \bibinfo{person}{Zhiqiang Zhang}, \bibinfo{person}{Lihong Gu}, {and} \bibinfo{person}{Jinjie Gu}.} \bibinfo{year}{2024}\natexlab{}.
\newblock \showarticletitle{DDCDR: A Disentangle-based Distillation Framework for Cross-Domain Recommendation}. In \bibinfo{booktitle}{\emph{Proceedings of the 30th ACM SIGKDD Conference on Knowledge Discovery and Data Mining}}. \bibinfo{pages}{4764--4773}.
\newblock


\bibitem[Badrinath et~al\mbox{.}(2025)]%
        {badrinath2025pinrec}
\bibfield{author}{\bibinfo{person}{Anirudhan Badrinath}, \bibinfo{person}{Prabhat Agarwal}, \bibinfo{person}{Laksh Bhasin}, \bibinfo{person}{Jaewon Yang}, \bibinfo{person}{Jiajing Xu}, {and} \bibinfo{person}{Charles Rosenberg}.} \bibinfo{year}{2025}\natexlab{}.
\newblock \showarticletitle{PinRec: Outcome-Conditioned, Multi-Token Generative Retrieval for Industry-Scale Recommendation Systems}.
\newblock \bibinfo{journal}{\emph{arXiv preprint arXiv:2504.10507}} (\bibinfo{year}{2025}).
\newblock


\bibitem[Bai et~al\mbox{.}(2022)]%
        {bai2022contrastive}
\bibfield{author}{\bibinfo{person}{Ting Bai}, \bibinfo{person}{Yudong Xiao}, \bibinfo{person}{Bin Wu}, \bibinfo{person}{Guojun Yang}, \bibinfo{person}{Hongyong Yu}, {and} \bibinfo{person}{Jian-Yun Nie}.} \bibinfo{year}{2022}\natexlab{}.
\newblock \showarticletitle{A contrastive sharing model for multi-task recommendation}. In \bibinfo{booktitle}{\emph{Proceedings of the ACM web conference 2022}}. \bibinfo{pages}{3239--3247}.
\newblock


\bibitem[Bai et~al\mbox{.}(2024)]%
        {bai2024gradcraft}
\bibfield{author}{\bibinfo{person}{Yimeng Bai}, \bibinfo{person}{Yang Zhang}, \bibinfo{person}{Fuli Feng}, \bibinfo{person}{Jing Lu}, \bibinfo{person}{Xiaoxue Zang}, \bibinfo{person}{Chenyi Lei}, {and} \bibinfo{person}{Yang Song}.} \bibinfo{year}{2024}\natexlab{}.
\newblock \showarticletitle{GradCraft: Elevating Multi-task Recommendations through Holistic Gradient Crafting}. In \bibinfo{booktitle}{\emph{Proceedings of the 30th ACM SIGKDD Conference on Knowledge Discovery and Data Mining}}. \bibinfo{pages}{4774--4783}.
\newblock


\bibitem[Baltescu et~al\mbox{.}(2022)]%
        {baltescu2022itemsage}
\bibfield{author}{\bibinfo{person}{Paul Baltescu}, \bibinfo{person}{Haoyu Chen}, \bibinfo{person}{Nikil Pancha}, \bibinfo{person}{Andrew Zhai}, \bibinfo{person}{Jure Leskovec}, {and} \bibinfo{person}{Charles Rosenberg}.} \bibinfo{year}{2022}\natexlab{}.
\newblock \showarticletitle{Itemsage: Learning product embeddings for shopping recommendations at pinterest}. In \bibinfo{booktitle}{\emph{Proceedings of the 28th ACM SIGKDD Conference on Knowledge Discovery and Data Mining}}. \bibinfo{pages}{2703--2711}.
\newblock


\bibitem[Baran et~al\mbox{.}(2023)]%
        {baran2023accelerating}
\bibfield{author}{\bibinfo{person}{Buket Baran}, \bibinfo{person}{Guilherme~Dinis Junior}, \bibinfo{person}{Antonina Danylenko}, \bibinfo{person}{Olayinka~S Folorunso}, \bibinfo{person}{G{\"o}sta Forsum}, \bibinfo{person}{Maksym Lefarov}, \bibinfo{person}{Lucas Maystre}, {and} \bibinfo{person}{Yu Zhao}.} \bibinfo{year}{2023}\natexlab{}.
\newblock \showarticletitle{Accelerating creator audience building through centralized exploration}. In \bibinfo{booktitle}{\emph{Proceedings of the 17th ACM Conference on Recommender Systems}}. \bibinfo{pages}{70--73}.
\newblock


\bibitem[Bauer et~al\mbox{.}(2024)]%
        {BauerEvaluation}
\bibfield{author}{\bibinfo{person}{Christine Bauer}, \bibinfo{person}{Eva Zangerle}, {and} \bibinfo{person}{Alan Said}.} \bibinfo{year}{2024}\natexlab{}.
\newblock \showarticletitle{Exploring the Landscape of Recommender Systems Evaluation: Practices and Perspectives}.
\newblock \bibinfo{journal}{\emph{ACM Trans. Recomm. Syst.}} \bibinfo{volume}{2}, \bibinfo{number}{1}, Article \bibinfo{articleno}{11} (\bibinfo{date}{March} \bibinfo{year}{2024}), \bibinfo{numpages}{31}~pages.
\newblock
\href{https://doi.org/10.1145/3629170}{doi:\nolinkurl{10.1145/3629170}}


\bibitem[Bendada et~al\mbox{.}(2023)]%
        {bendada2023scalable}
\bibfield{author}{\bibinfo{person}{Walid Bendada}, \bibinfo{person}{Guillaume Salha-Galvan}, \bibinfo{person}{Thomas Bouab{\c{c}}a}, {and} \bibinfo{person}{Tristan Cazenave}.} \bibinfo{year}{2023}\natexlab{}.
\newblock \showarticletitle{A scalable framework for automatic playlist continuation on music streaming services}. In \bibinfo{booktitle}{\emph{Proceedings of the 46th International ACM SIGIR Conference on Research and Development in Information Retrieval}}. \bibinfo{pages}{464--474}.
\newblock


\bibitem[Bi et~al\mbox{.}(2020)]%
        {bi2020heterogeneous}
\bibfield{author}{\bibinfo{person}{Ye Bi}, \bibinfo{person}{Liqiang Song}, \bibinfo{person}{Mengqiu Yao}, \bibinfo{person}{Zhenyu Wu}, \bibinfo{person}{Jianming Wang}, {and} \bibinfo{person}{Jing Xiao}.} \bibinfo{year}{2020}\natexlab{}.
\newblock \showarticletitle{A heterogeneous information network based cross domain insurance recommendation system for cold start users}. In \bibinfo{booktitle}{\emph{Proceedings of the 43rd international ACM SIGIR conference on research and development in information retrieval}}. \bibinfo{pages}{2211--2220}.
\newblock


\bibitem[Briand et~al\mbox{.}(2021)]%
        {briand2021semi}
\bibfield{author}{\bibinfo{person}{L{\'e}a Briand}, \bibinfo{person}{Guillaume Salha-Galvan}, \bibinfo{person}{Walid Bendada}, \bibinfo{person}{Mathieu Morlon}, {and} \bibinfo{person}{Viet-Anh Tran}.} \bibinfo{year}{2021}\natexlab{}.
\newblock \showarticletitle{A semi-personalized system for user cold start recommendation on music streaming apps}. In \bibinfo{booktitle}{\emph{Proceedings of the 27th ACM SIGKDD conference on knowledge discovery \& data mining}}. \bibinfo{pages}{2601--2609}.
\newblock


\bibitem[Cai et~al\mbox{.}(2022)]%
        {cai2022reloop}
\bibfield{author}{\bibinfo{person}{Guohao Cai}, \bibinfo{person}{Jieming Zhu}, \bibinfo{person}{Quanyu Dai}, \bibinfo{person}{Zhenhua Dong}, \bibinfo{person}{Xiuqiang He}, \bibinfo{person}{Ruiming Tang}, {and} \bibinfo{person}{Rui Zhang}.} \bibinfo{year}{2022}\natexlab{}.
\newblock \showarticletitle{Reloop: A self-correction continual learning loop for recommender systems}. In \bibinfo{booktitle}{\emph{Proceedings of the 45th International ACM SIGIR Conference on Research and Development in Information Retrieval}}. \bibinfo{pages}{2692--2697}.
\newblock


\bibitem[Cai et~al\mbox{.}(2023)]%
        {cai2023model}
\bibfield{author}{\bibinfo{person}{Tianchi Cai}, \bibinfo{person}{Shenliao Bao}, \bibinfo{person}{Jiyan Jiang}, \bibinfo{person}{Shiji Zhou}, \bibinfo{person}{Wenpeng Zhang}, \bibinfo{person}{Lihong Gu}, \bibinfo{person}{Jinjie Gu}, {and} \bibinfo{person}{Guannan Zhang}.} \bibinfo{year}{2023}\natexlab{}.
\newblock \showarticletitle{Model-free Reinforcement Learning with Stochastic Reward Stabilization for Recommender Systems}. In \bibinfo{booktitle}{\emph{Proceedings of the 46th International ACM SIGIR Conference on Research and Development in Information Retrieval}}. \bibinfo{pages}{2179--2183}.
\newblock


\bibitem[Celikik et~al\mbox{.}(2022)]%
        {celikik2022reusable}
\bibfield{author}{\bibinfo{person}{Marjan Celikik}, \bibinfo{person}{Ana Peleteiro~Ramallo}, {and} \bibinfo{person}{Jacek Wasilewski}.} \bibinfo{year}{2022}\natexlab{}.
\newblock \showarticletitle{Reusable Self-Attention Recommender Systems in Fashion Industry Applications}. In \bibinfo{booktitle}{\emph{Proceedings of the 16th ACM Conference on Recommender Systems}}. \bibinfo{pages}{448--451}.
\newblock


\bibitem[Chai et~al\mbox{.}(2022)]%
        {chai2022user}
\bibfield{author}{\bibinfo{person}{Zheng Chai}, \bibinfo{person}{Zhihong Chen}, \bibinfo{person}{Chenliang Li}, \bibinfo{person}{Rong Xiao}, \bibinfo{person}{Houyi Li}, \bibinfo{person}{Jiawei Wu}, \bibinfo{person}{Jingxu Chen}, {and} \bibinfo{person}{Haihong Tang}.} \bibinfo{year}{2022}\natexlab{}.
\newblock \showarticletitle{User-aware multi-interest learning for candidate matching in recommenders}. In \bibinfo{booktitle}{\emph{Proceedings of the 45th international ACM SIGIR conference on research and development in information retrieval}}. \bibinfo{pages}{1326--1335}.
\newblock


\bibitem[Chen et~al\mbox{.}(2024c)]%
        {chen2024missing}
\bibfield{author}{\bibinfo{person}{Gaode Chen}, \bibinfo{person}{Yuezihan Jiang}, \bibinfo{person}{Rui Huang}, \bibinfo{person}{Kuo Cai}, \bibinfo{person}{Yunze Luo}, \bibinfo{person}{Ruina Sun}, \bibinfo{person}{Qi Zhang}, \bibinfo{person}{Han Li}, {and} \bibinfo{person}{Kun Gai}.} \bibinfo{year}{2024}\natexlab{c}.
\newblock \showarticletitle{Missing Interest Modeling with Lifelong User Behavior Data for Retrieval Recommendation}. In \bibinfo{booktitle}{\emph{Proceedings of the 33rd ACM International Conference on Information and Knowledge Management}}. \bibinfo{pages}{4390--4396}.
\newblock


\bibitem[Chen et~al\mbox{.}(2024d)]%
        {chen2024multi}
\bibfield{author}{\bibinfo{person}{Gaode Chen}, \bibinfo{person}{Ruina Sun}, \bibinfo{person}{Yuezihan Jiang}, \bibinfo{person}{Jiangxia Cao}, \bibinfo{person}{Qi Zhang}, \bibinfo{person}{Jingjian Lin}, \bibinfo{person}{Han Li}, \bibinfo{person}{Kun Gai}, {and} \bibinfo{person}{Xinghua Zhang}.} \bibinfo{year}{2024}\natexlab{d}.
\newblock \showarticletitle{A Multi-modal Modeling Framework for Cold-start Short-video Recommendation}. In \bibinfo{booktitle}{\emph{Proceedings of the 18th ACM Conference on Recommender Systems}}. \bibinfo{pages}{391--400}.
\newblock


\bibitem[Chen et~al\mbox{.}(2024b)]%
        {chen2024macro}
\bibfield{author}{\bibinfo{person}{Hao Chen}, \bibinfo{person}{Yuanchen Bei}, \bibinfo{person}{Qijie Shen}, \bibinfo{person}{Yue Xu}, \bibinfo{person}{Sheng Zhou}, \bibinfo{person}{Wenbing Huang}, \bibinfo{person}{Feiran Huang}, \bibinfo{person}{Senzhang Wang}, {and} \bibinfo{person}{Xiao Huang}.} \bibinfo{year}{2024}\natexlab{b}.
\newblock \showarticletitle{Macro graph neural networks for online billion-scale recommender systems}. In \bibinfo{booktitle}{\emph{Proceedings of the ACM web conference 2024}}. \bibinfo{pages}{3598--3608}.
\newblock


\bibitem[Chen et~al\mbox{.}(2020)]%
        {chen2020improving}
\bibfield{author}{\bibinfo{person}{Suming~J Chen}, \bibinfo{person}{Zhen Qin}, \bibinfo{person}{Zac Wilson}, \bibinfo{person}{Brian Calaci}, \bibinfo{person}{Michael Rose}, \bibinfo{person}{Ryan Evans}, \bibinfo{person}{Sean Abraham}, \bibinfo{person}{Donald Metzler}, \bibinfo{person}{Sandeep Tata}, {and} \bibinfo{person}{Michael Colagrosso}.} \bibinfo{year}{2020}\natexlab{}.
\newblock \showarticletitle{Improving recommendation quality in google drive}. In \bibinfo{booktitle}{\emph{Proceedings of the 26th ACM SIGKDD international conference on knowledge discovery \& data mining}}. \bibinfo{pages}{2900--2908}.
\newblock


\bibitem[Chen et~al\mbox{.}(2024a)]%
        {10.1145/3627673.3680015}
\bibfield{author}{\bibinfo{person}{Weijun Chen}, \bibinfo{person}{Yuanchen Bei}, \bibinfo{person}{Qijie Shen}, \bibinfo{person}{Hao Chen}, \bibinfo{person}{Xiao Huang}, {and} \bibinfo{person}{Feiran Huang}.} \bibinfo{year}{2024}\natexlab{a}.
\newblock \showarticletitle{Feedback Reciprocal Graph Collaborative Filtering}. In \bibinfo{booktitle}{\emph{Proceedings of the 33rd ACM International Conference on Information and Knowledge Management}} (Boise, ID, USA) \emph{(\bibinfo{series}{CIKM '24})}. \bibinfo{publisher}{Association for Computing Machinery}, \bibinfo{address}{New York, NY, USA}, \bibinfo{pages}{4397–4405}.
\newblock
\showISBNx{9798400704369}
\href{https://doi.org/10.1145/3627673.3680015}{doi:\nolinkurl{10.1145/3627673.3680015}}


\bibitem[Chen et~al\mbox{.}(2024e)]%
        {chen2024shopping}
\bibfield{author}{\bibinfo{person}{Yankai Chen}, \bibinfo{person}{Quoc-Tuan Truong}, \bibinfo{person}{Xin Shen}, \bibinfo{person}{Jin Li}, {and} \bibinfo{person}{Irwin King}.} \bibinfo{year}{2024}\natexlab{e}.
\newblock \showarticletitle{Shopping trajectory representation learning with pre-training for e-commerce customer understanding and recommendation}. In \bibinfo{booktitle}{\emph{Proceedings of the 30th ACM SIGKDD Conference on Knowledge Discovery and Data Mining}}. \bibinfo{pages}{385--396}.
\newblock


\bibitem[Chen et~al\mbox{.}(2023)]%
        {chen2023contextual}
\bibfield{author}{\bibinfo{person}{Yan Chen}, \bibinfo{person}{Emilian Vankov}, \bibinfo{person}{Linas Baltrunas}, \bibinfo{person}{Preston Donovan}, \bibinfo{person}{Akash Mehta}, \bibinfo{person}{Benjamin Schroeder}, {and} \bibinfo{person}{Matthew Herman}.} \bibinfo{year}{2023}\natexlab{}.
\newblock \showarticletitle{Contextual multi-armed bandit for email layout recommendation}. In \bibinfo{booktitle}{\emph{Proceedings of the 17th ACM Conference on Recommender Systems}}. \bibinfo{pages}{400--402}.
\newblock


\bibitem[Chen et~al\mbox{.}(2021)]%
        {chen2021curriculum}
\bibfield{author}{\bibinfo{person}{Yudong Chen}, \bibinfo{person}{Xin Wang}, \bibinfo{person}{Miao Fan}, \bibinfo{person}{Jizhou Huang}, \bibinfo{person}{Shengwen Yang}, {and} \bibinfo{person}{Wenwu Zhu}.} \bibinfo{year}{2021}\natexlab{}.
\newblock \showarticletitle{Curriculum meta-learning for next POI recommendation}. In \bibinfo{booktitle}{\emph{Proceedings of the 27th ACM SIGKDD Conference on Knowledge Discovery \& Data Mining}}. \bibinfo{pages}{2692--2702}.
\newblock


\bibitem[Chen et~al\mbox{.}(2022)]%
        {chen2022co}
\bibfield{author}{\bibinfo{person}{Zhihong Chen}, \bibinfo{person}{Jiawei Wu}, \bibinfo{person}{Chenliang Li}, \bibinfo{person}{Jingxu Chen}, \bibinfo{person}{Rong Xiao}, {and} \bibinfo{person}{Binqiang Zhao}.} \bibinfo{year}{2022}\natexlab{}.
\newblock \showarticletitle{Co-training disentangled domain adaptation network for leveraging popularity bias in recommenders}. In \bibinfo{booktitle}{\emph{Proceedings of the 45th International ACM SIGIR conference on research and development in information retrieval}}. \bibinfo{pages}{60--69}.
\newblock


\bibitem[Dai et~al\mbox{.}(2023)]%
        {10.1145/3583780.3615218}
\bibfield{author}{\bibinfo{person}{Wei Dai}, \bibinfo{person}{Yingmin Su}, \bibinfo{person}{Xiaofeng Pan}, \bibinfo{person}{Yufeng Wang}, \bibinfo{person}{Zhenyu Zhu}, \bibinfo{person}{Nan Xu}, \bibinfo{person}{Chengjun Mao}, {and} \bibinfo{person}{Bo Cao}.} \bibinfo{year}{2023}\natexlab{}.
\newblock \showarticletitle{DPAN: Dynamic Preference-based and Attribute-aware Network for Relevant Recommendations}. In \bibinfo{booktitle}{\emph{Proceedings of the 32nd ACM International Conference on Information and Knowledge Management}} (Birmingham, United Kingdom) \emph{(\bibinfo{series}{CIKM '23})}. \bibinfo{publisher}{Association for Computing Machinery}, \bibinfo{address}{New York, NY, USA}, \bibinfo{pages}{3838–3842}.
\newblock
\showISBNx{9798400701245}
\href{https://doi.org/10.1145/3583780.3615218}{doi:\nolinkurl{10.1145/3583780.3615218}}


\bibitem[Deng et~al\mbox{.}(2025)]%
        {deng2025onerec}
\bibfield{author}{\bibinfo{person}{Jiaxin Deng}, \bibinfo{person}{Shiyao Wang}, \bibinfo{person}{Kuo Cai}, \bibinfo{person}{Lejian Ren}, \bibinfo{person}{Qigen Hu}, \bibinfo{person}{Weifeng Ding}, \bibinfo{person}{Qiang Luo}, {and} \bibinfo{person}{Guorui Zhou}.} \bibinfo{year}{2025}\natexlab{}.
\newblock \showarticletitle{Onerec: Unifying retrieve and rank with generative recommender and iterative preference alignment}.
\newblock \bibinfo{journal}{\emph{arXiv preprint arXiv:2502.18965}} (\bibinfo{year}{2025}).
\newblock


\bibitem[Deng et~al\mbox{.}(2024)]%
        {deng2024mmbee}
\bibfield{author}{\bibinfo{person}{Jiaxin Deng}, \bibinfo{person}{Shiyao Wang}, \bibinfo{person}{Yuchen Wang}, \bibinfo{person}{Jiansong Qi}, \bibinfo{person}{Liqin Zhao}, \bibinfo{person}{Guorui Zhou}, {and} \bibinfo{person}{Gaofeng Meng}.} \bibinfo{year}{2024}\natexlab{}.
\newblock \showarticletitle{MMBee: Live Streaming Gift-Sending Recommendations via Multi-Modal Fusion and Behaviour Expansion}. In \bibinfo{booktitle}{\emph{Proceedings of the 30th ACM SIGKDD Conference on Knowledge Discovery and Data Mining}}. \bibinfo{pages}{4896--4905}.
\newblock


\bibitem[Edizel et~al\mbox{.}(2024)]%
        {edizel2024towards}
\bibfield{author}{\bibinfo{person}{Bora Edizel}, \bibinfo{person}{Tim Sweetser}, \bibinfo{person}{Ashok Chandrashekar}, \bibinfo{person}{Kamilia Ahmadi}, {and} \bibinfo{person}{Puja Das}.} \bibinfo{year}{2024}\natexlab{}.
\newblock \showarticletitle{Towards understanding the gaps of offline and online evaluation metrics: Impact of series vs. movie recommendations}. In \bibinfo{booktitle}{\emph{Proceedings of the 18th ACM Conference on Recommender Systems}}. \bibinfo{pages}{844--846}.
\newblock


\bibitem[Fan et~al\mbox{.}(2024)]%
        {MovieLensIssue}
\bibfield{author}{\bibinfo{person}{Yu-Chen Fan}, \bibinfo{person}{Yitong Ji}, \bibinfo{person}{Jie Zhang}, {and} \bibinfo{person}{Aixin Sun}.} \bibinfo{year}{2024}\natexlab{}.
\newblock \showarticletitle{Our Model Achieves Excellent Performance on MovieLens: What Does It Mean?}
\newblock \bibinfo{journal}{\emph{ACM Trans. Inf. Syst.}} \bibinfo{volume}{42}, \bibinfo{number}{6}, Article \bibinfo{articleno}{159} (\bibinfo{date}{Oct.} \bibinfo{year}{2024}), \bibinfo{numpages}{25}~pages.
\newblock
\showISSN{1046-8188}
\href{https://doi.org/10.1145/3675163}{doi:\nolinkurl{10.1145/3675163}}


\bibitem[Fang et~al\mbox{.}(2023)]%
        {fang2023alleviating}
\bibfield{author}{\bibinfo{person}{Junpeng Fang}, \bibinfo{person}{Qing Cui}, \bibinfo{person}{Gongduo Zhang}, \bibinfo{person}{Caizhi Tang}, \bibinfo{person}{Lihong Gu}, \bibinfo{person}{Longfei Li}, \bibinfo{person}{Jinjie Gu}, \bibinfo{person}{Jun Zhou}, {and} \bibinfo{person}{Fei Wu}.} \bibinfo{year}{2023}\natexlab{}.
\newblock \showarticletitle{Alleviating Matching Bias in Marketing Recommendations}. In \bibinfo{booktitle}{\emph{Proceedings of the 46th International ACM SIGIR Conference on Research and Development in Information Retrieval}}. \bibinfo{pages}{3359--3363}.
\newblock


\bibitem[Farias et~al\mbox{.}(2023)]%
        {farias2023correcting}
\bibfield{author}{\bibinfo{person}{Vivek Farias}, \bibinfo{person}{Hao Li}, \bibinfo{person}{Tianyi Peng}, \bibinfo{person}{Xinyuyang Ren}, \bibinfo{person}{Huawei Zhang}, {and} \bibinfo{person}{Andrew Zheng}.} \bibinfo{year}{2023}\natexlab{}.
\newblock \showarticletitle{Correcting for interference in experiments: A case study at douyin}. In \bibinfo{booktitle}{\emph{Proceedings of the 17th ACM Conference on Recommender Systems}}. \bibinfo{pages}{455--466}.
\newblock


\bibitem[Feng et~al\mbox{.}(2021)]%
        {feng2021zero}
\bibfield{author}{\bibinfo{person}{Philip~J Feng}, \bibinfo{person}{Pingjun Pan}, \bibinfo{person}{Tingting Zhou}, \bibinfo{person}{Hongxiang Chen}, {and} \bibinfo{person}{Chuanjiang Luo}.} \bibinfo{year}{2021}\natexlab{}.
\newblock \showarticletitle{Zero shot on the cold-start problem: Model-agnostic interest learning for recommender systems}. In \bibinfo{booktitle}{\emph{Proceedings of the 30th ACM international conference on information \& knowledge management}}. \bibinfo{pages}{474--483}.
\newblock


\bibitem[Feng et~al\mbox{.}(2020)]%
        {feng2020atbrg}
\bibfield{author}{\bibinfo{person}{Yufei Feng}, \bibinfo{person}{Binbin Hu}, \bibinfo{person}{Fuyu Lv}, \bibinfo{person}{Qingwen Liu}, \bibinfo{person}{Zhiqiang Zhang}, {and} \bibinfo{person}{Wenwu Ou}.} \bibinfo{year}{2020}\natexlab{}.
\newblock \showarticletitle{Atbrg: Adaptive target-behavior relational graph network for effective recommendation}. In \bibinfo{booktitle}{\emph{Proceedings of the 43rd international ACM SIGIR conference on research and development in information retrieval}}. \bibinfo{pages}{2231--2240}.
\newblock


\bibitem[Firooz et~al\mbox{.}(2025)]%
        {firooz2025360brew}
\bibfield{author}{\bibinfo{person}{Hamed Firooz}, \bibinfo{person}{Maziar Sanjabi}, \bibinfo{person}{Adrian Englhardt}, \bibinfo{person}{Aman Gupta}, \bibinfo{person}{Ben Levine}, \bibinfo{person}{Dre Olgiati}, \bibinfo{person}{Gungor Polatkan}, \bibinfo{person}{Iuliia Melnychuk}, \bibinfo{person}{Karthik Ramgopal}, \bibinfo{person}{Kirill Talanine}, {et~al\mbox{.}}} \bibinfo{year}{2025}\natexlab{}.
\newblock \showarticletitle{360brew: A decoder-only foundation model for personalized ranking and recommendation}.
\newblock \bibinfo{journal}{\emph{arXiv preprint arXiv:2501.16450}} (\bibinfo{year}{2025}).
\newblock


\bibitem[Gao et~al\mbox{.}(2021)]%
        {gao2021learning}
\bibfield{author}{\bibinfo{person}{Weihao Gao}, \bibinfo{person}{Xiangjun Fan}, \bibinfo{person}{Chong Wang}, \bibinfo{person}{Jiankai Sun}, \bibinfo{person}{Kai Jia}, \bibinfo{person}{Wenzi Xiao}, \bibinfo{person}{Ruofan Ding}, \bibinfo{person}{Xingyan Bin}, \bibinfo{person}{Hui Yang}, {and} \bibinfo{person}{Xiaobing Liu}.} \bibinfo{year}{2021}\natexlab{}.
\newblock \showarticletitle{Learning an end-to-end structure for retrieval in large-scale recommendations}. In \bibinfo{booktitle}{\emph{Proceedings of the 30th ACM international conference on information \& knowledge management}}. \bibinfo{pages}{524--533}.
\newblock


\bibitem[Gong et~al\mbox{.}(2022)]%
        {gong2022real}
\bibfield{author}{\bibinfo{person}{Xudong Gong}, \bibinfo{person}{Qinlin Feng}, \bibinfo{person}{Yuan Zhang}, \bibinfo{person}{Jiangling Qin}, \bibinfo{person}{Weijie Ding}, \bibinfo{person}{Biao Li}, \bibinfo{person}{Peng Jiang}, {and} \bibinfo{person}{Kun Gai}.} \bibinfo{year}{2022}\natexlab{}.
\newblock \showarticletitle{Real-time short video recommendation on mobile devices}. In \bibinfo{booktitle}{\emph{Proceedings of the 31st ACM international conference on information \& knowledge management}}. \bibinfo{pages}{3103--3112}.
\newblock


\bibitem[Gong et~al\mbox{.}(2023)]%
        {gong2023unified}
\bibfield{author}{\bibinfo{person}{Yuqi Gong}, \bibinfo{person}{Xichen Ding}, \bibinfo{person}{Yehui Su}, \bibinfo{person}{Kaiming Shen}, \bibinfo{person}{Zhongyi Liu}, {and} \bibinfo{person}{Guannan Zhang}.} \bibinfo{year}{2023}\natexlab{}.
\newblock \showarticletitle{An unified search and recommendation foundation model for cold-start scenario}. In \bibinfo{booktitle}{\emph{Proceedings of the 32nd ACM International Conference on Information and Knowledge Management}}. \bibinfo{pages}{4595--4601}.
\newblock


\bibitem[Gou et~al\mbox{.}(2024)]%
        {gou2024controllable}
\bibfield{author}{\bibinfo{person}{Yanjie Gou}, \bibinfo{person}{Yuanzhou Yao}, \bibinfo{person}{Zhao Zhang}, \bibinfo{person}{Yiqing Wu}, \bibinfo{person}{Yi Hu}, \bibinfo{person}{Fuzhen Zhuang}, \bibinfo{person}{Jiangming Liu}, {and} \bibinfo{person}{Yongjun Xu}.} \bibinfo{year}{2024}\natexlab{}.
\newblock \showarticletitle{Controllable multi-behavior recommendation for in-game skins with large sequential model}. In \bibinfo{booktitle}{\emph{Proceedings of the 30th ACM SIGKDD Conference on Knowledge Discovery and Data Mining}}. \bibinfo{pages}{4986--4996}.
\newblock


\bibitem[Gr{\"u}n and Neufeld(2023)]%
        {grun2023transparently}
\bibfield{author}{\bibinfo{person}{Andreas Gr{\"u}n} {and} \bibinfo{person}{Xenija Neufeld}.} \bibinfo{year}{2023}\natexlab{}.
\newblock \showarticletitle{Transparently serving the public: Enhancing public service media values through exploration}. In \bibinfo{booktitle}{\emph{Proceedings of the 17th ACM Conference on Recommender Systems}}. \bibinfo{pages}{1045--1048}.
\newblock


\bibitem[Ha-Thuc et~al\mbox{.}(2021)]%
        {ha2021producer}
\bibfield{author}{\bibinfo{person}{Viet Ha-Thuc}, \bibinfo{person}{Matthew Wood}, \bibinfo{person}{Yunli Liu}, {and} \bibinfo{person}{Jagadeesan Sundaresan}.} \bibinfo{year}{2021}\natexlab{}.
\newblock \showarticletitle{From producer success to retention: A new role of search and recommendation systems on marketplaces}. In \bibinfo{booktitle}{\emph{Proceedings of the 44th International ACM SIGIR Conference on Research and Development in Information Retrieval}}. \bibinfo{pages}{2629--2630}.
\newblock


\bibitem[Han et~al\mbox{.}(2024)]%
        {10.1145/3627673.3679914}
\bibfield{author}{\bibinfo{person}{Ruidong Han}, \bibinfo{person}{Qianzhong Li}, \bibinfo{person}{He Jiang}, \bibinfo{person}{Rui Li}, \bibinfo{person}{Yurou Zhao}, \bibinfo{person}{Xiang Li}, {and} \bibinfo{person}{Wei Lin}.} \bibinfo{year}{2024}\natexlab{}.
\newblock \showarticletitle{Enhancing CTR Prediction through Sequential Recommendation Pre-training: Introducing the SRP4CTR framework}. In \bibinfo{booktitle}{\emph{Proceedings of the 33rd ACM International Conference on Information and Knowledge Management}} (Boise, ID, USA) \emph{(\bibinfo{series}{CIKM '24})}. \bibinfo{publisher}{Association for Computing Machinery}, \bibinfo{address}{New York, NY, USA}, \bibinfo{pages}{3777–3781}.
\newblock
\showISBNx{9798400704369}
\href{https://doi.org/10.1145/3627673.3679914}{doi:\nolinkurl{10.1145/3627673.3679914}}


\bibitem[Han et~al\mbox{.}(2025)]%
        {han2025mtgr}
\bibfield{author}{\bibinfo{person}{Ruidong Han}, \bibinfo{person}{Bin Yin}, \bibinfo{person}{Shangyu Chen}, \bibinfo{person}{He Jiang}, \bibinfo{person}{Fei Jiang}, \bibinfo{person}{Xiang Li}, \bibinfo{person}{Chi Ma}, \bibinfo{person}{Mincong Huang}, \bibinfo{person}{Xiaoguang Li}, \bibinfo{person}{Chunzhen Jing}, {et~al\mbox{.}}} \bibinfo{year}{2025}\natexlab{}.
\newblock \showarticletitle{MTGR: Industrial-Scale Generative Recommendation Framework in Meituan}.
\newblock \bibinfo{journal}{\emph{arXiv preprint arXiv:2505.18654}} (\bibinfo{year}{2025}).
\newblock


\bibitem[He et~al\mbox{.}(2020)]%
        {he2020contextual}
\bibfield{author}{\bibinfo{person}{Xu He}, \bibinfo{person}{Bo An}, \bibinfo{person}{Yanghua Li}, \bibinfo{person}{Haikai Chen}, \bibinfo{person}{Qingyu Guo}, \bibinfo{person}{Xin Li}, {and} \bibinfo{person}{Zhirong Wang}.} \bibinfo{year}{2020}\natexlab{}.
\newblock \showarticletitle{Contextual user browsing bandits for large-scale online mobile recommendation}. In \bibinfo{booktitle}{\emph{Proceedings of the 14th ACM Conference on Recommender Systems}}. \bibinfo{pages}{63--72}.
\newblock


\bibitem[Horn et~al\mbox{.}(2024)]%
        {horn2024more}
\bibfield{author}{\bibinfo{person}{Franklin Horn}, \bibinfo{person}{Aurelia Alston}, {and} \bibinfo{person}{Won~J You}.} \bibinfo{year}{2024}\natexlab{}.
\newblock \showarticletitle{" More to Read" at the Los Angeles Times: Solving a Cold Start Problem with LLMs to Improve Story Discovery}. In \bibinfo{booktitle}{\emph{Proceedings of the 18th ACM Conference on Recommender Systems}}. \bibinfo{pages}{742--744}.
\newblock


\bibitem[Hu et~al\mbox{.}(2023)]%
        {hu2023boss}
\bibfield{author}{\bibinfo{person}{Xiao Hu}, \bibinfo{person}{Yuan Cheng}, \bibinfo{person}{Zhi Zheng}, \bibinfo{person}{Yue Wang}, \bibinfo{person}{Xinxin Chi}, {and} \bibinfo{person}{Hengshu Zhu}.} \bibinfo{year}{2023}\natexlab{}.
\newblock \showarticletitle{BOSS: A bilateral occupational-suitability-aware recommender system for online recruitment}. In \bibinfo{booktitle}{\emph{Proceedings of the 29th ACM SIGKDD Conference on Knowledge Discovery and Data Mining}}. \bibinfo{pages}{4146--4155}.
\newblock


\bibitem[Huai et~al\mbox{.}(2023)]%
        {huai2023m2gnn}
\bibfield{author}{\bibinfo{person}{Zepeng Huai}, \bibinfo{person}{Yuji Yang}, \bibinfo{person}{Mengdi Zhang}, \bibinfo{person}{Zhongyi Zhang}, \bibinfo{person}{Yichun Li}, {and} \bibinfo{person}{Wei Wu}.} \bibinfo{year}{2023}\natexlab{}.
\newblock \showarticletitle{M2GNN: Metapath and multi-interest aggregated graph neural network for tag-based cross-domain recommendation}. In \bibinfo{booktitle}{\emph{Proceedings of the 46th International ACM SIGIR Conference on Research and Development in Information Retrieval}}. \bibinfo{pages}{1468--1477}.
\newblock


\bibitem[Huan et~al\mbox{.}(2022)]%
        {huan2022industrial}
\bibfield{author}{\bibinfo{person}{Zhaoxin Huan}, \bibinfo{person}{Gongduo Zhang}, \bibinfo{person}{Xiaolu Zhang}, \bibinfo{person}{Jun Zhou}, \bibinfo{person}{Qintong Wu}, \bibinfo{person}{Lihong Gu}, \bibinfo{person}{Jinjie Gu}, \bibinfo{person}{Yong He}, \bibinfo{person}{Yue Zhu}, {and} \bibinfo{person}{Linjian Mo}.} \bibinfo{year}{2022}\natexlab{}.
\newblock \showarticletitle{An industrial framework for cold-start recommendation in zero-shot scenarios}. In \bibinfo{booktitle}{\emph{Proceedings of the 45th International ACM SIGIR Conference on Research and Development in Information Retrieval}}. \bibinfo{pages}{3403--3407}.
\newblock


\bibitem[Huang et~al\mbox{.}(2024)]%
        {10.1145/3627673.3680055}
\bibfield{author}{\bibinfo{person}{Lei Huang}, \bibinfo{person}{Weitao Li}, \bibinfo{person}{Chenrui Zhang}, \bibinfo{person}{Jinpeng Wang}, \bibinfo{person}{Xianchun Yi}, {and} \bibinfo{person}{Sheng Chen}.} \bibinfo{year}{2024}\natexlab{}.
\newblock \showarticletitle{EXIT: An EXplicit Interest Transfer Framework for Cross-Domain Recommendation}. In \bibinfo{booktitle}{\emph{Proceedings of the 33rd ACM International Conference on Information and Knowledge Management}} (Boise, ID, USA) \emph{(\bibinfo{series}{CIKM '24})}. \bibinfo{publisher}{Association for Computing Machinery}, \bibinfo{address}{New York, NY, USA}, \bibinfo{pages}{4563–4570}.
\newblock
\showISBNx{9798400704369}
\href{https://doi.org/10.1145/3627673.3680055}{doi:\nolinkurl{10.1145/3627673.3680055}}


\bibitem[Huang et~al\mbox{.}(2025)]%
        {huang2025towards}
\bibfield{author}{\bibinfo{person}{Yanhua Huang}, \bibinfo{person}{Yuqi Chen}, \bibinfo{person}{Xiong Cao}, \bibinfo{person}{Rui Yang}, \bibinfo{person}{Mingliang Qi}, \bibinfo{person}{Yinghao Zhu}, \bibinfo{person}{Qingchang Han}, \bibinfo{person}{Yaowei Liu}, \bibinfo{person}{Zhaoyu Liu}, \bibinfo{person}{Xuefeng Yao}, {et~al\mbox{.}}} \bibinfo{year}{2025}\natexlab{}.
\newblock \showarticletitle{Towards Large-scale Generative Ranking}.
\newblock \bibinfo{journal}{\emph{arXiv preprint arXiv:2505.04180}} (\bibinfo{year}{2025}).
\newblock


\bibitem[Huang et~al\mbox{.}(2021)]%
        {huang2021sliding}
\bibfield{author}{\bibinfo{person}{Yanhua Huang}, \bibinfo{person}{Weikun Wang}, \bibinfo{person}{Lei Zhang}, {and} \bibinfo{person}{Ruiwen Xu}.} \bibinfo{year}{2021}\natexlab{}.
\newblock \showarticletitle{Sliding spectrum decomposition for diversified recommendation}. In \bibinfo{booktitle}{\emph{Proceedings of the 27th ACM SIGKDD conference on knowledge discovery \& data mining}}. \bibinfo{pages}{3041--3049}.
\newblock


\bibitem[Huangfu et~al\mbox{.}(2024)]%
        {10.1145/3627673.3680061}
\bibfield{author}{\bibinfo{person}{Zhigang Huangfu}, \bibinfo{person}{Binbin Hu}, \bibinfo{person}{Zhengwei Wu}, \bibinfo{person}{Fengyu Han}, \bibinfo{person}{Gong-Duo Zhang}, \bibinfo{person}{Gong-Duo Zhang}, \bibinfo{person}{Lihong Gu}, \bibinfo{person}{Lihong Gu}, \bibinfo{person}{Zhiqiang Zhang}, {and} \bibinfo{person}{Zhiqiang Zhang}.} \bibinfo{year}{2024}\natexlab{}.
\newblock \showarticletitle{To Explore or Exploit? A Gradient-informed Framework to Address the Feedback Loop for Graph based Recommendation}. In \bibinfo{booktitle}{\emph{Proceedings of the 33rd ACM International Conference on Information and Knowledge Management}} (Boise, ID, USA) \emph{(\bibinfo{series}{CIKM '24})}. \bibinfo{publisher}{Association for Computing Machinery}, \bibinfo{address}{New York, NY, USA}, \bibinfo{pages}{4579–4586}.
\newblock
\showISBNx{9798400704369}
\href{https://doi.org/10.1145/3627673.3680061}{doi:\nolinkurl{10.1145/3627673.3680061}}


\bibitem[Jeunen et~al\mbox{.}(2024)]%
        {jeunen2024multi}
\bibfield{author}{\bibinfo{person}{Olivier Jeunen}, \bibinfo{person}{Jatin Mandav}, \bibinfo{person}{Ivan Potapov}, \bibinfo{person}{Nakul Agarwal}, \bibinfo{person}{Sourabh Vaid}, \bibinfo{person}{Wenzhe Shi}, {and} \bibinfo{person}{Aleksei Ustimenko}.} \bibinfo{year}{2024}\natexlab{}.
\newblock \showarticletitle{Multi-objective recommendation via multivariate policy learning}. In \bibinfo{booktitle}{\emph{Proceedings of the 18th ACM Conference on Recommender Systems}}. \bibinfo{pages}{712--721}.
\newblock


\bibitem[Ji et~al\mbox{.}(2021)]%
        {10.1145/3459637.3482292}
\bibfield{author}{\bibinfo{person}{Luo Ji}, \bibinfo{person}{Qi Qin}, \bibinfo{person}{Bingqing Han}, {and} \bibinfo{person}{Hongxia Yang}.} \bibinfo{year}{2021}\natexlab{}.
\newblock \showarticletitle{Reinforcement Learning to Optimize Lifetime Value in Cold-Start Recommendation}. In \bibinfo{booktitle}{\emph{Proceedings of the 30th ACM International Conference on Information \& Knowledge Management}} (Virtual Event, Queensland, Australia) \emph{(\bibinfo{series}{CIKM '21})}. \bibinfo{publisher}{Association for Computing Machinery}, \bibinfo{address}{New York, NY, USA}, \bibinfo{pages}{782–791}.
\newblock
\showISBNx{9781450384469}
\href{https://doi.org/10.1145/3459637.3482292}{doi:\nolinkurl{10.1145/3459637.3482292}}


\bibitem[Ji et~al\mbox{.}(2023)]%
        {JiDataLeakage}
\bibfield{author}{\bibinfo{person}{Yitong Ji}, \bibinfo{person}{Aixin Sun}, \bibinfo{person}{Jie Zhang}, {and} \bibinfo{person}{Chenliang Li}.} \bibinfo{year}{2023}\natexlab{}.
\newblock \showarticletitle{A Critical Study on Data Leakage in Recommender System Offline Evaluation}.
\newblock \bibinfo{journal}{\emph{{ACM} Trans. Inf. Syst.}} \bibinfo{volume}{41}, \bibinfo{number}{3} (\bibinfo{year}{2023}), \bibinfo{pages}{75:1--75:27}.
\newblock
\href{https://doi.org/10.1145/3569930}{doi:\nolinkurl{10.1145/3569930}}


\bibitem[Jian et~al\mbox{.}(2023)]%
        {jian2023practical}
\bibfield{author}{\bibinfo{person}{Daohong Jian}, \bibinfo{person}{Yang Bao}, \bibinfo{person}{Jun Zhou}, {and} \bibinfo{person}{Hua Wu}.} \bibinfo{year}{2023}\natexlab{}.
\newblock \showarticletitle{A Practical Online Allocation Framework at Industry-scale in Constrained Recommendation}. In \bibinfo{booktitle}{\emph{Proceedings of the 46th International ACM SIGIR Conference on Research and Development in Information Retrieval}}. \bibinfo{pages}{3270--3274}.
\newblock


\bibitem[Jiang et~al\mbox{.}(2022)]%
        {jiang2022triangle}
\bibfield{author}{\bibinfo{person}{Wensen Jiang}, \bibinfo{person}{Yizhu Jiao}, \bibinfo{person}{Qingqin Wang}, \bibinfo{person}{Chuanming Liang}, \bibinfo{person}{Lijie Guo}, \bibinfo{person}{Yao Zhang}, \bibinfo{person}{Zhijun Sun}, \bibinfo{person}{Yun Xiong}, {and} \bibinfo{person}{Yangyong Zhu}.} \bibinfo{year}{2022}\natexlab{}.
\newblock \showarticletitle{Triangle graph interest network for click-through rate prediction}. In \bibinfo{booktitle}{\emph{Proceedings of the fifteenth ACM international conference on web search and data mining}}. \bibinfo{pages}{401--409}.
\newblock


\bibitem[Jiang et~al\mbox{.}(2024a)]%
        {jiang2024prompt}
\bibfield{author}{\bibinfo{person}{Yuezihan Jiang}, \bibinfo{person}{Gaode Chen}, \bibinfo{person}{Wenhan Zhang}, \bibinfo{person}{Jingchi Wang}, \bibinfo{person}{Yinjie Jiang}, \bibinfo{person}{Qi Zhang}, \bibinfo{person}{Jingjian Lin}, \bibinfo{person}{Peng Jiang}, {and} \bibinfo{person}{Kaigui Bian}.} \bibinfo{year}{2024}\natexlab{a}.
\newblock \showarticletitle{Prompt Tuning for Item Cold-start Recommendation}. In \bibinfo{booktitle}{\emph{Proceedings of the 18th ACM Conference on Recommender Systems}}. \bibinfo{pages}{411--421}.
\newblock


\bibitem[Jiang et~al\mbox{.}(2024b)]%
        {jiang2024mmgcl}
\bibfield{author}{\bibinfo{person}{Yuezihan Jiang}, \bibinfo{person}{Changyu Li}, \bibinfo{person}{Gaode Chen}, \bibinfo{person}{Peiyi Li}, \bibinfo{person}{Qi Zhang}, \bibinfo{person}{Jingjian Lin}, \bibinfo{person}{Peng Jiang}, \bibinfo{person}{Fei Sun}, {and} \bibinfo{person}{Wentao Zhang}.} \bibinfo{year}{2024}\natexlab{b}.
\newblock \showarticletitle{MMGCL: Meta Knowledge-Enhanced Multi-view Graph Contrastive Learning for Recommendations}. In \bibinfo{booktitle}{\emph{Proceedings of the 18th ACM Conference on Recommender Systems}}. \bibinfo{pages}{538--548}.
\newblock


\bibitem[Joglekar et~al\mbox{.}(2020)]%
        {joglekar2020neural}
\bibfield{author}{\bibinfo{person}{Manas~R Joglekar}, \bibinfo{person}{Cong Li}, \bibinfo{person}{Mei Chen}, \bibinfo{person}{Taibai Xu}, \bibinfo{person}{Xiaoming Wang}, \bibinfo{person}{Jay~K Adams}, \bibinfo{person}{Pranav Khaitan}, \bibinfo{person}{Jiahui Liu}, {and} \bibinfo{person}{Quoc~V Le}.} \bibinfo{year}{2020}\natexlab{}.
\newblock \showarticletitle{Neural input search for large scale recommendation models}. In \bibinfo{booktitle}{\emph{Proceedings of the 26th ACM SIGKDD International Conference on Knowledge Discovery \& Data Mining}}. \bibinfo{pages}{2387--2397}.
\newblock


\bibitem[Kekuda et~al\mbox{.}(2024)]%
        {kekuda2024embedding}
\bibfield{author}{\bibinfo{person}{Akshay Kekuda}, \bibinfo{person}{Yuyang Zhang}, {and} \bibinfo{person}{Arun Udayashankar}.} \bibinfo{year}{2024}\natexlab{}.
\newblock \showarticletitle{Embedding based retrieval for long tail search queries in ecommerce}. In \bibinfo{booktitle}{\emph{Proceedings of the 18th ACM Conference on Recommender Systems}}. \bibinfo{pages}{771--774}.
\newblock


\bibitem[Khrylchenko et~al\mbox{.}(2025)]%
        {khrylchenko2025scaling}
\bibfield{author}{\bibinfo{person}{Kirill Khrylchenko}, \bibinfo{person}{Artem Matveev}, \bibinfo{person}{Sergei Makeev}, {and} \bibinfo{person}{Vladimir Baikalov}.} \bibinfo{year}{2025}\natexlab{}.
\newblock \showarticletitle{Scaling Recommender Transformers to One Billion Parameters}.
\newblock \bibinfo{journal}{\emph{arXiv preprint arXiv:2507.15994}} (\bibinfo{year}{2025}).
\newblock


\bibitem[Klimashevskaia et~al\mbox{.}(2023)]%
        {klimashevskaia2023evaluating}
\bibfield{author}{\bibinfo{person}{Anastasiia Klimashevskaia}, \bibinfo{person}{Mehdi Elahi}, \bibinfo{person}{Dietmar Jannach}, \bibinfo{person}{Lars Skj{\ae}rven}, \bibinfo{person}{Astrid Tessem}, {and} \bibinfo{person}{Christoph Trattner}.} \bibinfo{year}{2023}\natexlab{}.
\newblock \showarticletitle{Evaluating the effects of calibrated popularity bias mitigation: a field study}. In \bibinfo{booktitle}{\emph{Proceedings of the 17th ACM Conference on Recommender Systems}}. \bibinfo{pages}{1084--1089}.
\newblock


\bibitem[Koneru et~al\mbox{.}(2024)]%
        {koneru2024enhancing}
\bibfield{author}{\bibinfo{person}{Venkata~Harshit Koneru}, \bibinfo{person}{Xenija Neufeld}, \bibinfo{person}{Sebastian Loth}, {and} \bibinfo{person}{Andreas Gr{\"u}n}.} \bibinfo{year}{2024}\natexlab{}.
\newblock \showarticletitle{Enhancing Recommendation Quality of the SASRec Model by Mitigating Popularity Bias}. In \bibinfo{booktitle}{\emph{Proceedings of the 18th ACM Conference on Recommender Systems}}. \bibinfo{pages}{781--783}.
\newblock


\bibitem[Kouki et~al\mbox{.}(2020)]%
        {kouki2020lab}
\bibfield{author}{\bibinfo{person}{Pigi Kouki}, \bibinfo{person}{Ilias Fountalis}, \bibinfo{person}{Nikolaos Vasiloglou}, \bibinfo{person}{Xiquan Cui}, \bibinfo{person}{Edo Liberty}, {and} \bibinfo{person}{Khalifeh Al~Jadda}.} \bibinfo{year}{2020}\natexlab{}.
\newblock \showarticletitle{From the lab to production: A case study of session-based recommendations in the home-improvement domain}. In \bibinfo{booktitle}{\emph{Proceedings of the 14th ACM conference on recommender systems}}. \bibinfo{pages}{140--149}.
\newblock


\bibitem[Kruse et~al\mbox{.}(2023)]%
        {kruse2023creating}
\bibfield{author}{\bibinfo{person}{Johannes Kruse}, \bibinfo{person}{Kasper Lindskow}, \bibinfo{person}{Michael~Riis Andersen}, {and} \bibinfo{person}{Jes Frellsen}.} \bibinfo{year}{2023}\natexlab{}.
\newblock \showarticletitle{Creating the next generation of news experience on ekstrabladet. dk with recommender systems}. In \bibinfo{booktitle}{\emph{Proceedings of the 17th ACM Conference on Recommender Systems}}. \bibinfo{pages}{1067--1070}.
\newblock


\bibitem[Kung et~al\mbox{.}(2024)]%
        {kung2024improving}
\bibfield{author}{\bibinfo{person}{Pau Perng-Hwa Kung}, \bibinfo{person}{Zihao Fan}, \bibinfo{person}{Tong Zhao}, \bibinfo{person}{Yozen Liu}, \bibinfo{person}{Zhixin Lai}, \bibinfo{person}{Jiahui Shi}, \bibinfo{person}{Yan Wu}, \bibinfo{person}{Jun Yu}, \bibinfo{person}{Neil Shah}, {and} \bibinfo{person}{Ganesh Venkataraman}.} \bibinfo{year}{2024}\natexlab{}.
\newblock \showarticletitle{Improving embedding-based retrieval in friend recommendation with ann query expansion}. In \bibinfo{booktitle}{\emph{Proceedings of the 47th International ACM SIGIR Conference on Research and Development in Information Retrieval}}. \bibinfo{pages}{2930--2934}.
\newblock


\bibitem[Lan et~al\mbox{.}(2025)]%
        {lan2025next}
\bibfield{author}{\bibinfo{person}{Yu-Ting Lan}, \bibinfo{person}{Yang Huo}, \bibinfo{person}{Yi Shen}, \bibinfo{person}{Xiao Yang}, {and} \bibinfo{person}{Zuotao Liu}.} \bibinfo{year}{2025}\natexlab{}.
\newblock \showarticletitle{Next-User Retrieval: Enhancing Cold-Start Recommendations via Generative Next-User Modeling}.
\newblock \bibinfo{journal}{\emph{arXiv preprint arXiv:2506.15267}} (\bibinfo{year}{2025}).
\newblock


\bibitem[Lang et~al\mbox{.}(2021)]%
        {lang2021architecture}
\bibfield{author}{\bibinfo{person}{Lang Lang}, \bibinfo{person}{Zhenlong Zhu}, \bibinfo{person}{Xuanye Liu}, \bibinfo{person}{Jianxin Zhao}, \bibinfo{person}{Jixing Xu}, {and} \bibinfo{person}{Minghui Shan}.} \bibinfo{year}{2021}\natexlab{}.
\newblock \showarticletitle{Architecture and operation adaptive network for online recommendations}. In \bibinfo{booktitle}{\emph{Proceedings of the 27th ACM SIGKDD conference on knowledge discovery \& data mining}}. \bibinfo{pages}{3139--3149}.
\newblock


\bibitem[Le et~al\mbox{.}(2023)]%
        {le2023cec}
\bibfield{author}{\bibinfo{person}{Ran Le}, \bibinfo{person}{Guo-qing Jiang}, \bibinfo{person}{Xiufeng Shu}, \bibinfo{person}{Ruidong Han}, \bibinfo{person}{Qianzhong Li}, \bibinfo{person}{Yacheng Li}, \bibinfo{person}{Xiang Li}, {and} \bibinfo{person}{Wei Lin}.} \bibinfo{year}{2023}\natexlab{}.
\newblock \showarticletitle{CEC: Towards Learning Global Optimized Recommendation through Causality Enhanced Conversion Model}. In \bibinfo{booktitle}{\emph{Proceedings of the 46th International ACM SIGIR Conference on Research and Development in Information Retrieval}}. \bibinfo{pages}{1879--1883}.
\newblock


\bibitem[Lei et~al\mbox{.}(2021)]%
        {lei2021semi}
\bibfield{author}{\bibinfo{person}{Chenyi Lei}, \bibinfo{person}{Yong Liu}, \bibinfo{person}{Lingzi Zhang}, \bibinfo{person}{Guoxin Wang}, \bibinfo{person}{Haihong Tang}, \bibinfo{person}{Houqiang Li}, {and} \bibinfo{person}{Chunyan Miao}.} \bibinfo{year}{2021}\natexlab{}.
\newblock \showarticletitle{Semi: A sequential multi-modal information transfer network for e-commerce micro-video recommendations}. In \bibinfo{booktitle}{\emph{Proceedings of the 27th ACM SIGKDD Conference on Knowledge Discovery \& Data Mining}}. \bibinfo{pages}{3161--3171}.
\newblock


\bibitem[Li et~al\mbox{.}(2024d)]%
        {li2024contextual}
\bibfield{author}{\bibinfo{person}{Fan Li}, \bibinfo{person}{Xu Si}, \bibinfo{person}{Shisong Tang}, \bibinfo{person}{Dingmin Wang}, \bibinfo{person}{Kunyan Han}, \bibinfo{person}{Bing Han}, \bibinfo{person}{Guorui Zhou}, \bibinfo{person}{Yang Song}, {and} \bibinfo{person}{Hechang Chen}.} \bibinfo{year}{2024}\natexlab{d}.
\newblock \showarticletitle{Contextual distillation model for diversified recommendation}. In \bibinfo{booktitle}{\emph{Proceedings of the 30th ACM SIGKDD Conference on Knowledge Discovery and Data Mining}}. \bibinfo{pages}{5307--5316}.
\newblock


\bibitem[Li et~al\mbox{.}(2021)]%
        {li2021path}
\bibfield{author}{\bibinfo{person}{Houyi Li}, \bibinfo{person}{Zhihong Chen}, \bibinfo{person}{Chenliang Li}, \bibinfo{person}{Rong Xiao}, \bibinfo{person}{Hongbo Deng}, \bibinfo{person}{Peng Zhang}, \bibinfo{person}{Yongchao Liu}, {and} \bibinfo{person}{Haihong Tang}.} \bibinfo{year}{2021}\natexlab{}.
\newblock \showarticletitle{Path-based deep network for candidate item matching in recommenders}. In \bibinfo{booktitle}{\emph{Proceedings of the 44th International ACM SIGIR Conference on Research and Development in Information Retrieval}}. \bibinfo{pages}{1493--1502}.
\newblock


\bibitem[Li et~al\mbox{.}(2024e)]%
        {li2024modem}
\bibfield{author}{\bibinfo{person}{Jiang Li}, \bibinfo{person}{Zhen Zhang}, \bibinfo{person}{Xiang Feng}, \bibinfo{person}{Muyang Li}, \bibinfo{person}{Yongqi Liu}, {and} \bibinfo{person}{Lantao Hu}.} \bibinfo{year}{2024}\natexlab{e}.
\newblock \showarticletitle{MODEM: Decoupling User Behavior for Shared-Account Video Recommendations on Large Screen Devices}. In \bibinfo{booktitle}{\emph{Proceedings of the 18th ACM Conference on Recommender Systems}}. \bibinfo{pages}{907--911}.
\newblock


\bibitem[Li et~al\mbox{.}(2024a)]%
        {li2024modeling}
\bibfield{author}{\bibinfo{person}{Nian Li}, \bibinfo{person}{Xin Ban}, \bibinfo{person}{Cheng Ling}, \bibinfo{person}{Chen Gao}, \bibinfo{person}{Lantao Hu}, \bibinfo{person}{Peng Jiang}, \bibinfo{person}{Kun Gai}, \bibinfo{person}{Yong Li}, {and} \bibinfo{person}{Qingmin Liao}.} \bibinfo{year}{2024}\natexlab{a}.
\newblock \showarticletitle{Modeling user fatigue for sequential recommendation}. In \bibinfo{booktitle}{\emph{Proceedings of the 47th International ACM SIGIR Conference on Research and Development in Information Retrieval}}. \bibinfo{pages}{996--1005}.
\newblock


\bibitem[Li et~al\mbox{.}(2020)]%
        {li2020purs}
\bibfield{author}{\bibinfo{person}{Pan Li}, \bibinfo{person}{Maofei Que}, \bibinfo{person}{Zhichao Jiang}, \bibinfo{person}{Yao Hu}, {and} \bibinfo{person}{Alexander Tuzhilin}.} \bibinfo{year}{2020}\natexlab{}.
\newblock \showarticletitle{PURS: personalized unexpected recommender system for improving user satisfaction}. In \bibinfo{booktitle}{\emph{Proceedings of the 14th ACM conference on recommender systems}}. \bibinfo{pages}{279--288}.
\newblock


\bibitem[Li et~al\mbox{.}(2023b)]%
        {li2023stan}
\bibfield{author}{\bibinfo{person}{Wanda Li}, \bibinfo{person}{Wenhao Zheng}, \bibinfo{person}{Xuanji Xiao}, {and} \bibinfo{person}{Suhang Wang}.} \bibinfo{year}{2023}\natexlab{b}.
\newblock \showarticletitle{Stan: stage-adaptive network for multi-task recommendation by learning user lifecycle-based representation}. In \bibinfo{booktitle}{\emph{Proceedings of the 17th ACM Conference on Recommender Systems}}. \bibinfo{pages}{602--612}.
\newblock


\bibitem[Li et~al\mbox{.}(2024f)]%
        {li2024scene}
\bibfield{author}{\bibinfo{person}{Wenhao Li}, \bibinfo{person}{Jie Zhou}, \bibinfo{person}{Chuan Luo}, \bibinfo{person}{Chao Tang}, \bibinfo{person}{Kun Zhang}, {and} \bibinfo{person}{Shixiong Zhao}.} \bibinfo{year}{2024}\natexlab{f}.
\newblock \showarticletitle{Scene-wise Adaptive Network for Dynamic Cold-start Scenes Optimization in CTR Prediction}. In \bibinfo{booktitle}{\emph{Proceedings of the 18th ACM Conference on Recommender Systems}}. \bibinfo{pages}{370--379}.
\newblock


\bibitem[Li et~al\mbox{.}(2024c)]%
        {li2024recent}
\bibfield{author}{\bibinfo{person}{Yang Li}, \bibinfo{person}{Kangbo Liu}, \bibinfo{person}{Ranjan Satapathy}, \bibinfo{person}{Suhang Wang}, {and} \bibinfo{person}{Erik Cambria}.} \bibinfo{year}{2024}\natexlab{c}.
\newblock \showarticletitle{Recent developments in recommender systems: A survey}.
\newblock \bibinfo{journal}{\emph{IEEE Computational Intelligence Magazine}} \bibinfo{volume}{19}, \bibinfo{number}{2} (\bibinfo{year}{2024}), \bibinfo{pages}{78--95}.
\newblock


\bibitem[Li et~al\mbox{.}(2023a)]%
        {li2023autoopt}
\bibfield{author}{\bibinfo{person}{Yujun Li}, \bibinfo{person}{Xing Tang}, \bibinfo{person}{Bo Chen}, \bibinfo{person}{Yimin Huang}, \bibinfo{person}{Ruiming Tang}, {and} \bibinfo{person}{Zhenguo Li}.} \bibinfo{year}{2023}\natexlab{a}.
\newblock \showarticletitle{AutoOpt: automatic hyperparameter scheduling and optimization for deep click-through rate prediction}. In \bibinfo{booktitle}{\emph{Proceedings of the 17th ACM Conference on Recommender Systems}}. \bibinfo{pages}{183--194}.
\newblock


\bibitem[Li et~al\mbox{.}(2024b)]%
        {10.1145/3627673.3680028}
\bibfield{author}{\bibinfo{person}{Zelong Li}, \bibinfo{person}{Yan Liang}, \bibinfo{person}{Ming Wang}, \bibinfo{person}{Sungro Yoon}, \bibinfo{person}{Jiaying Shi}, \bibinfo{person}{Xin Shen}, \bibinfo{person}{Xiang He}, \bibinfo{person}{Chenwei Zhang}, \bibinfo{person}{Wenyi Wu}, \bibinfo{person}{Hanbo Wang}, \bibinfo{person}{Jin Li}, \bibinfo{person}{Jim Chan}, {and} \bibinfo{person}{Yongfeng Zhang}.} \bibinfo{year}{2024}\natexlab{b}.
\newblock \showarticletitle{Explainable and Coherent Complement Recommendation Based on Large Language Models}. In \bibinfo{booktitle}{\emph{Proceedings of the 33rd ACM International Conference on Information and Knowledge Management}} (Boise, ID, USA) \emph{(\bibinfo{series}{CIKM '24})}. \bibinfo{publisher}{Association for Computing Machinery}, \bibinfo{address}{New York, NY, USA}, \bibinfo{pages}{4678–4685}.
\newblock
\showISBNx{9798400704369}
\href{https://doi.org/10.1145/3627673.3680028}{doi:\nolinkurl{10.1145/3627673.3680028}}


\bibitem[Lichtenberg et~al\mbox{.}(2024)]%
        {lichtenberg2024ranking}
\bibfield{author}{\bibinfo{person}{Jan~Malte Lichtenberg}, \bibinfo{person}{Giuseppe Di~Benedetto}, {and} \bibinfo{person}{Matteo Ruffini}.} \bibinfo{year}{2024}\natexlab{}.
\newblock \showarticletitle{Ranking across different content types: The robust beauty of multinomial blending}. In \bibinfo{booktitle}{\emph{Proceedings of the 18th ACM Conference on Recommender Systems}}. \bibinfo{pages}{823--825}.
\newblock


\bibitem[Lin et~al\mbox{.}(2024)]%
        {lin2024bootstrapping}
\bibfield{author}{\bibinfo{person}{Hongtao Lin}, \bibinfo{person}{Haoyu Chen}, \bibinfo{person}{Jaewon Yang}, {and} \bibinfo{person}{Jiajing Xu}.} \bibinfo{year}{2024}\natexlab{}.
\newblock \showarticletitle{Bootstrapping Conditional Retrieval for User-to-Item Recommendations}. In \bibinfo{booktitle}{\emph{Proceedings of the 18th ACM Conference on Recommender Systems}}. \bibinfo{pages}{755--757}.
\newblock


\bibitem[Lin et~al\mbox{.}(2023b)]%
        {lin2023exploring}
\bibfield{author}{\bibinfo{person}{Shaochuan Lin}, \bibinfo{person}{Jiayan Pei}, \bibinfo{person}{Taotao Zhou}, \bibinfo{person}{Hengxu He}, \bibinfo{person}{Jia Jia}, {and} \bibinfo{person}{Ning Hu}.} \bibinfo{year}{2023}\natexlab{b}.
\newblock \showarticletitle{Exploring the spatiotemporal features of online food recommendation service}. In \bibinfo{booktitle}{\emph{Proceedings of the 46th International ACM SIGIR Conference on Research and Development in Information Retrieval}}. \bibinfo{pages}{3354--3358}.
\newblock


\bibitem[Lin et~al\mbox{.}(2023a)]%
        {lin2023tree}
\bibfield{author}{\bibinfo{person}{Xiao Lin}, \bibinfo{person}{Xiaokai Chen}, \bibinfo{person}{Linfeng Song}, \bibinfo{person}{Jingwei Liu}, \bibinfo{person}{Biao Li}, {and} \bibinfo{person}{Peng Jiang}.} \bibinfo{year}{2023}\natexlab{a}.
\newblock \showarticletitle{Tree based progressive regression model for watch-time prediction in short-video recommendation}. In \bibinfo{booktitle}{\emph{Proceedings of the 29th ACM SIGKDD Conference on Knowledge Discovery and Data Mining}}. \bibinfo{pages}{4497--4506}.
\newblock


\bibitem[Lin et~al\mbox{.}(2022)]%
        {lin2022feature}
\bibfield{author}{\bibinfo{person}{Zihan Lin}, \bibinfo{person}{Hui Wang}, \bibinfo{person}{Jingshu Mao}, \bibinfo{person}{Wayne~Xin Zhao}, \bibinfo{person}{Cheng Wang}, \bibinfo{person}{Peng Jiang}, {and} \bibinfo{person}{Ji-Rong Wen}.} \bibinfo{year}{2022}\natexlab{}.
\newblock \showarticletitle{Feature-aware diversified re-ranking with disentangled representations for relevant recommendation}. In \bibinfo{booktitle}{\emph{Proceedings of the 28th ACM SIGKDD Conference on Knowledge Discovery and Data Mining}}. \bibinfo{pages}{3327--3335}.
\newblock


\bibitem[Lindstrom et~al\mbox{.}(2024)]%
        {lindstrom2024encouraging}
\bibfield{author}{\bibinfo{person}{Henrik Lindstrom}, \bibinfo{person}{Humberto~Jesus Corona~Pampin}, \bibinfo{person}{Enrico Palumbo}, {and} \bibinfo{person}{Alva Liu}.} \bibinfo{year}{2024}\natexlab{}.
\newblock \showarticletitle{Encouraging Exploration in Spotify Search through Query Recommendations}. In \bibinfo{booktitle}{\emph{Proceedings of the 18th ACM Conference on Recommender Systems}}. \bibinfo{pages}{775--777}.
\newblock


\bibitem[Liu et~al\mbox{.}(2020)]%
        {liu2020autogroup}
\bibfield{author}{\bibinfo{person}{Bin Liu}, \bibinfo{person}{Niannan Xue}, \bibinfo{person}{Huifeng Guo}, \bibinfo{person}{Ruiming Tang}, \bibinfo{person}{Stefanos Zafeiriou}, \bibinfo{person}{Xiuqiang He}, {and} \bibinfo{person}{Zhenguo Li}.} \bibinfo{year}{2020}\natexlab{}.
\newblock \showarticletitle{AutoGroup: Automatic feature grouping for modelling explicit high-order feature interactions in CTR prediction}. In \bibinfo{booktitle}{\emph{Proceedings of the 43rd international ACM SIGIR conference on research and development in information retrieval}}. \bibinfo{pages}{199--208}.
\newblock


\bibitem[Liu et~al\mbox{.}(2024a)]%
        {liu2024unified}
\bibfield{author}{\bibinfo{person}{Jinhan Liu}, \bibinfo{person}{Qiyu Chen}, \bibinfo{person}{Junjie Xu}, \bibinfo{person}{Junjie Li}, \bibinfo{person}{Baoli Li}, {and} \bibinfo{person}{Sulong Xu}.} \bibinfo{year}{2024}\natexlab{a}.
\newblock \showarticletitle{A Unified Search and Recommendation Framework Based on Multi-Scenario Learning for Ranking in E-commerce}. In \bibinfo{booktitle}{\emph{Proceedings of the 47th International ACM SIGIR Conference on Research and Development in Information Retrieval}}. \bibinfo{pages}{2890--2894}.
\newblock


\bibitem[Liu et~al\mbox{.}(2022)]%
        {liu2022multi}
\bibfield{author}{\bibinfo{person}{Junning Liu}, \bibinfo{person}{Xinjian Li}, \bibinfo{person}{Bo An}, \bibinfo{person}{Zijie Xia}, {and} \bibinfo{person}{Xu Wang}.} \bibinfo{year}{2022}\natexlab{}.
\newblock \showarticletitle{Multi-faceted hierarchical multi-task learning for recommender systems}. In \bibinfo{booktitle}{\emph{Proceedings of the 31st ACM International Conference on Information \& Knowledge Management}}. \bibinfo{pages}{3332--3341}.
\newblock


\bibitem[Liu et~al\mbox{.}(2023b)]%
        {liu2023multitask}
\bibfield{author}{\bibinfo{person}{Qingyun Liu}, \bibinfo{person}{Zhe Zhao}, \bibinfo{person}{Liang Liu}, \bibinfo{person}{Zhen Zhang}, \bibinfo{person}{Junjie Shan}, \bibinfo{person}{Yuening Li}, \bibinfo{person}{Shuchao Bi}, \bibinfo{person}{Lichan Hong}, {and} \bibinfo{person}{Ed~H Chi}.} \bibinfo{year}{2023}\natexlab{b}.
\newblock \showarticletitle{Multitask Ranking System for Immersive Feed and No More Clicks: A Case Study of Short-Form Video Recommendation}. In \bibinfo{booktitle}{\emph{Proceedings of the 32nd ACM International Conference on Information and Knowledge Management}}. \bibinfo{pages}{4709--4716}.
\newblock


\bibitem[Liu et~al\mbox{.}(2023a)]%
        {10.1145/3583780.3615200}
\bibfield{author}{\bibinfo{person}{Shaohua Liu}, \bibinfo{person}{Yu Qi}, \bibinfo{person}{Gen Li}, \bibinfo{person}{Mingjian Chen}, \bibinfo{person}{Teng Zhang}, \bibinfo{person}{Jia Cheng}, {and} \bibinfo{person}{Jun Lei}.} \bibinfo{year}{2023}\natexlab{a}.
\newblock \showarticletitle{STGIN: Spatial-Temporal Graph Interaction Network for Large-scale POI Recommendation}. In \bibinfo{booktitle}{\emph{Proceedings of the 32nd ACM International Conference on Information and Knowledge Management}} (Birmingham, United Kingdom) \emph{(\bibinfo{series}{CIKM '23})}. \bibinfo{publisher}{Association for Computing Machinery}, \bibinfo{address}{New York, NY, USA}, \bibinfo{pages}{4120–4124}.
\newblock
\showISBNx{9798400701245}
\href{https://doi.org/10.1145/3583780.3615200}{doi:\nolinkurl{10.1145/3583780.3615200}}


\bibitem[Liu et~al\mbox{.}(2024b)]%
        {liu2024self}
\bibfield{author}{\bibinfo{person}{Zhiqiang Liu}, \bibinfo{person}{Xiaoxiao Xu}, \bibinfo{person}{Jiaqi Yu}, \bibinfo{person}{Han Xu}, \bibinfo{person}{Lantao Hu}, \bibinfo{person}{Han Li}, {and} \bibinfo{person}{Kun Gai}.} \bibinfo{year}{2024}\natexlab{b}.
\newblock \showarticletitle{A Self-Adaptive Fairness Constraint Framework for Industrial Recommender System}. In \bibinfo{booktitle}{\emph{Proceedings of the 33rd ACM International Conference on Information and Knowledge Management}}. \bibinfo{pages}{4726--4733}.
\newblock


\bibitem[Lu et~al\mbox{.}(2022)]%
        {lu2022deep}
\bibfield{author}{\bibinfo{person}{Chengqiang Lu}, \bibinfo{person}{Mingyang Yin}, \bibinfo{person}{Shuheng Shen}, \bibinfo{person}{Luo Ji}, \bibinfo{person}{Qi Liu}, {and} \bibinfo{person}{Hongxia Yang}.} \bibinfo{year}{2022}\natexlab{}.
\newblock \showarticletitle{Deep unified representation for heterogeneous recommendation}. In \bibinfo{booktitle}{\emph{Proceedings of the ACM Web Conference 2022}}. \bibinfo{pages}{2141--2152}.
\newblock


\bibitem[Lu et~al\mbox{.}(2015)]%
        {lu2015recommender}
\bibfield{author}{\bibinfo{person}{Jie Lu}, \bibinfo{person}{Dianshuang Wu}, \bibinfo{person}{Mingsong Mao}, \bibinfo{person}{Wei Wang}, {and} \bibinfo{person}{Guangquan Zhang}.} \bibinfo{year}{2015}\natexlab{}.
\newblock \showarticletitle{Recommender system application developments: a survey}.
\newblock \bibinfo{journal}{\emph{Decision support systems}}  \bibinfo{volume}{74} (\bibinfo{year}{2015}), \bibinfo{pages}{12--32}.
\newblock


\bibitem[Lv et~al\mbox{.}(2023)]%
        {lv2023deep}
\bibfield{author}{\bibinfo{person}{Yimin Lv}, \bibinfo{person}{Shuli Wang}, \bibinfo{person}{Beihong Jin}, \bibinfo{person}{Yisong Yu}, \bibinfo{person}{Yapeng Zhang}, \bibinfo{person}{Jian Dong}, \bibinfo{person}{Yongkang Wang}, \bibinfo{person}{Xingxing Wang}, {and} \bibinfo{person}{Dong Wang}.} \bibinfo{year}{2023}\natexlab{}.
\newblock \showarticletitle{Deep situation-aware interaction network for click-through rate prediction}. In \bibinfo{booktitle}{\emph{Proceedings of the 17th ACM Conference on Recommender Systems}}. \bibinfo{pages}{171--182}.
\newblock


\bibitem[Ma et~al\mbox{.}(2024)]%
        {10.1145/3627673.3680065}
\bibfield{author}{\bibinfo{person}{Jianxing Ma}, \bibinfo{person}{Zhibo Xiao}, \bibinfo{person}{Luwei Yang}, \bibinfo{person}{Hansheng Xue}, \bibinfo{person}{Xuanzhou Liu}, \bibinfo{person}{Wen Jiang}, \bibinfo{person}{Wei Ning}, {and} \bibinfo{person}{Guannan Zhang}.} \bibinfo{year}{2024}\natexlab{}.
\newblock \showarticletitle{Modeling User Intent Beyond Trigger: Incorporating Uncertainty for Trigger-Induced Recommendation}. In \bibinfo{booktitle}{\emph{Proceedings of the 33rd ACM International Conference on Information and Knowledge Management}} (Boise, ID, USA) \emph{(\bibinfo{series}{CIKM '24})}. \bibinfo{publisher}{Association for Computing Machinery}, \bibinfo{address}{New York, NY, USA}, \bibinfo{pages}{4743–4751}.
\newblock
\showISBNx{9798400704369}
\href{https://doi.org/10.1145/3627673.3680065}{doi:\nolinkurl{10.1145/3627673.3680065}}


\bibitem[Ma et~al\mbox{.}(2022)]%
        {ma2022two}
\bibfield{author}{\bibinfo{person}{Siyong Ma}, \bibinfo{person}{Puja Das}, \bibinfo{person}{Sofia~Maria Nikolakaki}, \bibinfo{person}{Qifeng Chen}, {and} \bibinfo{person}{Humeyra Topcu~Altintas}.} \bibinfo{year}{2022}\natexlab{}.
\newblock \showarticletitle{Two-layer bandit optimization for recommendations}. In \bibinfo{booktitle}{\emph{Proceedings of the 16th ACM Conference on Recommender Systems}}. \bibinfo{pages}{509--511}.
\newblock


\bibitem[McInerney et~al\mbox{.}(2020)]%
        {mcinerney2020counterfactual}
\bibfield{author}{\bibinfo{person}{James McInerney}, \bibinfo{person}{Brian Brost}, \bibinfo{person}{Praveen Chandar}, \bibinfo{person}{Rishabh Mehrotra}, {and} \bibinfo{person}{Benjamin Carterette}.} \bibinfo{year}{2020}\natexlab{}.
\newblock \showarticletitle{Counterfactual evaluation of slate recommendations with sequential reward interactions}. In \bibinfo{booktitle}{\emph{Proceedings of the 26th ACM SIGKDD International Conference on Knowledge Discovery \& Data Mining}}. \bibinfo{pages}{1779--1788}.
\newblock


\bibitem[Mei et~al\mbox{.}(2022)]%
        {mei2022lightweight}
\bibfield{author}{\bibinfo{person}{M~Jeffrey Mei}, \bibinfo{person}{Cole Zuber}, {and} \bibinfo{person}{Yasaman Khazaeni}.} \bibinfo{year}{2022}\natexlab{}.
\newblock \showarticletitle{A lightweight transformer for next-item product recommendation}. In \bibinfo{booktitle}{\emph{Proceedings of the 16th ACM Conference on Recommender Systems}}. \bibinfo{pages}{546--549}.
\newblock


\bibitem[Meisburger et~al\mbox{.}(2023)]%
        {meisburger2023bolt}
\bibfield{author}{\bibinfo{person}{Nicholas Meisburger}, \bibinfo{person}{Vihan Lakshman}, \bibinfo{person}{Benito Geordie}, \bibinfo{person}{Joshua Engels}, \bibinfo{person}{David~Torres Ramos}, \bibinfo{person}{Pratik Pranav}, \bibinfo{person}{Benjamin Coleman}, \bibinfo{person}{Benjamin Meisburger}, \bibinfo{person}{Shubh Gupta}, \bibinfo{person}{Yashwanth Adunukota}, {et~al\mbox{.}}} \bibinfo{year}{2023}\natexlab{}.
\newblock \showarticletitle{BOLT: An Automated Deep Learning Framework for Training and Deploying Large-Scale Search and Recommendation Models on Commodity CPU Hardware}. In \bibinfo{booktitle}{\emph{Proceedings of the 32nd ACM International Conference on Information and Knowledge Management}}. \bibinfo{pages}{4738--4744}.
\newblock


\bibitem[Mondal et~al\mbox{.}(2022)]%
        {mondal2022aspire}
\bibfield{author}{\bibinfo{person}{Abhirup Mondal}, \bibinfo{person}{Anirban Majumder}, {and} \bibinfo{person}{Vineet Chaoji}.} \bibinfo{year}{2022}\natexlab{}.
\newblock \showarticletitle{Aspire: Air shipping recommendation for e-commerce products via causal inference framework}. In \bibinfo{booktitle}{\emph{Proceedings of the 28th ACM SIGKDD Conference on Knowledge Discovery and Data Mining}}. \bibinfo{pages}{3584--3592}.
\newblock


\bibitem[Nazari et~al\mbox{.}(2022)]%
        {nazari2022choice}
\bibfield{author}{\bibinfo{person}{Zahra Nazari}, \bibinfo{person}{Praveen Chandar}, \bibinfo{person}{Ghazal Fazelnia}, \bibinfo{person}{Catherine~M Edwards}, \bibinfo{person}{Benjamin Carterette}, {and} \bibinfo{person}{Mounia Lalmas}.} \bibinfo{year}{2022}\natexlab{}.
\newblock \showarticletitle{Choice of implicit signal matters: Accounting for user aspirations in podcast recommendations}. In \bibinfo{booktitle}{\emph{Proceedings of the ACM Web Conference 2022}}. \bibinfo{pages}{2433--2441}.
\newblock


\bibitem[Nie et~al\mbox{.}(2024)]%
        {nie2024hybrid}
\bibfield{author}{\bibinfo{person}{Guangtao Nie}, \bibinfo{person}{Rong Zhi}, \bibinfo{person}{Xiaofan Yan}, \bibinfo{person}{Yufan Du}, \bibinfo{person}{Xiangyang Zhang}, \bibinfo{person}{Jianwei Chen}, \bibinfo{person}{Mi Zhou}, \bibinfo{person}{Hongshen Chen}, \bibinfo{person}{Tianhao Li}, \bibinfo{person}{Ziguang Cheng}, {et~al\mbox{.}}} \bibinfo{year}{2024}\natexlab{}.
\newblock \showarticletitle{A hybrid multi-agent conversational recommender system with llm and search engine in e-commerce}. In \bibinfo{booktitle}{\emph{Proceedings of the 18th ACM Conference on Recommender Systems}}. \bibinfo{pages}{745--747}.
\newblock


\bibitem[Nie et~al\mbox{.}(2022)]%
        {nie2022mic}
\bibfield{author}{\bibinfo{person}{Ping Nie}, \bibinfo{person}{Yujie Lu}, \bibinfo{person}{Shengyu Zhang}, \bibinfo{person}{Ming Zhao}, \bibinfo{person}{Ruobing Xie}, \bibinfo{person}{William~Yang Wang}, {and} \bibinfo{person}{Yi Ren}.} \bibinfo{year}{2022}\natexlab{}.
\newblock \showarticletitle{MIC: model-agnostic integrated cross-channel recommender}. In \bibinfo{booktitle}{\emph{Proceedings of the 31st ACM International Conference on Information \& Knowledge Management}}. \bibinfo{pages}{3400--3409}.
\newblock


\bibitem[Pan et~al\mbox{.}(2022)]%
        {10.1145/3511808.3557101}
\bibfield{author}{\bibinfo{person}{Xingyu Pan}, \bibinfo{person}{Yushuo Chen}, \bibinfo{person}{Changxin Tian}, \bibinfo{person}{Zihan Lin}, \bibinfo{person}{Jinpeng Wang}, \bibinfo{person}{He Hu}, {and} \bibinfo{person}{Wayne~Xin Zhao}.} \bibinfo{year}{2022}\natexlab{}.
\newblock \showarticletitle{Multimodal Meta-Learning for Cold-Start Sequential Recommendation}. In \bibinfo{booktitle}{\emph{Proceedings of the 31st ACM International Conference on Information \& Knowledge Management}} (Atlanta, GA, USA) \emph{(\bibinfo{series}{CIKM '22})}. \bibinfo{publisher}{Association for Computing Machinery}, \bibinfo{address}{New York, NY, USA}, \bibinfo{pages}{3421–3430}.
\newblock
\showISBNx{9781450392365}
\href{https://doi.org/10.1145/3511808.3557101}{doi:\nolinkurl{10.1145/3511808.3557101}}


\bibitem[Pan et~al\mbox{.}(2023)]%
        {pan2023learning}
\bibfield{author}{\bibinfo{person}{Yunzhu Pan}, \bibinfo{person}{Nian Li}, \bibinfo{person}{Chen Gao}, \bibinfo{person}{Jianxin Chang}, \bibinfo{person}{Yanan Niu}, \bibinfo{person}{Yang Song}, \bibinfo{person}{Depeng Jin}, {and} \bibinfo{person}{Yong Li}.} \bibinfo{year}{2023}\natexlab{}.
\newblock \showarticletitle{Learning and optimization of implicit negative feedback for industrial short-video recommender system}. In \bibinfo{booktitle}{\emph{Proceedings of the 32nd ACM International Conference on Information and Knowledge Management}}. \bibinfo{pages}{4787--4793}.
\newblock


\bibitem[Pande et~al\mbox{.}(2023)]%
        {pande2023personalized}
\bibfield{author}{\bibinfo{person}{Amit Pande}, \bibinfo{person}{Kunal Ghosh}, {and} \bibinfo{person}{Rankyung Park}.} \bibinfo{year}{2023}\natexlab{}.
\newblock \showarticletitle{Personalized Category Frequency prediction for Buy It Again recommendations}. In \bibinfo{booktitle}{\emph{Proceedings of the 17th ACM Conference on Recommender Systems}}. \bibinfo{pages}{730--736}.
\newblock


\bibitem[Park et~al\mbox{.}(2024)]%
        {park2024slh}
\bibfield{author}{\bibinfo{person}{Rankyung Park}, \bibinfo{person}{Amit Pande}, \bibinfo{person}{David Relyea}, \bibinfo{person}{Pushkar Chennu}, {and} \bibinfo{person}{Prathyusha Kanmanth~Reddy}.} \bibinfo{year}{2024}\natexlab{}.
\newblock \showarticletitle{SLH-BIA: Short-Long Hawkes Process for Buy It Again Recommendations at Scale}. In \bibinfo{booktitle}{\emph{Proceedings of the 47th International ACM SIGIR Conference on Research and Development in Information Retrieval}}. \bibinfo{pages}{2965--2969}.
\newblock


\bibitem[Ploshkin et~al\mbox{.}(2025)]%
        {Yambda5B}
\bibfield{author}{\bibinfo{person}{A. Ploshkin}, \bibinfo{person}{V. Tytskiy}, \bibinfo{person}{A. Pismenny}, \bibinfo{person}{V. Baikalov}, \bibinfo{person}{E. Taychinov}, \bibinfo{person}{A. Permiakov}, \bibinfo{person}{D. Burlakov}, \bibinfo{person}{E. Krofto}, {and} \bibinfo{person}{N. Savushkin}.} \bibinfo{year}{2025}\natexlab{}.
\newblock \showarticletitle{Yambda-5B - {A} Large-Scale Multi-modal Dataset for Ranking And Retrieval}.
\newblock \bibinfo{journal}{\emph{CoRR}}  \bibinfo{volume}{abs/2505.22238} (\bibinfo{year}{2025}).
\newblock
\showeprint[arXiv]{2505.22238}
\href{https://doi.org/10.48550/ARXIV.2505.22238}{doi:\nolinkurl{10.48550/ARXIV.2505.22238}}


\bibitem[Qian et~al\mbox{.}(2022)]%
        {qian2022intelligent}
\bibfield{author}{\bibinfo{person}{Xufeng Qian}, \bibinfo{person}{Yue Xu}, \bibinfo{person}{Fuyu Lv}, \bibinfo{person}{Shengyu Zhang}, \bibinfo{person}{Ziwen Jiang}, \bibinfo{person}{Qingwen Liu}, \bibinfo{person}{Xiaoyi Zeng}, \bibinfo{person}{Tat-Seng Chua}, {and} \bibinfo{person}{Fei Wu}.} \bibinfo{year}{2022}\natexlab{}.
\newblock \showarticletitle{Intelligent request strategy design in recommender system}. In \bibinfo{booktitle}{\emph{Proceedings of the 28th ACM SIGKDD Conference on Knowledge Discovery and Data Mining}}. \bibinfo{pages}{3772--3782}.
\newblock


\bibitem[Qin et~al\mbox{.}(2023)]%
        {qin2023learning}
\bibfield{author}{\bibinfo{person}{Jiarui Qin}, \bibinfo{person}{Jiachen Zhu}, \bibinfo{person}{Yankai Liu}, \bibinfo{person}{Junchao Gao}, \bibinfo{person}{Jianjie Ying}, \bibinfo{person}{Chaoxiong Liu}, \bibinfo{person}{Ding Wang}, \bibinfo{person}{Junlan Feng}, \bibinfo{person}{Chao Deng}, \bibinfo{person}{Xiaozheng Wang}, {et~al\mbox{.}}} \bibinfo{year}{2023}\natexlab{}.
\newblock \showarticletitle{Learning to distinguish multi-user coupling behaviors for tv recommendation}. In \bibinfo{booktitle}{\emph{Proceedings of the sixteenth ACM international conference on web search and data mining}}. \bibinfo{pages}{204--212}.
\newblock


\bibitem[Qin et~al\mbox{.}(2021)]%
        {qin2021bootstrapping}
\bibfield{author}{\bibinfo{person}{Zhen Qin}, \bibinfo{person}{Honglei Zhuang}, \bibinfo{person}{Rolf Jagerman}, \bibinfo{person}{Xinyu Qian}, \bibinfo{person}{Po Hu}, \bibinfo{person}{Dan~Chary Chen}, \bibinfo{person}{Xuanhui Wang}, \bibinfo{person}{Michael Bendersky}, {and} \bibinfo{person}{Marc Najork}.} \bibinfo{year}{2021}\natexlab{}.
\newblock \showarticletitle{Bootstrapping recommendations at chrome web store}. In \bibinfo{booktitle}{\emph{Proceedings of the 27th ACM SIGKDD Conference on Knowledge Discovery \& Data Mining}}. \bibinfo{pages}{3483--3491}.
\newblock


\bibitem[Qu et~al\mbox{.}(2022)]%
        {qu2022single}
\bibfield{author}{\bibinfo{person}{Liang Qu}, \bibinfo{person}{Yonghong Ye}, \bibinfo{person}{Ningzhi Tang}, \bibinfo{person}{Lixin Zhang}, \bibinfo{person}{Yuhui Shi}, {and} \bibinfo{person}{Hongzhi Yin}.} \bibinfo{year}{2022}\natexlab{}.
\newblock \showarticletitle{Single-shot embedding dimension search in recommender system}. In \bibinfo{booktitle}{\emph{Proceedings of the 45th International ACM SIGIR conference on research and development in Information Retrieval}}. \bibinfo{pages}{513--522}.
\newblock


\bibitem[Rangadurai et~al\mbox{.}(2022)]%
        {rangadurai2022nxtpost}
\bibfield{author}{\bibinfo{person}{Kaushik Rangadurai}, \bibinfo{person}{Yiqun Liu}, \bibinfo{person}{Siddarth Malreddy}, \bibinfo{person}{Xiaoyi Liu}, \bibinfo{person}{Piyush Maheshwari}, \bibinfo{person}{Vishwanath Sangale}, {and} \bibinfo{person}{Fedor Borisyuk}.} \bibinfo{year}{2022}\natexlab{}.
\newblock \showarticletitle{Nxtpost: User to post recommendations in facebook groups}. In \bibinfo{booktitle}{\emph{Proceedings of the 28th ACM SIGKDD Conference on Knowledge Discovery and Data Mining}}. \bibinfo{pages}{3792--3800}.
\newblock


\bibitem[Ren et~al\mbox{.}(2024)]%
        {ren2024non}
\bibfield{author}{\bibinfo{person}{Yuxin Ren}, \bibinfo{person}{Qiya Yang}, \bibinfo{person}{Yichun Wu}, \bibinfo{person}{Wei Xu}, \bibinfo{person}{Yalong Wang}, {and} \bibinfo{person}{Zhiqiang Zhang}.} \bibinfo{year}{2024}\natexlab{}.
\newblock \showarticletitle{Non-autoregressive generative models for reranking recommendation}. In \bibinfo{booktitle}{\emph{Proceedings of the 30th ACM SIGKDD Conference on Knowledge Discovery and Data Mining}}. \bibinfo{pages}{5625--5634}.
\newblock


\bibitem[Ren et~al\mbox{.}(2023)]%
        {ren2023greenseq}
\bibfield{author}{\bibinfo{person}{Yankun Ren}, \bibinfo{person}{Xinxing Yang}, \bibinfo{person}{Xingyu Lu}, \bibinfo{person}{Longfei Li}, \bibinfo{person}{Jun Zhou}, \bibinfo{person}{Jinjie Gu}, {and} \bibinfo{person}{Guannan Zhang}.} \bibinfo{year}{2023}\natexlab{}.
\newblock \showarticletitle{GreenSeq: automatic design of green networks for sequential recommendation systems}. In \bibinfo{booktitle}{\emph{Proceedings of the 46th International ACM SIGIR Conference on Research and Development in Information Retrieval}}. \bibinfo{pages}{3364--3368}.
\newblock


\bibitem[Sagtani et~al\mbox{.}(2023)]%
        {sagtani2023quantifying}
\bibfield{author}{\bibinfo{person}{Hitesh Sagtani}, \bibinfo{person}{Madan~Gopal Jhawar}, \bibinfo{person}{Akshat Gupta}, {and} \bibinfo{person}{Rishabh Mehrotra}.} \bibinfo{year}{2023}\natexlab{}.
\newblock \showarticletitle{Quantifying and leveraging user fatigue for interventions in recommender systems}. In \bibinfo{booktitle}{\emph{Proceedings of the 46th International ACM SIGIR Conference on Research and Development in Information Retrieval}}. \bibinfo{pages}{2293--2297}.
\newblock


\bibitem[Shao et~al\mbox{.}(2024)]%
        {shao2024optimizing}
\bibfield{author}{\bibinfo{person}{Yuan Shao}, \bibinfo{person}{Bibang Liu}, \bibinfo{person}{Sourabh Bansod}, \bibinfo{person}{Arnab Bhadury}, \bibinfo{person}{Mingyan Gao}, {and} \bibinfo{person}{Yaping Zhang}.} \bibinfo{year}{2024}\natexlab{}.
\newblock \showarticletitle{Optimizing for Participation in Recommender System}. In \bibinfo{booktitle}{\emph{Proceedings of the 18th ACM Conference on Recommender Systems}}. \bibinfo{pages}{806--808}.
\newblock


\bibitem[Shen et~al\mbox{.}(2021)]%
        {10.1145/3459637.3481948}
\bibfield{author}{\bibinfo{person}{Qijie Shen}, \bibinfo{person}{Wanjie Tao}, \bibinfo{person}{Jing Zhang}, \bibinfo{person}{Hong Wen}, \bibinfo{person}{Zulong Chen}, {and} \bibinfo{person}{Quan Lu}.} \bibinfo{year}{2021}\natexlab{}.
\newblock \showarticletitle{SAR-Net: A Scenario-Aware Ranking Network for Personalized Fair Recommendation in Hundreds of Travel Scenarios}. In \bibinfo{booktitle}{\emph{Proceedings of the 30th ACM International Conference on Information \& Knowledge Management}} (Virtual Event, Queensland, Australia) \emph{(\bibinfo{series}{CIKM '21})}. \bibinfo{publisher}{Association for Computing Machinery}, \bibinfo{address}{New York, NY, USA}, \bibinfo{pages}{4094–4103}.
\newblock
\showISBNx{9781450384469}
\href{https://doi.org/10.1145/3459637.3481948}{doi:\nolinkurl{10.1145/3459637.3481948}}


\bibitem[Shen et~al\mbox{.}(2022)]%
        {shen2022deep}
\bibfield{author}{\bibinfo{person}{Qijie Shen}, \bibinfo{person}{Hong Wen}, \bibinfo{person}{Wanjie Tao}, \bibinfo{person}{Jing Zhang}, \bibinfo{person}{Fuyu Lv}, \bibinfo{person}{Zulong Chen}, {and} \bibinfo{person}{Zhao Li}.} \bibinfo{year}{2022}\natexlab{}.
\newblock \showarticletitle{Deep interest highlight network for click-through rate prediction in trigger-induced recommendation}. In \bibinfo{booktitle}{\emph{Proceedings of the ACM web conference 2022}}. \bibinfo{pages}{422--430}.
\newblock


\bibitem[Shi et~al\mbox{.}(2023)]%
        {shi2023embedding}
\bibfield{author}{\bibinfo{person}{Jiahui Shi}, \bibinfo{person}{Vivek Chaurasiya}, \bibinfo{person}{Yozen Liu}, \bibinfo{person}{Shubham Vij}, \bibinfo{person}{Yan Wu}, \bibinfo{person}{Satya Kanduri}, \bibinfo{person}{Neil Shah}, \bibinfo{person}{Peicheng Yu}, \bibinfo{person}{Nik Srivastava}, \bibinfo{person}{Lei Shi}, {et~al\mbox{.}}} \bibinfo{year}{2023}\natexlab{}.
\newblock \showarticletitle{Embedding based retrieval in friend recommendation}. In \bibinfo{booktitle}{\emph{Proceedings of the 46th International ACM SIGIR Conference on Research and Development in Information Retrieval}}. \bibinfo{pages}{3330--3334}.
\newblock


\bibitem[Shi et~al\mbox{.}(2021)]%
        {shi2021wg4rec}
\bibfield{author}{\bibinfo{person}{Shaoyun Shi}, \bibinfo{person}{Weizhi Ma}, \bibinfo{person}{Zhen Wang}, \bibinfo{person}{Min Zhang}, \bibinfo{person}{Kun Fang}, \bibinfo{person}{Jingfang Xu}, \bibinfo{person}{Yiqun Liu}, {and} \bibinfo{person}{Shaoping Ma}.} \bibinfo{year}{2021}\natexlab{}.
\newblock \showarticletitle{Wg4rec: Modeling textual content with word graph for news recommendation}. In \bibinfo{booktitle}{\emph{Proceedings of the 30th ACM international conference on information \& knowledge management}}. \bibinfo{pages}{1651--1660}.
\newblock


\bibitem[Si et~al\mbox{.}(2024)]%
        {si2024twin}
\bibfield{author}{\bibinfo{person}{Zihua Si}, \bibinfo{person}{Lin Guan}, \bibinfo{person}{ZhongXiang Sun}, \bibinfo{person}{Xiaoxue Zang}, \bibinfo{person}{Jing Lu}, \bibinfo{person}{Yiqun Hui}, \bibinfo{person}{Xingchao Cao}, \bibinfo{person}{Zeyu Yang}, \bibinfo{person}{Yichen Zheng}, \bibinfo{person}{Dewei Leng}, {et~al\mbox{.}}} \bibinfo{year}{2024}\natexlab{}.
\newblock \showarticletitle{Twin v2: Scaling ultra-long user behavior sequence modeling for enhanced ctr prediction at kuaishou}. In \bibinfo{booktitle}{\emph{Proceedings of the 33rd ACM International Conference on Information and Knowledge Management}}. \bibinfo{pages}{4890--4897}.
\newblock


\bibitem[Sierag and Zielnicki(2022)]%
        {sierag2022client}
\bibfield{author}{\bibinfo{person}{Dirk Sierag} {and} \bibinfo{person}{Kevin Zielnicki}.} \bibinfo{year}{2022}\natexlab{}.
\newblock \showarticletitle{Client Time Series Model: a Multi-Target Recommender System based on Temporally-Masked Encoders}. In \bibinfo{booktitle}{\emph{Proceedings of the 16th ACM Conference on Recommender Systems}}. \bibinfo{pages}{512--515}.
\newblock


\bibitem[Song et~al\mbox{.}(2022)]%
        {song2022friend}
\bibfield{author}{\bibinfo{person}{Xiran Song}, \bibinfo{person}{Jianxun Lian}, \bibinfo{person}{Hong Huang}, \bibinfo{person}{Mingqi Wu}, \bibinfo{person}{Hai Jin}, {and} \bibinfo{person}{Xing Xie}.} \bibinfo{year}{2022}\natexlab{}.
\newblock \showarticletitle{Friend recommendations with self-rescaling graph neural networks}. In \bibinfo{booktitle}{\emph{Proceedings of the 28th ACM SIGKDD conference on knowledge discovery and data mining}}. \bibinfo{pages}{3909--3919}.
\newblock


\bibitem[Spi\v{s}\'{a}k et~al\mbox{.}(2024)]%
        {spivsak2024interpretability}
\bibfield{author}{\bibinfo{person}{Martin Spi\v{s}\'{a}k}, \bibinfo{person}{Radek Bartyzal}, \bibinfo{person}{Anton\'{\i}n Hoskovec}, {and} \bibinfo{person}{Ladislav Pe\v{s}ka}.} \bibinfo{year}{2024}\natexlab{}.
\newblock \showarticletitle{On Interpretability of Linear Autoencoders}. In \bibinfo{booktitle}{\emph{Proceedings of the 18th ACM Conference on Recommender Systems}} (Bari, Italy) \emph{(\bibinfo{series}{RecSys '24})}. \bibinfo{publisher}{Association for Computing Machinery}, \bibinfo{address}{New York, NY, USA}, \bibinfo{pages}{975–980}.
\newblock
\showISBNx{9798400705052}
\href{https://doi.org/10.1145/3640457.3688179}{doi:\nolinkurl{10.1145/3640457.3688179}}


\bibitem[Su et~al\mbox{.}(2024)]%
        {su2024rpaf}
\bibfield{author}{\bibinfo{person}{Shuo Su}, \bibinfo{person}{Xiaoshuang Chen}, \bibinfo{person}{Yao Wang}, \bibinfo{person}{Yulin Wu}, \bibinfo{person}{Ziqiang Zhang}, \bibinfo{person}{Kaiqiao Zhan}, \bibinfo{person}{Ben Wang}, {and} \bibinfo{person}{Kun Gai}.} \bibinfo{year}{2024}\natexlab{}.
\newblock \showarticletitle{RPAF: A Reinforcement Prediction-Allocation Framework for Cache Allocation in Large-Scale Recommender Systems}. In \bibinfo{booktitle}{\emph{Proceedings of the 18th ACM Conference on Recommender Systems}}. \bibinfo{pages}{670--679}.
\newblock


\bibitem[Sun(2023)]%
        {FreshlookSun23}
\bibfield{author}{\bibinfo{person}{Aixin Sun}.} \bibinfo{year}{2023}\natexlab{}.
\newblock \showarticletitle{Take a Fresh Look at Recommender Systems from an Evaluation Standpoint}. In \bibinfo{booktitle}{\emph{Proceedings of the 46th International {ACM} {SIGIR} Conference on Research and Development in Information Retrieval, {SIGIR} 2023, Taipei, Taiwan, July 23-27, 2023}}, \bibfield{editor}{\bibinfo{person}{Hsin{-}Hsi Chen}, \bibinfo{person}{Wei{-}Jou~(Edward) Duh}, \bibinfo{person}{Hen{-}Hsen Huang}, \bibinfo{person}{Makoto~P. Kato}, \bibinfo{person}{Josiane Mothe}, {and} \bibinfo{person}{Barbara Poblete}} (Eds.). \bibinfo{publisher}{{ACM}}, \bibinfo{pages}{2629--2638}.
\newblock
\href{https://doi.org/10.1145/3539618.3591931}{doi:\nolinkurl{10.1145/3539618.3591931}}


\bibitem[Sun et~al\mbox{.}(2020)]%
        {sun2020neighbor}
\bibfield{author}{\bibinfo{person}{Jianing Sun}, \bibinfo{person}{Yingxue Zhang}, \bibinfo{person}{Wei Guo}, \bibinfo{person}{Huifeng Guo}, \bibinfo{person}{Ruiming Tang}, \bibinfo{person}{Xiuqiang He}, \bibinfo{person}{Chen Ma}, {and} \bibinfo{person}{Mark Coates}.} \bibinfo{year}{2020}\natexlab{}.
\newblock \showarticletitle{Neighbor interaction aware graph convolution networks for recommendation}. In \bibinfo{booktitle}{\emph{Proceedings of the 43rd international ACM SIGIR conference on research and development in information retrieval}}. \bibinfo{pages}{1289--1298}.
\newblock


\bibitem[Tang et~al\mbox{.}(2020)]%
        {tang2020progressive}
\bibfield{author}{\bibinfo{person}{Hongyan Tang}, \bibinfo{person}{Junning Liu}, \bibinfo{person}{Ming Zhao}, {and} \bibinfo{person}{Xudong Gong}.} \bibinfo{year}{2020}\natexlab{}.
\newblock \showarticletitle{Progressive layered extraction (ple): A novel multi-task learning (mtl) model for personalized recommendations}. In \bibinfo{booktitle}{\emph{Proceedings of the 14th ACM conference on recommender systems}}. \bibinfo{pages}{269--278}.
\newblock


\bibitem[Tang et~al\mbox{.}(2024)]%
        {tang2024touch}
\bibfield{author}{\bibinfo{person}{Xing Tang}, \bibinfo{person}{Yang Qiao}, \bibinfo{person}{Fuyuan Lyu}, \bibinfo{person}{Dugang Liu}, {and} \bibinfo{person}{Xiuqiang He}.} \bibinfo{year}{2024}\natexlab{}.
\newblock \showarticletitle{Touch the core: Exploring task dependence among hybrid targets for recommendation}. In \bibinfo{booktitle}{\emph{Proceedings of the 18th ACM Conference on Recommender Systems}}. \bibinfo{pages}{329--339}.
\newblock


\bibitem[Tian et~al\mbox{.}(2024)]%
        {tian2024reland}
\bibfield{author}{\bibinfo{person}{Changxin Tian}, \bibinfo{person}{Binbin Hu}, \bibinfo{person}{Chunjing Gan}, \bibinfo{person}{Haoyu Chen}, \bibinfo{person}{Zhuo Zhang}, \bibinfo{person}{Li Yu}, \bibinfo{person}{Ziqi Liu}, \bibinfo{person}{Zhiqiang Zhang}, \bibinfo{person}{Jun Zhou}, {and} \bibinfo{person}{Jiawei Chen}.} \bibinfo{year}{2024}\natexlab{}.
\newblock \showarticletitle{ReLand: Integrating Large Language Models' Insights into Industrial Recommenders via a Controllable Reasoning Pool}. In \bibinfo{booktitle}{\emph{Proceedings of the 18th ACM Conference on Recommender Systems}}. \bibinfo{pages}{63--73}.
\newblock


\bibitem[Tong et~al\mbox{.}(2023)]%
        {tong2023navigating}
\bibfield{author}{\bibinfo{person}{Ding Tong}, \bibinfo{person}{Qifeng Qiao}, \bibinfo{person}{Ting-Po Lee}, \bibinfo{person}{James McInerney}, {and} \bibinfo{person}{Justin Basilico}.} \bibinfo{year}{2023}\natexlab{}.
\newblock \showarticletitle{Navigating the feedback loop in recommender systems: Insights and strategies from industry practice}. In \bibinfo{booktitle}{\emph{Proceedings of the 17th ACM Conference on Recommender Systems}}. \bibinfo{pages}{1058--1061}.
\newblock


\bibitem[Tu et~al\mbox{.}(2023)]%
        {tu2023disentangled}
\bibfield{author}{\bibinfo{person}{Ke Tu}, \bibinfo{person}{Wei Qu}, \bibinfo{person}{Zhengwei Wu}, \bibinfo{person}{Zhiqiang Zhang}, \bibinfo{person}{Zhongyi Liu}, \bibinfo{person}{Yiming Zhao}, \bibinfo{person}{Le Wu}, \bibinfo{person}{Jun Zhou}, {and} \bibinfo{person}{Guannan Zhang}.} \bibinfo{year}{2023}\natexlab{}.
\newblock \showarticletitle{Disentangled Interest importance aware Knowledge Graph Neural Network for Fund Recommendation}. In \bibinfo{booktitle}{\emph{Proceedings of the 32nd ACM International Conference on Information and Knowledge Management}}. \bibinfo{pages}{2482--2491}.
\newblock


\bibitem[Wan et~al\mbox{.}(2024)]%
        {wan2024larr}
\bibfield{author}{\bibinfo{person}{Zhizhong Wan}, \bibinfo{person}{Bin Yin}, \bibinfo{person}{Junjie Xie}, \bibinfo{person}{Fei Jiang}, \bibinfo{person}{Xiang Li}, {and} \bibinfo{person}{Wei Lin}.} \bibinfo{year}{2024}\natexlab{}.
\newblock \showarticletitle{LARR: Large Language Model Aided Real-time Scene Recommendation with Semantic Understanding}. In \bibinfo{booktitle}{\emph{Proceedings of the 18th ACM Conference on Recommender Systems}}. \bibinfo{pages}{23--32}.
\newblock


\bibitem[Wang et~al\mbox{.}(2021)]%
        {10.1145/3459637.3481923}
\bibfield{author}{\bibinfo{person}{Haishuai Wang}, \bibinfo{person}{Zhao Li}, \bibinfo{person}{Xuanwu Liu}, \bibinfo{person}{Donghui Ding}, \bibinfo{person}{Zehong Hu}, \bibinfo{person}{Peng Zhang}, \bibinfo{person}{Chuan Zhou}, {and} \bibinfo{person}{Jiajun Bu}.} \bibinfo{year}{2021}\natexlab{}.
\newblock \showarticletitle{Fulfillment-Time-Aware Personalized Ranking for On-Demand Food Recommendation}. In \bibinfo{booktitle}{\emph{Proceedings of the 30th ACM International Conference on Information \& Knowledge Management}} (Virtual Event, Queensland, Australia) \emph{(\bibinfo{series}{CIKM '21})}. \bibinfo{publisher}{Association for Computing Machinery}, \bibinfo{address}{New York, NY, USA}, \bibinfo{pages}{4184–4192}.
\newblock
\showISBNx{9781450384469}
\href{https://doi.org/10.1145/3459637.3481923}{doi:\nolinkurl{10.1145/3459637.3481923}}


\bibitem[Wang et~al\mbox{.}(2020)]%
        {wang2020m2grl}
\bibfield{author}{\bibinfo{person}{Menghan Wang}, \bibinfo{person}{Yujie Lin}, \bibinfo{person}{Guli Lin}, \bibinfo{person}{Keping Yang}, {and} \bibinfo{person}{Xiao-ming Wu}.} \bibinfo{year}{2020}\natexlab{}.
\newblock \showarticletitle{M2GRL: A multi-task multi-view graph representation learning framework for web-scale recommender systems}. In \bibinfo{booktitle}{\emph{Proceedings of the 26th ACM SIGKDD international conference on knowledge discovery \& data mining}}. \bibinfo{pages}{2349--2358}.
\newblock


\bibitem[Wang et~al\mbox{.}(2024b)]%
        {wang2024future}
\bibfield{author}{\bibinfo{person}{Xiaobei Wang}, \bibinfo{person}{Shuchang Liu}, \bibinfo{person}{Xueliang Wang}, \bibinfo{person}{Qingpeng Cai}, \bibinfo{person}{Lantao Hu}, \bibinfo{person}{Han Li}, \bibinfo{person}{Peng Jiang}, \bibinfo{person}{Kun Gai}, {and} \bibinfo{person}{Guangming Xie}.} \bibinfo{year}{2024}\natexlab{b}.
\newblock \showarticletitle{Future impact decomposition in request-level recommendations}. In \bibinfo{booktitle}{\emph{Proceedings of the 30th ACM SIGKDD Conference on Knowledge Discovery and Data Mining}}. \bibinfo{pages}{5905--5916}.
\newblock


\bibitem[Wang et~al\mbox{.}(2024a)]%
        {wang2024not}
\bibfield{author}{\bibinfo{person}{Yuan Wang}, \bibinfo{person}{Zhiyu Li}, \bibinfo{person}{Changshuo Zhang}, \bibinfo{person}{Sirui Chen}, \bibinfo{person}{Xiao Zhang}, \bibinfo{person}{Jun Xu}, {and} \bibinfo{person}{Quan Lin}.} \bibinfo{year}{2024}\natexlab{a}.
\newblock \showarticletitle{Do Not Wait: Learning Re-Ranking Model Without User Feedback At Serving Time in E-Commerce}. In \bibinfo{booktitle}{\emph{Proceedings of the 18th ACM Conference on Recommender Systems}}. \bibinfo{pages}{896--901}.
\newblock


\bibitem[Wang et~al\mbox{.}(2022)]%
        {wang2022recommending}
\bibfield{author}{\bibinfo{person}{Yuyan Wang}, \bibinfo{person}{Long Tao}, {and} \bibinfo{person}{Xian~Xing Zhang}.} \bibinfo{year}{2022}\natexlab{}.
\newblock \showarticletitle{Recommending for a multi-sided marketplace with heterogeneous contents}. In \bibinfo{booktitle}{\emph{Proceedings of the 16th ACM Conference on Recommender Systems}}. \bibinfo{pages}{456--459}.
\newblock


\bibitem[Wang et~al\mbox{.}(2023a)]%
        {wang2023diversity}
\bibfield{author}{\bibinfo{person}{Zihong Wang}, \bibinfo{person}{Yingxia Shao}, \bibinfo{person}{Jiyuan He}, \bibinfo{person}{Jinbao Liu}, \bibinfo{person}{Shitao Xiao}, \bibinfo{person}{Tao Feng}, {and} \bibinfo{person}{Ming Liu}.} \bibinfo{year}{2023}\natexlab{a}.
\newblock \showarticletitle{Diversity-aware deep ranking network for recommendation}. In \bibinfo{booktitle}{\emph{Proceedings of the 32nd ACM International Conference on Information and Knowledge Management}}. \bibinfo{pages}{2564--2573}.
\newblock


\bibitem[Wang et~al\mbox{.}(2023b)]%
        {wang2023industrial}
\bibfield{author}{\bibinfo{person}{Zongyi Wang}, \bibinfo{person}{Yanyan Zou}, \bibinfo{person}{Anyu Dai}, \bibinfo{person}{Linfang Hou}, \bibinfo{person}{Nan Qiao}, \bibinfo{person}{Luobao Zou}, \bibinfo{person}{Mian Ma}, \bibinfo{person}{Zhuoye Ding}, {and} \bibinfo{person}{Sulong Xu}.} \bibinfo{year}{2023}\natexlab{b}.
\newblock \showarticletitle{An industrial framework for personalized serendipitous recommendation in E-commerce}. In \bibinfo{booktitle}{\emph{Proceedings of the 17th ACM Conference on Recommender Systems}}. \bibinfo{pages}{1015--1018}.
\newblock


\bibitem[Wilm et~al\mbox{.}(2023)]%
        {wilm2023scaling}
\bibfield{author}{\bibinfo{person}{Timo Wilm}, \bibinfo{person}{Philipp Normann}, \bibinfo{person}{Sophie Baumeister}, {and} \bibinfo{person}{Paul-Vincent Kobow}.} \bibinfo{year}{2023}\natexlab{}.
\newblock \showarticletitle{Scaling session-based transformer recommendations using optimized negative sampling and loss functions}. In \bibinfo{booktitle}{\emph{Proceedings of the 17th ACM Conference on Recommender Systems}}. \bibinfo{pages}{1023--1026}.
\newblock


\bibitem[Wilm et~al\mbox{.}(2024)]%
        {wilm2024pareto}
\bibfield{author}{\bibinfo{person}{Timo Wilm}, \bibinfo{person}{Philipp Normann}, {and} \bibinfo{person}{Felix Stepprath}.} \bibinfo{year}{2024}\natexlab{}.
\newblock \showarticletitle{Pareto Front Approximation for Multi-Objective Session-Based Recommender Systems}. In \bibinfo{booktitle}{\emph{Proceedings of the 18th ACM Conference on Recommender Systems}}. \bibinfo{pages}{809--812}.
\newblock


\bibitem[Wu et~al\mbox{.}(2020)]%
        {MindDataset}
\bibfield{author}{\bibinfo{person}{Fangzhao Wu}, \bibinfo{person}{Ying Qiao}, \bibinfo{person}{Jiun-Hung Chen}, \bibinfo{person}{Chuhan Wu}, \bibinfo{person}{Tao Qi}, \bibinfo{person}{Jianxun Lian}, \bibinfo{person}{Danyang Liu}, \bibinfo{person}{Xing Xie}, \bibinfo{person}{Jianfeng Gao}, \bibinfo{person}{Winnie Wu}, {and} \bibinfo{person}{Ming Zhou}.} \bibinfo{year}{2020}\natexlab{}.
\newblock \showarticletitle{{MIND}: A Large-scale Dataset for News Recommendation}. In \bibinfo{booktitle}{\emph{Proceedings of the 58th Annual Meeting of the Association for Computational Linguistics}}, \bibfield{editor}{\bibinfo{person}{Dan Jurafsky}, \bibinfo{person}{Joyce Chai}, \bibinfo{person}{Natalie Schluter}, {and} \bibinfo{person}{Joel Tetreault}} (Eds.). \bibinfo{publisher}{Association for Computational Linguistics}, \bibinfo{address}{Online}, \bibinfo{pages}{3597--3606}.
\newblock
\href{https://doi.org/10.18653/v1/2020.acl-main.331}{doi:\nolinkurl{10.18653/v1/2020.acl-main.331}}


\bibitem[Wu and Grbovic(2020)]%
        {wu2020airbnb}
\bibfield{author}{\bibinfo{person}{Liang Wu} {and} \bibinfo{person}{Mihajlo Grbovic}.} \bibinfo{year}{2020}\natexlab{}.
\newblock \showarticletitle{How Airbnb tells you will enjoy sunset sailing in Barcelona? Recommendation in a two-sided travel marketplace}. In \bibinfo{booktitle}{\emph{Proceedings of the 43rd International ACM SIGIR Conference on Research and Development in Information Retrieval}}. \bibinfo{pages}{2387--2396}.
\newblock


\bibitem[Wu et~al\mbox{.}(2023)]%
        {wu2023ruel}
\bibfield{author}{\bibinfo{person}{Ning Wu}, \bibinfo{person}{Ming Gong}, \bibinfo{person}{Linjun Shou}, \bibinfo{person}{Jian Pei}, {and} \bibinfo{person}{Daxin Jiang}.} \bibinfo{year}{2023}\natexlab{}.
\newblock \showarticletitle{Ruel: Retrieval-augmented user representation with edge browser logs for sequential recommendation}. In \bibinfo{booktitle}{\emph{Proceedings of the 32nd ACM International Conference on Information and Knowledge Management}}. \bibinfo{pages}{4871--4878}.
\newblock


\bibitem[Wu et~al\mbox{.}(2024)]%
        {wu2024learned}
\bibfield{author}{\bibinfo{person}{Yi Wu}, \bibinfo{person}{Daryl Chang}, \bibinfo{person}{Jennifer She}, \bibinfo{person}{Zhe Zhao}, \bibinfo{person}{Li Wei}, {and} \bibinfo{person}{Lukasz Heldt}.} \bibinfo{year}{2024}\natexlab{}.
\newblock \showarticletitle{Learned Ranking Function: From Short-term Behavior Predictions to Long-term User Satisfaction}. In \bibinfo{booktitle}{\emph{Proceedings of the 18th ACM Conference on Recommender Systems}}. \bibinfo{pages}{1004--1009}.
\newblock


\bibitem[Xi et~al\mbox{.}(2024)]%
        {xi2024towards}
\bibfield{author}{\bibinfo{person}{Yunjia Xi}, \bibinfo{person}{Weiwen Liu}, \bibinfo{person}{Jianghao Lin}, \bibinfo{person}{Xiaoling Cai}, \bibinfo{person}{Hong Zhu}, \bibinfo{person}{Jieming Zhu}, \bibinfo{person}{Bo Chen}, \bibinfo{person}{Ruiming Tang}, \bibinfo{person}{Weinan Zhang}, {and} \bibinfo{person}{Yong Yu}.} \bibinfo{year}{2024}\natexlab{}.
\newblock \showarticletitle{Towards open-world recommendation with knowledge augmentation from large language models}. In \bibinfo{booktitle}{\emph{Proceedings of the 18th ACM Conference on Recommender Systems}}. \bibinfo{pages}{12--22}.
\newblock


\bibitem[Xian et~al\mbox{.}(2021)]%
        {xian2021ex3}
\bibfield{author}{\bibinfo{person}{Yikun Xian}, \bibinfo{person}{Tong Zhao}, \bibinfo{person}{Jin Li}, \bibinfo{person}{Jim Chan}, \bibinfo{person}{Andrey Kan}, \bibinfo{person}{Jun Ma}, \bibinfo{person}{Xin~Luna Dong}, \bibinfo{person}{Christos Faloutsos}, \bibinfo{person}{George Karypis}, \bibinfo{person}{Shan Muthukrishnan}, {et~al\mbox{.}}} \bibinfo{year}{2021}\natexlab{}.
\newblock \showarticletitle{Ex3: Explainable attribute-aware item-set recommendations}. In \bibinfo{booktitle}{\emph{Proceedings of the 15th ACM Conference on Recommender Systems}}. \bibinfo{pages}{484--494}.
\newblock


\bibitem[Xiao et~al\mbox{.}(2022)]%
        {xiao2022training}
\bibfield{author}{\bibinfo{person}{Shitao Xiao}, \bibinfo{person}{Zheng Liu}, \bibinfo{person}{Yingxia Shao}, \bibinfo{person}{Tao Di}, \bibinfo{person}{Bhuvan Middha}, \bibinfo{person}{Fangzhao Wu}, {and} \bibinfo{person}{Xing Xie}.} \bibinfo{year}{2022}\natexlab{}.
\newblock \showarticletitle{Training large-scale news recommenders with pretrained language models in the loop}. In \bibinfo{booktitle}{\emph{Proceedings of the 28th ACM SIGKDD Conference on Knowledge Discovery and Data Mining}}. \bibinfo{pages}{4215--4225}.
\newblock


\bibitem[Xiao et~al\mbox{.}(2024)]%
        {xiao2024deep}
\bibfield{author}{\bibinfo{person}{Zhibo Xiao}, \bibinfo{person}{Luwei Yang}, \bibinfo{person}{Tao Zhang}, \bibinfo{person}{Wen Jiang}, \bibinfo{person}{Wei Ning}, {and} \bibinfo{person}{Yujiu Yang}.} \bibinfo{year}{2024}\natexlab{}.
\newblock \showarticletitle{Deep evolutional instant interest network for ctr prediction in trigger-induced recommendation}. In \bibinfo{booktitle}{\emph{Proceedings of the 17th ACM International Conference on Web Search and Data Mining}}. \bibinfo{pages}{846--854}.
\newblock


\bibitem[Xie et~al\mbox{.}(2021b)]%
        {xie2021real}
\bibfield{author}{\bibinfo{person}{Ruobing Xie}, \bibinfo{person}{Rui Wang}, \bibinfo{person}{Shaoliang Zhang}, \bibinfo{person}{Zhihong Yang}, \bibinfo{person}{Feng Xia}, {and} \bibinfo{person}{Leyu Lin}.} \bibinfo{year}{2021}\natexlab{b}.
\newblock \showarticletitle{Real-time relevant recommendation suggestion}. In \bibinfo{booktitle}{\emph{Proceedings of the 14th ACM International Conference on Web Search and Data Mining}}. \bibinfo{pages}{112--120}.
\newblock


\bibitem[Xie et~al\mbox{.}(2021a)]%
        {xie2021causcf}
\bibfield{author}{\bibinfo{person}{Xu Xie}, \bibinfo{person}{Zhaoyang Liu}, \bibinfo{person}{Shiwen Wu}, \bibinfo{person}{Fei Sun}, \bibinfo{person}{Cihang Liu}, \bibinfo{person}{Jiawei Chen}, \bibinfo{person}{Jinyang Gao}, \bibinfo{person}{Bin Cui}, {and} \bibinfo{person}{Bolin Ding}.} \bibinfo{year}{2021}\natexlab{a}.
\newblock \showarticletitle{CausCF: Causal collaborative filtering for recommendation effect estimation}. In \bibinfo{booktitle}{\emph{Proceedings of the 30th ACM International Conference on Information \& Knowledge Management}}. \bibinfo{pages}{4253--4263}.
\newblock


\bibitem[Xu et~al\mbox{.}(2020a)]%
        {xu2020privileged}
\bibfield{author}{\bibinfo{person}{Chen Xu}, \bibinfo{person}{Quan Li}, \bibinfo{person}{Junfeng Ge}, \bibinfo{person}{Jinyang Gao}, \bibinfo{person}{Xiaoyong Yang}, \bibinfo{person}{Changhua Pei}, \bibinfo{person}{Fei Sun}, \bibinfo{person}{Jian Wu}, \bibinfo{person}{Hanxiao Sun}, {and} \bibinfo{person}{Wenwu Ou}.} \bibinfo{year}{2020}\natexlab{a}.
\newblock \showarticletitle{Privileged features distillation at taobao recommendations}. In \bibinfo{booktitle}{\emph{Proceedings of the 26th ACM SIGKDD International Conference on Knowledge Discovery \& Data Mining}}. \bibinfo{pages}{2590--2598}.
\newblock


\bibitem[Xu et~al\mbox{.}(2020c)]%
        {xu2020gemini}
\bibfield{author}{\bibinfo{person}{Jixing Xu}, \bibinfo{person}{Zhenlong Zhu}, \bibinfo{person}{Jianxin Zhao}, \bibinfo{person}{Xuanye Liu}, \bibinfo{person}{Minghui Shan}, {and} \bibinfo{person}{Jiecheng Guo}.} \bibinfo{year}{2020}\natexlab{c}.
\newblock \showarticletitle{Gemini: A novel and universal heterogeneous graph information fusing framework for online recommendations}. In \bibinfo{booktitle}{\emph{Proceedings of the 26th ACM SIGKDD international conference on knowledge discovery \& data mining}}. \bibinfo{pages}{3356--3365}.
\newblock


\bibitem[Xu et~al\mbox{.}(2024a)]%
        {10.1145/3627673.3680037}
\bibfield{author}{\bibinfo{person}{Lanling Xu}, \bibinfo{person}{Zhen Tian}, \bibinfo{person}{Bingqian Li}, \bibinfo{person}{Junjie Zhang}, \bibinfo{person}{Daoyuan Wang}, \bibinfo{person}{Hongyu Wang}, \bibinfo{person}{Jinpeng Wang}, \bibinfo{person}{Sheng Chen}, {and} \bibinfo{person}{Wayne~Xin Zhao}.} \bibinfo{year}{2024}\natexlab{a}.
\newblock \showarticletitle{Sequence-level Semantic Representation Fusion for Recommender Systems}. In \bibinfo{booktitle}{\emph{Proceedings of the 33rd ACM International Conference on Information and Knowledge Management}} (Boise, ID, USA) \emph{(\bibinfo{series}{CIKM '24})}. \bibinfo{publisher}{Association for Computing Machinery}, \bibinfo{address}{New York, NY, USA}, \bibinfo{pages}{5015–5022}.
\newblock
\showISBNx{9798400704369}
\href{https://doi.org/10.1145/3627673.3680037}{doi:\nolinkurl{10.1145/3627673.3680037}}


\bibitem[Xu et~al\mbox{.}(2023)]%
        {xu2023optimizing}
\bibfield{author}{\bibinfo{person}{Ruiyang Xu}, \bibinfo{person}{Jalaj Bhandari}, \bibinfo{person}{Dmytro Korenkevych}, \bibinfo{person}{Fan Liu}, \bibinfo{person}{Yuchen He}, \bibinfo{person}{Alex Nikulkov}, {and} \bibinfo{person}{Zheqing Zhu}.} \bibinfo{year}{2023}\natexlab{}.
\newblock \showarticletitle{Optimizing long-term value for auction-based recommender systems via on-policy reinforcement learning}. In \bibinfo{booktitle}{\emph{Proceedings of the 17th ACM Conference on Recommender Systems}}. \bibinfo{pages}{955--962}.
\newblock


\bibitem[Xu et~al\mbox{.}(2024b)]%
        {xu2024rethinking}
\bibfield{author}{\bibinfo{person}{Wujiang Xu}, \bibinfo{person}{Qitian Wu}, \bibinfo{person}{Runzhong Wang}, \bibinfo{person}{Mingming Ha}, \bibinfo{person}{Qiongxu Ma}, \bibinfo{person}{Linxun Chen}, \bibinfo{person}{Bing Han}, {and} \bibinfo{person}{Junchi Yan}.} \bibinfo{year}{2024}\natexlab{b}.
\newblock \showarticletitle{Rethinking cross-domain sequential recommendation under open-world assumptions}. In \bibinfo{booktitle}{\emph{Proceedings of the ACM Web Conference 2024}}. \bibinfo{pages}{3173--3184}.
\newblock


\bibitem[Xu et~al\mbox{.}(2020b)]%
        {xu2020recommender}
\bibfield{author}{\bibinfo{person}{Zhe Xu}, \bibinfo{person}{Chang Men}, \bibinfo{person}{Peng Li}, \bibinfo{person}{Bicheng Jin}, \bibinfo{person}{Ge Li}, \bibinfo{person}{Yue Yang}, \bibinfo{person}{Chunyang Liu}, \bibinfo{person}{Ben Wang}, {and} \bibinfo{person}{Xiaohu Qie}.} \bibinfo{year}{2020}\natexlab{b}.
\newblock \showarticletitle{When recommender systems meet fleet management: Practical study in online driver repositioning system}. In \bibinfo{booktitle}{\emph{Proceedings of the web conference 2020}}. \bibinfo{pages}{2220--2229}.
\newblock


\bibitem[Yang et~al\mbox{.}(2023)]%
        {yang2023graph}
\bibfield{author}{\bibinfo{person}{Shuai Yang}, \bibinfo{person}{Lixin Zhang}, \bibinfo{person}{Feng Xia}, {and} \bibinfo{person}{Leyu Lin}.} \bibinfo{year}{2023}\natexlab{}.
\newblock \showarticletitle{Graph exploration matters: improving both individual-level and system-level diversity in wechat feed recommendation}. In \bibinfo{booktitle}{\emph{Proceedings of the 32nd ACM International Conference on Information and Knowledge Management}}. \bibinfo{pages}{4901--4908}.
\newblock


\bibitem[Yang et~al\mbox{.}(2024b)]%
        {yang2024enhancing}
\bibfield{author}{\bibinfo{person}{Yunfei Yang}, \bibinfo{person}{Zhenghao Qi}, \bibinfo{person}{Honghuan Wu}, \bibinfo{person}{Qi Song}, \bibinfo{person}{Tieyao Zhang}, \bibinfo{person}{Hao Li}, \bibinfo{person}{Yimin Tu}, \bibinfo{person}{Kaiqiao Zhan}, {and} \bibinfo{person}{Ben Wang}.} \bibinfo{year}{2024}\natexlab{b}.
\newblock \showarticletitle{Enhancing Playback Performance in Video Recommender Systems with an On-Device Gating and Ranking Framework}. In \bibinfo{booktitle}{\emph{Proceedings of the 33rd ACM International Conference on Information and Knowledge Management}}. \bibinfo{pages}{5031--5037}.
\newblock


\bibitem[Yang et~al\mbox{.}(2024a)]%
        {yang2024mlora}
\bibfield{author}{\bibinfo{person}{Zhiming Yang}, \bibinfo{person}{Haining Gao}, \bibinfo{person}{Dehong Gao}, \bibinfo{person}{Luwei Yang}, \bibinfo{person}{Libin Yang}, \bibinfo{person}{Xiaoyan Cai}, \bibinfo{person}{Wei Ning}, {and} \bibinfo{person}{Guannan Zhang}.} \bibinfo{year}{2024}\natexlab{a}.
\newblock \showarticletitle{Mlora: Multi-domain low-rank adaptive network for ctr prediction}. In \bibinfo{booktitle}{\emph{Proceedings of the 18th ACM Conference on Recommender Systems}}. \bibinfo{pages}{287--297}.
\newblock


\bibitem[Yao et~al\mbox{.}(2024)]%
        {yao2024user}
\bibfield{author}{\bibinfo{person}{Fan Yao}, \bibinfo{person}{Yiming Liao}, \bibinfo{person}{Mingzhe Wu}, \bibinfo{person}{Chuanhao Li}, \bibinfo{person}{Yan Zhu}, \bibinfo{person}{James Yang}, \bibinfo{person}{Jingzhou Liu}, \bibinfo{person}{Qifan Wang}, \bibinfo{person}{Haifeng Xu}, {and} \bibinfo{person}{Hongning Wang}.} \bibinfo{year}{2024}\natexlab{}.
\newblock \showarticletitle{User welfare optimization in recommender systems with competing content creators}. In \bibinfo{booktitle}{\emph{Proceedings of the 30th ACM SIGKDD Conference on Knowledge Discovery and Data Mining}}. \bibinfo{pages}{3874--3885}.
\newblock


\bibitem[Ye et~al\mbox{.}(2023)]%
        {ye2023transformer}
\bibfield{author}{\bibinfo{person}{Wenting Ye}, \bibinfo{person}{Hongfei Yang}, \bibinfo{person}{Shuai Zhao}, \bibinfo{person}{Haoyang Fang}, \bibinfo{person}{Xingjian Shi}, {and} \bibinfo{person}{Naveen Neppalli}.} \bibinfo{year}{2023}\natexlab{}.
\newblock \showarticletitle{A Transformer-Based Substitute Recommendation Model Incorporating Weakly Supervised Customer Behavior Data}. In \bibinfo{booktitle}{\emph{Proceedings of the 46th International ACM SIGIR Conference on Research and Development in Information Retrieval}}. \bibinfo{pages}{3325--3329}.
\newblock


\bibitem[Yi et~al\mbox{.}(2023)]%
        {yi2023online}
\bibfield{author}{\bibinfo{person}{Xinyang Yi}, \bibinfo{person}{Shao-Chuan Wang}, \bibinfo{person}{Ruining He}, \bibinfo{person}{Hariharan Chandrasekaran}, \bibinfo{person}{Charles Wu}, \bibinfo{person}{Lukasz Heldt}, \bibinfo{person}{Lichan Hong}, \bibinfo{person}{Minmin Chen}, {and} \bibinfo{person}{Ed~H Chi}.} \bibinfo{year}{2023}\natexlab{}.
\newblock \showarticletitle{Online matching: A real-time bandit system for large-scale recommendations}. In \bibinfo{booktitle}{\emph{Proceedings of the 17th ACM Conference on Recommender Systems}}. \bibinfo{pages}{403--414}.
\newblock


\bibitem[Yin et~al\mbox{.}(2023)]%
        {yin2023heterogeneous}
\bibfield{author}{\bibinfo{person}{Bin Yin}, \bibinfo{person}{Junjie Xie}, \bibinfo{person}{Yu Qin}, \bibinfo{person}{Zixiang Ding}, \bibinfo{person}{Zhichao Feng}, \bibinfo{person}{Xiang Li}, {and} \bibinfo{person}{Wei Lin}.} \bibinfo{year}{2023}\natexlab{}.
\newblock \showarticletitle{Heterogeneous knowledge fusion: A novel approach for personalized recommendation via llm}. In \bibinfo{booktitle}{\emph{Proceedings of the 17th ACM Conference on Recommender Systems}}. \bibinfo{pages}{599--601}.
\newblock


\bibitem[Yuan et~al\mbox{.}(2023)]%
        {yuan2023hydrus}
\bibfield{author}{\bibinfo{person}{Zhiyu Yuan}, \bibinfo{person}{Kai Ren}, \bibinfo{person}{Gang Wang}, {and} \bibinfo{person}{Xin Miao}.} \bibinfo{year}{2023}\natexlab{}.
\newblock \showarticletitle{Hydrus: Improving Personalized Quality of Experience in Short-form Video Services}. In \bibinfo{booktitle}{\emph{Proceedings of the 46th International ACM SIGIR Conference on Research and Development in Information Retrieval}}. \bibinfo{pages}{1127--1136}.
\newblock


\bibitem[Zhai et~al\mbox{.}(2024)]%
        {zhai2024actions}
\bibfield{author}{\bibinfo{person}{Jiaqi Zhai}, \bibinfo{person}{Lucy Liao}, \bibinfo{person}{Xing Liu}, \bibinfo{person}{Yueming Wang}, \bibinfo{person}{Rui Li}, \bibinfo{person}{Xuan Cao}, \bibinfo{person}{Leon Gao}, \bibinfo{person}{Zhaojie Gong}, \bibinfo{person}{Fangda Gu}, \bibinfo{person}{Michael He}, {et~al\mbox{.}}} \bibinfo{year}{2024}\natexlab{}.
\newblock \showarticletitle{Actions speak louder than words: Trillion-parameter sequential transducers for generative recommendations}.
\newblock \bibinfo{journal}{\emph{arXiv preprint arXiv:2402.17152}} (\bibinfo{year}{2024}).
\newblock


\bibitem[Zhan et~al\mbox{.}(2022)]%
        {zhan2022deconfounding}
\bibfield{author}{\bibinfo{person}{Ruohan Zhan}, \bibinfo{person}{Changhua Pei}, \bibinfo{person}{Qiang Su}, \bibinfo{person}{Jianfeng Wen}, \bibinfo{person}{Xueliang Wang}, \bibinfo{person}{Guanyu Mu}, \bibinfo{person}{Dong Zheng}, \bibinfo{person}{Peng Jiang}, {and} \bibinfo{person}{Kun Gai}.} \bibinfo{year}{2022}\natexlab{}.
\newblock \showarticletitle{Deconfounding duration bias in watch-time prediction for video recommendation}. In \bibinfo{booktitle}{\emph{Proceedings of the 28th ACM SIGKDD conference on knowledge discovery and data mining}}. \bibinfo{pages}{4472--4481}.
\newblock


\bibitem[Zhang et~al\mbox{.}(2023c)]%
        {zhang2023shark}
\bibfield{author}{\bibinfo{person}{Beichuan Zhang}, \bibinfo{person}{Chenggen Sun}, \bibinfo{person}{Jianchao Tan}, \bibinfo{person}{Xinjun Cai}, \bibinfo{person}{Jun Zhao}, \bibinfo{person}{Mengqi Miao}, \bibinfo{person}{Kang Yin}, \bibinfo{person}{Chengru Song}, \bibinfo{person}{Na Mou}, {and} \bibinfo{person}{Yang Song}.} \bibinfo{year}{2023}\natexlab{c}.
\newblock \showarticletitle{Shark: A lightweight model compression approach for large-scale recommender systems}. In \bibinfo{booktitle}{\emph{Proceedings of the 32nd ACM International Conference on Information and Knowledge Management}}. \bibinfo{pages}{4930--4937}.
\newblock


\bibitem[Zhang et~al\mbox{.}(2023b)]%
        {zhang2023multi}
\bibfield{author}{\bibinfo{person}{Cong Zhang}, \bibinfo{person}{Dongyang Liu}, \bibinfo{person}{Lin Zuo}, \bibinfo{person}{Junlan Feng}, \bibinfo{person}{Chao Deng}, \bibinfo{person}{Jian Sun}, \bibinfo{person}{Haitao Zeng}, {and} \bibinfo{person}{Yaohong Zhao}.} \bibinfo{year}{2023}\natexlab{b}.
\newblock \showarticletitle{Multi-gate Mixture-of-Contrastive-Experts with Graph-based Gating Mechanism for TV Recommendation}. In \bibinfo{booktitle}{\emph{Proceedings of the 32nd ACM International Conference on Information and Knowledge Management}}. \bibinfo{pages}{4938--4944}.
\newblock


\bibitem[Zhang et~al\mbox{.}(2024e)]%
        {10.1145/3627673.3680067}
\bibfield{author}{\bibinfo{person}{Feng Zhang}, \bibinfo{person}{Yulin Xu}, \bibinfo{person}{Hongjie Chen}, \bibinfo{person}{Xu Yuan}, \bibinfo{person}{QingWen Liu}, {and} \bibinfo{person}{YuNing Jiang}.} \bibinfo{year}{2024}\natexlab{e}.
\newblock \showarticletitle{Effective Utilization of Large-scale Unobserved Data in Recommendation Systems}. In \bibinfo{booktitle}{\emph{Proceedings of the 33rd ACM International Conference on Information and Knowledge Management}} (Boise, ID, USA) \emph{(\bibinfo{series}{CIKM '24})}. \bibinfo{publisher}{Association for Computing Machinery}, \bibinfo{address}{New York, NY, USA}, \bibinfo{pages}{5070–5077}.
\newblock
\showISBNx{9798400704369}
\href{https://doi.org/10.1145/3627673.3680067}{doi:\nolinkurl{10.1145/3627673.3680067}}


\bibitem[Zhang et~al\mbox{.}(2022a)]%
        {zhang2022multi}
\bibfield{author}{\bibinfo{person}{Qihua Zhang}, \bibinfo{person}{Junning Liu}, \bibinfo{person}{Yuzhuo Dai}, \bibinfo{person}{Yiyan Qi}, \bibinfo{person}{Yifan Yuan}, \bibinfo{person}{Kunlun Zheng}, \bibinfo{person}{Fan Huang}, {and} \bibinfo{person}{Xianfeng Tan}.} \bibinfo{year}{2022}\natexlab{a}.
\newblock \showarticletitle{Multi-task fusion via reinforcement learning for long-term user satisfaction in recommender systems}. In \bibinfo{booktitle}{\emph{Proceedings of the 28th ACM SIGKDD conference on knowledge discovery and data mining}}. \bibinfo{pages}{4510--4520}.
\newblock


\bibitem[Zhang et~al\mbox{.}(2024b)]%
        {zhang2024enhanced}
\bibfield{author}{\bibinfo{person}{Qiang Zhang}, \bibinfo{person}{Zhipeng Teng}, \bibinfo{person}{Disheng Wu}, {and} \bibinfo{person}{Jiayin Wang}.} \bibinfo{year}{2024}\natexlab{b}.
\newblock \showarticletitle{An enhanced batch query architecture in real-time recommendation}. In \bibinfo{booktitle}{\emph{Proceedings of the 33rd ACM International Conference on Information and Knowledge Management}}. \bibinfo{pages}{5078--5085}.
\newblock


\bibitem[Zhang et~al\mbox{.}(2019)]%
        {zhang2019deep}
\bibfield{author}{\bibinfo{person}{Shuai Zhang}, \bibinfo{person}{Lina Yao}, \bibinfo{person}{Aixin Sun}, {and} \bibinfo{person}{Yi Tay}.} \bibinfo{year}{2019}\natexlab{}.
\newblock \showarticletitle{Deep learning based recommender system: A survey and new perspectives}.
\newblock \bibinfo{journal}{\emph{ACM computing surveys (CSUR)}} \bibinfo{volume}{52}, \bibinfo{number}{1} (\bibinfo{year}{2019}), \bibinfo{pages}{1--38}.
\newblock


\bibitem[Zhang et~al\mbox{.}(2023f)]%
        {zhang2023collaborative}
\bibfield{author}{\bibinfo{person}{Wei Zhang}, \bibinfo{person}{Pengye Zhang}, \bibinfo{person}{Bo Zhang}, \bibinfo{person}{Xingxing Wang}, {and} \bibinfo{person}{Dong Wang}.} \bibinfo{year}{2023}\natexlab{f}.
\newblock \showarticletitle{A collaborative transfer learning framework for cross-domain recommendation}. In \bibinfo{booktitle}{\emph{Proceedings of the 29th ACM SIGKDD Conference on Knowledge Discovery and Data Mining}}. \bibinfo{pages}{5576--5585}.
\newblock


\bibitem[Zhang et~al\mbox{.}(2023e)]%
        {zhang2023constrained}
\bibfield{author}{\bibinfo{person}{Xingyi Zhang}, \bibinfo{person}{Shuliang Xu}, \bibinfo{person}{Wenqing Lin}, {and} \bibinfo{person}{Sibo Wang}.} \bibinfo{year}{2023}\natexlab{e}.
\newblock \showarticletitle{Constrained social community recommendation}. In \bibinfo{booktitle}{\emph{Proceedings of the 29th ACM SIGKDD conference on knowledge discovery and data mining}}. \bibinfo{pages}{5586--5596}.
\newblock


\bibitem[Zhang et~al\mbox{.}(2023a)]%
        {zhang2023leveraging}
\bibfield{author}{\bibinfo{person}{Yang Zhang}, \bibinfo{person}{Yimeng Bai}, \bibinfo{person}{Jianxin Chang}, \bibinfo{person}{Xiaoxue Zang}, \bibinfo{person}{Song Lu}, \bibinfo{person}{Jing Lu}, \bibinfo{person}{Fuli Feng}, \bibinfo{person}{Yanan Niu}, {and} \bibinfo{person}{Yang Song}.} \bibinfo{year}{2023}\natexlab{a}.
\newblock \showarticletitle{Leveraging watch-time feedback for short-video recommendations: A causal labeling framework}. In \bibinfo{booktitle}{\emph{Proceedings of the 32nd ACM International Conference on Information and Knowledge Management}}. \bibinfo{pages}{4952--4959}.
\newblock


\bibitem[Zhang et~al\mbox{.}(2024c)]%
        {zhang2024self}
\bibfield{author}{\bibinfo{person}{Yin Zhang}, \bibinfo{person}{Ruoxi Wang}, \bibinfo{person}{Xiang Li}, \bibinfo{person}{Tiansheng Yao}, \bibinfo{person}{Andrew Evdokimov}, \bibinfo{person}{Jonathan Valverde}, \bibinfo{person}{Yuan Gao}, \bibinfo{person}{Jerry Zhang}, \bibinfo{person}{Evan Ettinger}, \bibinfo{person}{Ed~H Chi}, {et~al\mbox{.}}} \bibinfo{year}{2024}\natexlab{c}.
\newblock \showarticletitle{Self-Auxiliary Distillation for Sample Efficient Learning in Google-Scale Recommenders}. In \bibinfo{booktitle}{\emph{Proceedings of the 18th ACM Conference on Recommender Systems}}. \bibinfo{pages}{829--831}.
\newblock


\bibitem[Zhang et~al\mbox{.}(2022b)]%
        {zhang2022scenario}
\bibfield{author}{\bibinfo{person}{Yuanliang Zhang}, \bibinfo{person}{Xiaofeng Wang}, \bibinfo{person}{Jinxin Hu}, \bibinfo{person}{Ke Gao}, \bibinfo{person}{Chenyi Lei}, {and} \bibinfo{person}{Fei Fang}.} \bibinfo{year}{2022}\natexlab{b}.
\newblock \showarticletitle{Scenario-adaptive and self-supervised model for multi-scenario personalized recommendation}. In \bibinfo{booktitle}{\emph{Proceedings of the 31st ACM International Conference on Information \& Knowledge Management}}. \bibinfo{pages}{3674--3683}.
\newblock


\bibitem[Zhang et~al\mbox{.}(2024d)]%
        {zhang2024unified}
\bibfield{author}{\bibinfo{person}{Yuting Zhang}, \bibinfo{person}{Yiqing Wu}, \bibinfo{person}{Ruidong Han}, \bibinfo{person}{Ying Sun}, \bibinfo{person}{Yongchun Zhu}, \bibinfo{person}{Xiang Li}, \bibinfo{person}{Wei Lin}, \bibinfo{person}{Fuzhen Zhuang}, \bibinfo{person}{Zhulin An}, {and} \bibinfo{person}{Yongjun Xu}.} \bibinfo{year}{2024}\natexlab{d}.
\newblock \showarticletitle{Unified Dual-Intent Translation for Joint Modeling of Search and Recommendation}. In \bibinfo{booktitle}{\emph{Proceedings of the 30th ACM SIGKDD Conference on Knowledge Discovery and Data Mining}}. \bibinfo{pages}{6291--6300}.
\newblock


\bibitem[Zhang et~al\mbox{.}(2023d)]%
        {zhang2023modeling}
\bibfield{author}{\bibinfo{person}{Yuting Zhang}, \bibinfo{person}{Yiqing Wu}, \bibinfo{person}{Ran Le}, \bibinfo{person}{Yongchun Zhu}, \bibinfo{person}{Fuzhen Zhuang}, \bibinfo{person}{Ruidong Han}, \bibinfo{person}{Xiang Li}, \bibinfo{person}{Wei Lin}, \bibinfo{person}{Zhulin An}, {and} \bibinfo{person}{Yongjun Xu}.} \bibinfo{year}{2023}\natexlab{d}.
\newblock \showarticletitle{Modeling dual period-varying preferences for takeaway recommendation}. In \bibinfo{booktitle}{\emph{Proceedings of the 29th ACM SIGKDD Conference on Knowledge Discovery and Data Mining}}. \bibinfo{pages}{5628--5638}.
\newblock


\bibitem[Zhang et~al\mbox{.}(2024a)]%
        {zhang2024co}
\bibfield{author}{\bibinfo{person}{Zhen Zhang}, \bibinfo{person}{Qingyun Liu}, \bibinfo{person}{Yuening Li}, \bibinfo{person}{Sourabh Bansod}, \bibinfo{person}{Mingyan Gao}, \bibinfo{person}{Yaping Zhang}, \bibinfo{person}{Zhe Zhao}, \bibinfo{person}{Lichan Hong}, \bibinfo{person}{Ed~H Chi}, \bibinfo{person}{Shuchao Bi}, {et~al\mbox{.}}} \bibinfo{year}{2024}\natexlab{a}.
\newblock \showarticletitle{Co-optimize Content Generation and Consumption in a Large Scale Video Recommendation System}. In \bibinfo{booktitle}{\emph{Proceedings of the 18th ACM Conference on Recommender Systems}}. \bibinfo{pages}{762--764}.
\newblock


\bibitem[Zhao et~al\mbox{.}(2023b)]%
        {zhao2023uncovering}
\bibfield{author}{\bibinfo{person}{Haiyuan Zhao}, \bibinfo{person}{Lei Zhang}, \bibinfo{person}{Jun Xu}, \bibinfo{person}{Guohao Cai}, \bibinfo{person}{Zhenhua Dong}, {and} \bibinfo{person}{Ji-Rong Wen}.} \bibinfo{year}{2023}\natexlab{b}.
\newblock \showarticletitle{Uncovering user interest from biased and noised watch time in video recommendation}. In \bibinfo{booktitle}{\emph{Proceedings of the 17th ACM Conference on Recommender Systems}}. \bibinfo{pages}{528--539}.
\newblock


\bibitem[Zhao et~al\mbox{.}(2023a)]%
        {zhao2023m5}
\bibfield{author}{\bibinfo{person}{Pengyu Zhao}, \bibinfo{person}{Xin Gao}, \bibinfo{person}{Chunxu Xu}, {and} \bibinfo{person}{Liang Chen}.} \bibinfo{year}{2023}\natexlab{a}.
\newblock \showarticletitle{M5: Multi-Modal Multi-Interest Multi-Scenario Matching for Over-the-Top Recommendation}. In \bibinfo{booktitle}{\emph{Proceedings of the 29th ACM SIGKDD Conference on Knowledge Discovery and Data Mining}}. \bibinfo{pages}{5650--5659}.
\newblock


\bibitem[Zhao et~al\mbox{.}(2024)]%
        {zhao2024breaking}
\bibfield{author}{\bibinfo{person}{Qian Zhao}, \bibinfo{person}{Hao Qian}, \bibinfo{person}{Ziqi Liu}, \bibinfo{person}{Gong-Duo Zhang}, {and} \bibinfo{person}{Lihong Gu}.} \bibinfo{year}{2024}\natexlab{}.
\newblock \showarticletitle{Breaking the barrier: utilizing large language models for industrial recommendation systems through an inferential knowledge graph}. In \bibinfo{booktitle}{\emph{Proceedings of the 33rd ACM International Conference on Information and Knowledge Management}}. \bibinfo{pages}{5086--5093}.
\newblock


\bibitem[Zhao et~al\mbox{.}(2022)]%
        {zhao2022improving}
\bibfield{author}{\bibinfo{person}{Xu Zhao}, \bibinfo{person}{Yi Ren}, \bibinfo{person}{Ying Du}, \bibinfo{person}{Shenzheng Zhang}, {and} \bibinfo{person}{Nian Wang}.} \bibinfo{year}{2022}\natexlab{}.
\newblock \showarticletitle{Improving item cold-start recommendation via model-agnostic conditional variational autoencoder}. In \bibinfo{booktitle}{\emph{Proceedings of the 45th International ACM SIGIR Conference on Research and Development in Information Retrieval}}. \bibinfo{pages}{2595--2600}.
\newblock


\bibitem[Zheng et~al\mbox{.}(2024)]%
        {zheng2024mirror}
\bibfield{author}{\bibinfo{person}{Zhi Zheng}, \bibinfo{person}{Xiao Hu}, \bibinfo{person}{Shanshan Gao}, \bibinfo{person}{Hengshu Zhu}, {and} \bibinfo{person}{Hui Xiong}.} \bibinfo{year}{2024}\natexlab{}.
\newblock \showarticletitle{Mirror: A multi-view reciprocal recommender system for online recruitment}. In \bibinfo{booktitle}{\emph{Proceedings of the 47th International ACM SIGIR Conference on Research and Development in Information Retrieval}}. \bibinfo{pages}{543--552}.
\newblock


\bibitem[Zhu et~al\mbox{.}(2022a)]%
        {10.1145/3511808.3557126}
\bibfield{author}{\bibinfo{person}{Fanwei Zhu}, \bibinfo{person}{Zulong Chen}, \bibinfo{person}{Fan Zhang}, \bibinfo{person}{Jiazhen Lou}, \bibinfo{person}{Hong Wen}, \bibinfo{person}{Shui Liu}, \bibinfo{person}{Qi Rao}, \bibinfo{person}{Tengfei Yuan}, \bibinfo{person}{Shenghua Ni}, \bibinfo{person}{Jinxin Hu}, \bibinfo{person}{Fuzhen Sun}, {and} \bibinfo{person}{Quan Lu}.} \bibinfo{year}{2022}\natexlab{a}.
\newblock \showarticletitle{SASNet: Stage-aware Sequential Matching for Online Travel Recommendation}. In \bibinfo{booktitle}{\emph{Proceedings of the 31st ACM International Conference on Information \& Knowledge Management}} (Atlanta, GA, USA) \emph{(\bibinfo{series}{CIKM '22})}. \bibinfo{publisher}{Association for Computing Machinery}, \bibinfo{address}{New York, NY, USA}, \bibinfo{pages}{3725–3734}.
\newblock
\showISBNx{9781450392365}
\href{https://doi.org/10.1145/3511808.3557126}{doi:\nolinkurl{10.1145/3511808.3557126}}


\bibitem[Zhu et~al\mbox{.}(2022b)]%
        {zhu2022spherical}
\bibfield{author}{\bibinfo{person}{Wenqiao Zhu}, \bibinfo{person}{Yesheng Xu}, \bibinfo{person}{Xin Huang}, \bibinfo{person}{Qiyang Min}, {and} \bibinfo{person}{Xun Zhou}.} \bibinfo{year}{2022}\natexlab{b}.
\newblock \showarticletitle{Spherical Graph Embedding for Item Retrieval in Recommendation System}. In \bibinfo{booktitle}{\emph{Proceedings of the 31st ACM International Conference on Information \& Knowledge Management}}. \bibinfo{pages}{4752--4756}.
\newblock


\bibitem[Zhu et~al\mbox{.}(2024)]%
        {zhu2024interest}
\bibfield{author}{\bibinfo{person}{Yongchun Zhu}, \bibinfo{person}{Jingwu Chen}, \bibinfo{person}{Ling Chen}, \bibinfo{person}{Yitan Li}, \bibinfo{person}{Feng Zhang}, {and} \bibinfo{person}{Zuotao Liu}.} \bibinfo{year}{2024}\natexlab{}.
\newblock \showarticletitle{Interest clock: Time perception in real-time streaming recommendation system}. In \bibinfo{booktitle}{\emph{Proceedings of the 47th International ACM SIGIR Conference on Research and Development in Information Retrieval}}. \bibinfo{pages}{2915--2919}.
\newblock


\bibitem[Zou et~al\mbox{.}(2024)]%
        {zou2024hesitation}
\bibfield{author}{\bibinfo{person}{Kuan Zou}, \bibinfo{person}{Aixin Sun}, \bibinfo{person}{Xuemeng Jiang}, \bibinfo{person}{Yitong Ji}, \bibinfo{person}{Hao Zhang}, \bibinfo{person}{Jing Wang}, {and} \bibinfo{person}{Ruijie Guo}.} \bibinfo{year}{2024}\natexlab{}.
\newblock \showarticletitle{Hesitation and Tolerance in Recommender Systems}.
\newblock \bibinfo{journal}{\emph{arXiv preprint arXiv:2412.09950}} (\bibinfo{year}{2024}).
\newblock


\end{thebibliography}



\end{document}